\documentclass[aps,prb,twocolumn,groupedaddress,showpacs,showkeys]{revtex4}

\usepackage{graphicx}  
\usepackage{dcolumn}   
\usepackage{bm}        
\usepackage{multirow}  
\usepackage{color}     
\usepackage{rotating}  
\usepackage{amsmath}
\usepackage{amssymb}
\usepackage{relsize}
\usepackage{braket}

\newcommand{\be}{\begin{equation}}
\newcommand{\ee}{\end{equation}}

\newcommand{\beq}{\begin{eqnarray}}
\newcommand{\eeq}{\end{eqnarray}}
\newcommand{\ba}{\begin{array}}
\newcommand{\ea}{\end{array}}

\def\H1{\widehat{H}_1}

\renewcommand{\i}{\ensuremath{\mathrm{i}}}
\newcommand{\e}{\ensuremath{\mathrm{e}}}
\renewcommand{\d}{\ensuremath{\mathrm{d}}}

\newcommand{\diag}{\mathrm{diag}}
\begin{document}

\title{Magnetic properties of HgTe quantum wells}
\author{Benedikt Scharf, Alex Matos-Abiague, and Jaroslav Fabian}
\affiliation{Institute for Theoretical Physics, University of Regensburg, 93040 Regensburg, Germany}
\date{\today}

\begin{abstract}
Using analytical formulas as well as a finite-difference scheme, we investigate the magnetic field dependence of the energy spectra and magnetic edge states of HgTe/CdTe-based quantum wells in the presence of perpendicular magnetic fields and hard walls, for the band-structure parameters corresponding to the normal and inverted regimes. Whereas one cannot find counterpropagating, spin-polarized states in the normal regime, below the crossover point between the uppermost (electron-like) valence and lowest (hole-like) conduction Landau levels, one can still observe such states at finite magnetic fields in the inverted regime, although these states are no longer protected by time-reversal symmetry. Furthermore, the bulk magnetization and susceptibility in HgTe quantum wells are studied, in particular their dependence on the magnetic field, chemical potential, and carrier densities. We find that for fixed chemical potentials as well as for fixed carrier densities, the magnetization and magnetic susceptibility in both the normal and the inverted regimes exhibit de Haas-van Alphen oscillations, whose amplitude decreases with increasing temperature. Moreover, if the band structure is inverted, the ground-state magnetization (and consequently also the ground-state susceptibility) is discontinuous at the crossover point between the uppermost valence and lowest conduction Landau levels. At finite temperatures and/or doping, this discontinuity is canceled by the contribution from the electrons and holes and the total magnetization and susceptibility are continuous. In the normal regime, this discontinuity of the ground-state magnetization does not arise and the magnetization is continuous for zero as well as finite temperatures.
\end{abstract}


\maketitle

\section{Introduction}\label{Sec:Intro}
In recent years, much attention has been devoted to the field of topological insulators, which are materials insulating in the bulk, but which possess dissipationless conducting states at their edge (two-dimensional topological insulators) or surface (three-dimensional topological insulators).\cite{Hasan2010:RMP,Qi2011:RMP} Since the introduction of the concept of two-dimensional topological insulators---often referred to as quantum spin Hall (QSH) insulators---and their first prediction in graphene,\cite{Kane2005:PRL,Kane2005:PRL2} several other systems have been proposed theoretically to exhibit QSH states, such as inverted HgTe/CdTe quantum-well structures,\cite{Bernevig2006:Science} GaAs under shear strain,\cite{Bernevig2006:PRL} two-dimensional bismuth,\cite{Murakami2006:PRL} or inverted InAs/GaSb/AlSb Type-II semiconductor quantum wells.\cite{Liu2008:PRL} Experimentally, the QSH state has first been observed in inverted HgTe quantum wells,\cite{Koenig2007:Science,Koenig2008:JPSJ,Buettner2011:NatPhys,Bruene2012:NatPhys} where one can tune the band structure by fabricating quantum wells with different thicknesses.
\cite{Liu2008:PRL}
Similarly to the quantum Hall (QH) state, which can be characterized by Chern numbers,\cite{Thouless1982:PRL,Kohmoto1985:AoP} the QSH state can also be described by a topological invariant, in this case the $Z_2$ invariant.\cite{Kane2005:PRL,Fu2007:PRB} This invariant describes whether one deals with a trivial insulator, that is, an insulator without edge states protected by time-reversal symmetry, or a QSH insulator. One of the most prominent features of QSH insulators is the existence of dissipationless helical edge states, that is, edge states whose spin orientation is determined by the direction of the electron momentum and are protected from backscattering.\cite{Wu2006:PRL,Xu2006:PRB} Thus, at a given edge, one can find a pair of counterpropagating, spin-polarized edge states, a fact whose experimental verification has only very recently been reported.\cite{Bruene2012:NatPhys} Since those counterpropagating, spin-polarized edge states are robust against time-reversal invariant perturbations such as scattering by nonmagnetic impurities, they are promising for applications within the field of spintronics,\cite{Zutic2004:RMP,Fabian2007:APS} the central theme of which is the generation and control of nonequilibrium electron spin in solids.

At the center of the QSH state are relativistic corrections, which can---if strong enough---lead to band inversion, that is, a situation where the normal order of the conduction and valence bands is inverted.\cite{Chadi1972:PRB,Zhu2012:PRB} By fabricating HgTe quantum wells with a thickness larger than the critical thickness $d_c\approx6.3$ nm, such an inverted band structure can be created in HgTe/CdTe quantum-well structures. In fact, materials with band inversion have been studied for some time\cite{Dyakonov1981:JETPL} and another interesting prediction---different from the QSH state---has been that the combination of two materials with mutually inverted band structures can lead to the formation of interface states which---depending on the material parameters---can possess a linear two-dimensional spectrum.\cite{Volkov1985:JETPL,Pankratov1987:SolidStateCommunications}

Following the observation of the QSH state in HgTe-based quantum wells, much effort has been invested in the theoretical investigation of the properties of two-dimensional topological insulators, their edge states, and possible applications. Examples include the extension of the low-energy Hamiltonian introduced in Ref.~\onlinecite{Bernevig2006:Science} to account for additional spin-orbit terms due to out-of-plane inversion breaking in HgTe quantum wells\cite{Rothe2010:NJoP} as well as studies on how helical edge states and bulk states interact in two-dimensional topological insulators.\cite{Reinthaler2012:PRB} The effect of magnetic fields on transport in inverted HgTe quantum wells has been treated in Refs.~\onlinecite{Tkachov2010:PRL,Tkachov2012:PhysicaE,Chen2012:PRB}. Moreover, the effect of finite sizes on the QSH edge states in HgTe quantum wells has been investigated and it has been shown that for small widths the edge states of the opposite sides in a finite system can overlap and produce a gap in the spectrum.\cite{Zhou2008:PRL} Based on this coupling of the wave functions from opposite edges, a spin transistor based on a constriction made of HgTe has been proposed.\cite{Krueckl2011:PRL} Finite-size effects in topological insulators have not only been studied for HgTe, but also in three-dimensional topological insulators, in particular the crossover to QSH insulators in thin films.\cite{Linder2009:PRB,Liu2010:PRB,Lu2010:PRB}

Our purpose is to present a systematic study of the effect a perpendicular magnetic field has on the energy spectrum and magnetic edge states of HgTe/CdTe quantum wells (as described by the Hamiltonian introduced in Ref.~\onlinecite{Bernevig2006:Science}) in the normal as well as in the inverted regime. In particular, we present an analytical solution for the magnetic edge states confined by a hard-wall potential in the spirit of Refs.~\onlinecite{Grigoryan2009:PRB,Badalyan2009:PRL}, where the problem of spin edge states and magnetic spin edge states in two-dimensional electron gases with hard walls and spin-orbit coupling has been solved analytically. Complementary to this procedure, we also make use of a numerical scheme based on the method of finite differences. Furthermore, the magnetic properties of HgTe quantum wells are investigated within this model, again for both, the normal and inverted regimes.

The manuscript is organized as follows: Section~\ref{Sec:Model} gives a short overview of the effective model used to describe the HgTe quantum well. In Sec.~\ref{Sec:MES}, following the presentation of two methods to calculate the energy spectrum and eigenstates, an analytical and a finite-differences method, the evolution of QSH and QH states with increasing magnetic fields is discussed. The second part of the manuscript, Sec.~\ref{Sec:Mag}, is devoted to the discussion of the magnetic properties of this system. Finally, the manuscript is concluded by a brief summary.

\section{Model}\label{Sec:Model}
Our model is based on the two-dimensional effective Hamiltonian of HgTe/CdTe quantum wells derived from the Kane model by Bernevig \emph{et al}.\cite{Bernevig2006:Science} This effective $4\times4$ Hamiltonian captures the essential physics in HgTe/CdTe quantum wells at low energies and describes the spin-degenerate electron-like ($E$) and heavy hole-like ($H$) states $\ket{E\uparrow}$, $\ket{H\uparrow}$, $\ket{E\downarrow}$, and $\ket{H\downarrow}$ near the $\Gamma$ point. The effect of a magnetic field $\mathbf{B}(\mathbf{r})$ can be included in this model by adding a Zeeman term\cite{Buettner2011:NatPhys} and promoting the components of the wave vector to operators, that is, $k_i\rightarrow\hat{\pi}_i/\hbar$, where $i$ denotes the in-plane coordinates $x$ or $y$ of the quantum well, $\hat{\pi}_i=\hat{p}_i+eA_i(\mathbf{r})$ the kinetic momentum operators, $\hat{p}_i$ the momentum operators, $\mathbf{A}(\mathbf{r})$ the magnetic vector potential, and $e=|e|$ the elementary charge.

In our model, we consider a constant magnetic field perpendicular to the quantum well, that is, $\mathbf{B}=B\mathbf{e}_z$ with $B>0$ (throughout this manuscript). Since hard walls will be added in Secs.~\ref{Sec:MES_analytical} and~\ref{Sec:MES_FD} to confine the system in the $y$-direction, it is convenient to choose the gauge
\begin{equation}\label{gauge}
\mathbf{A}(\mathbf{r})=-By\mathbf{e}_x,
\end{equation}
for which the effective Hamiltonian reads as
\begin{equation}\label{effective_Hamiltonian}
\begin{aligned}
\hat{H}=&\mathcal{C}\mathbf{1}+\mathcal{M}\Gamma_5-\frac{\mathcal{D}\mathbf{1}+\mathcal{B}\Gamma_5}{\hbar^2}\left[\left(\hat{p}_x-\frac{\hbar y}{l_{\mathsmaller{B}}^2}\right)^2+\hat{p}_y^2\right]\\
&+\frac{\mathcal{A}\Gamma_1}{\hbar}\left(\hat{p}_x-\frac{\hbar y}{l_{\mathsmaller{B}}^2}\right)+\frac{\mathcal{A}\Gamma_2}{\hbar}\hat{p}_y+\frac{\mu_{\mathsmaller{B}}B\Gamma^z_g}{2},
\end{aligned}
\end{equation}
with the system parameters $\mathcal{A}$, $\mathcal{B}$, $\mathcal{C}$, $\mathcal{D}$, and $\mathcal{M}$, the magnetic length $l_{\mathsmaller{B}}=\sqrt{\hbar/e|B|}=\sqrt{\hbar/eB}$, the Bohr magneton $\mu_{\mathsmaller{B}}$, and the $4\times4$ unity matrix $\mathbf{1}$. For the basis order $\ket{E\uparrow}$, $\ket{H\uparrow}$, $\ket{E\downarrow}$, $\ket{H\downarrow}$, the remaining $4\times4$ matrices are given by
\begin{equation}\label{matrices}
\begin{aligned}
\Gamma_1=\left(\begin{array}{cc}
         \sigma_x & 0 \\
         0 & -\sigma_x \\
         \end{array}\right),
\Gamma_2=\left(\begin{array}{cc}
         -\sigma_y & 0 \\
         0 & -\sigma_y \\
         \end{array}\right),\\
\Gamma_5=\left(\begin{array}{cc}
         \sigma_z & 0 \\
         0 & \sigma_z \\
         \end{array}\right),
\Gamma^z_g=\left(\begin{array}{cc}
         \sigma_g & 0 \\
         0 & -\sigma_g \\
         \end{array}\right),
\end{aligned}
\end{equation}
where $\sigma_x$, $\sigma_y$, and $\sigma_z$ denote the Pauli matrices and $\sigma_g=\diag(g_{\mathsmaller{\mathrm{e}}},g_{\mathsmaller{\mathrm{h}}})$ contains the effective (out-of-plane) g-factors $g_{\mathsmaller{\mathrm{e}}}$ and $g_{\mathsmaller{\mathrm{h}}}$ of the $E$ and $H$ bands, respectively.

The material parameters introduced above, $\mathcal{A}$, $\mathcal{B}$, $\mathcal{C}$, $\mathcal{D}$, and $\mathcal{M}$, are expansion parameters that, like $g_{\mathsmaller{\mathrm{e}}}$ and $g_{\mathsmaller{\mathrm{h}}}$, depend on the quantum-well thickness $d$.\cite{Bernevig2006:Science,Koenig2008:JPSJ} Thus, the quantum-well thickness can be used to tune the band structure. Here, $\mathcal{A}$ describes the coupling between the electron-like and hole-like bands, $\mathcal{C}$ and $\mathcal{D}$ describe a standard parabolic dispersion of all bands, whereas $\mathcal{M}$ and $\mathcal{B}$ determine whether the band structure is inverted or not: If the thickness of the quantum well is smaller than the critical thickness, $d_c\approx6.3$ nm, the band structure is normal and $\mathcal{M}/\mathcal{B}<0$, while, for a quantum-well thickness above $d_c$, the band structure is inverted and $\mathcal{M}/\mathcal{B}>0$.

In some cases, a reduced form of Eq.~(\ref{effective_Hamiltonian}) can be used. For relatively strong magnetic fields, the terms quadratic with the kinetic momentum in Eq.~(\ref{effective_Hamiltonian}) are small near the $\Gamma$ point and can be omitted, as can the contribution from the Zeeman term, that is, $\mathcal{B}=\mathcal{D}=0$ and $g_\mathsmaller{\mathrm{e/h}}=0$.\cite{Schmidt2009:PRB,Tkachov2010:PRL}

\section{Magnetic edge states}\label{Sec:MES}
\subsection{Analytical solution}\label{Sec:MES_analytical}
In this section, we discuss the analytical solution---which in many ways resembles the calculation of the spin edge states in two-dimensional electron gases with spin-orbit coupling\cite{Grigoryan2009:PRB}---of the model system described by Eq.~(\ref{effective_Hamiltonian}) for several different geometries: (i) bulk, that is, an infinite system, (ii) a semi-infinite system confined to $y>0$, and (iii) a finite strip with the width $w$ in $y$-direction. For all these cases, we apply periodic boundary conditions in $x$-direction. The confinement can be described by adding the infinite hard-wall potentials
\begin{equation}\label{confining_potential_semiinfinite}
V(y)=\left\{\begin{array}{ll}
            0 & \mathrm{for}\quad y>0\\
	    \infty & \mathrm{elsewhere}
            \end{array}\right.
\end{equation}
in (ii) and
\begin{equation}\label{confining_potential_finite_strip}
V(y)=\left\{\begin{array}{ll}
            0 & \mathrm{for}\quad \left|y\right|<w/2\\
	    \infty & \mathrm{elsewhere}
            \end{array}\right.
\end{equation}
in (iii).

In order to determine the solutions for cases (i)-(iii), we first need to find the general solution to the differential equation given by the free Schr\"{o}dinger equation
\begin{equation}\label{SG}
\hat{H}\Psi(x,y)=E\Psi(x,y),
\end{equation}
where $\Psi(x,y)$ is a four-component spinor. By imposing the appropriate boundary conditions along the $y$-direction on this general solution, we can obtain the solutions for each of the cases considered. Since translational invariance along the $x$-direction as well as the spin direction are preserved by $\hat{H}$ and $\hat{H}+V(y)\mathbf{1}$, respectively, the wave vector in $x$-direction, $k$, and the spin orientation, $s=\uparrow/\downarrow$, are good quantum numbers in each of the three cases, which naturally suggests the ansatz
\begin{equation}\label{general_ansatz}
\Psi^\uparrow_k(x,y)=\frac{\e^{\i kx}}{\sqrt{L}}\left(\begin{array}{c} f_\uparrow(\xi)\\ g_\uparrow(\xi)\\ 0\\ 0\\ \end{array}\right),\;
\Psi^\downarrow_k(x,y)=\frac{\e^{\i kx}}{\sqrt{L}}\left(\begin{array}{c} 0\\ 0\\ f_\downarrow(\xi)\\ g_\downarrow(\xi)\\ \end{array}\right),
\end{equation}
where $L$ is the length of the strip in $x$-direction and where, for convenience, we have introduced the transformation $\xi=\xi(y)=\sqrt{2}\left(y-l_{\mathsmaller{B}}^2k\right)/l_{\mathsmaller{B}}$.

Inserting the ansatz~(\ref{general_ansatz}) for spin-up electrons into Eq.~(\ref{SG}), we obtain the following system of differential equations:
\begin{equation}\label{effective_1D_SE_up}
\begin{aligned}
\left[\mathcal{C}-E-\frac{2\mathcal{D}}{l_{\mathsmaller{B}}^2}\left(\frac{\xi^2}{4}-\partial_\xi^2\right)\right]\left(\begin{array}{c}f_\uparrow(\xi)\\ g_\uparrow(\xi)\end{array}\right) & \\ +\left[\mathcal{M}-\frac{2\mathcal{B}}{l_{\mathsmaller{B}}^2}\left(\frac{\xi^2}{4}-\partial_\xi^2\right)\right]\left(\begin{array}{c}f_\uparrow(\xi)\\ -g_\uparrow(\xi)\end{array}\right) & \\
-\frac{\sqrt{2}\mathcal{A}}{l_{\mathsmaller{B}}}\left(\begin{array}{c}\left(\frac{\xi}{2}-\partial_\xi\right)g_\uparrow(\xi)\\ \left(\frac{\xi}{2}+\partial_\xi\right)f_\uparrow(\xi)\end{array}\right)+\frac{\mu_{\mathsmaller{B}}B}{2}\left(\begin{array}{c}g_{\mathsmaller{\mathrm{e}}}f_\uparrow(\xi)\\ g_{\mathsmaller{\mathrm{h}}}g_\uparrow(\xi)\end{array}\right) & =0.
\end{aligned}
\end{equation}

Due to the specific form of Eq.~(\ref{effective_1D_SE_up}), its solution can be conveniently written in terms of the parabolic cylindrical functions $D_\nu(\xi)$, which satisfy the following recurrence relations:\cite{Olver2010}
\begin{equation}\label{reccurence_1}
\left(\frac{\xi}{2}\pm\partial_\xi\right)D_\nu(\xi)=\left\{\begin{array}{c}\nu D_{\nu-1}(\xi)\\D_{\nu+1}(\xi)\end{array}\right.,
\end{equation}
\begin{equation}\label{reccurence_2}
\left(\frac{\xi^2}{4}-\partial_\xi^2\right)D_\nu(\xi)=\left(\nu+\frac{1}{2}\right)D_\nu(\xi).
\end{equation}
With the heavy hole-like component $g_\uparrow(\xi)$ coupled to the electron-like component $f_\uparrow(\xi)$ by the raising operator and the opposite coupling described by the lowering operator, one type of solution is of the form
\begin{equation}\label{ansatz_xi}
f_\uparrow(\xi)=v_1D_\nu(\xi)\quad\text{and}\quad g_\uparrow(\xi)=v_2D_{\nu-1}(\xi),
\end{equation}
where $v_1$ and $v_2$ are complex numbers, which are to be determined by solving the system of linear equations obtained from inserting this ansatz into Eq.~(\ref{effective_1D_SE_up}). This system has non-trivial solutions for
\begin{equation}\label{index_up}
\nu=\nu_{\pm}^\uparrow=\frac{l_{\mathsmaller{B}}^2}{2}\left[F(1)\pm\sqrt{F^2(1)+\frac{G_{\mathsmaller{\mathrm{e}}}(1)G_{\mathsmaller{\mathrm{h}}}(1)}{\mathcal{B}^2-\mathcal{D}^2}}\right],
\end{equation}
where
\begin{equation}\label{func_F}
\begin{aligned}
F\left(s\right)=&s\,\frac{\mu_{\mathsmaller{B}}B}{4}\left(\frac{g_{\mathsmaller{\mathrm{e}}}}{\mathcal{D}+\mathcal{B}}+\frac{g_{\mathsmaller{\mathrm{h}}}}{\mathcal{D}-\mathcal{B}}\right)\\
&-\frac{\mathcal{A}^2-2\left[\mathcal{M}\mathcal{B}+\mathcal{D}\left(E-\mathcal{C}\right)\right]}{2\left(\mathcal{B}^2-\mathcal{D}^2\right)}
\end{aligned}
\end{equation}
and
\begin{equation}\label{func_G}
\begin{aligned}
G_{\mathsmaller{\mathrm{e/h}}}\left(s\right)=s\left(\frac{g_{\mathsmaller{\mathrm{e/h}}}\mu_{\mathsmaller{B}}B}{2}-\frac{\mathcal{B}\pm\mathcal{D}}{l_{\mathsmaller{B}}^2}\right)-\left(E-\mathcal{C}\right)\pm\mathcal{M}.
\end{aligned}
\end{equation}
By determining those non-trivial solutions, for $\mathcal{A}\neq0$ we find the two (non-normalized) solutions
\begin{equation}\label{spin_up_solution_xi}
\chi_{\pm}^\uparrow(\xi)=\left(\sqrt{2}\mathcal{A}D_{\nu_{\pm}^\uparrow}(\xi)/l_{\mathsmaller{B}},c_{\pm}^\uparrow D_{\nu_{\pm}^\uparrow-1}(\xi)\right)^T
\end{equation}
to Eq.~(\ref{effective_1D_SE_up}) with
\begin{equation}\label{coefficient_up}
c_{\pm}^\uparrow=\mathcal{M}-\left(E-\mathcal{C}\right)-\frac{2\left(\mathcal{B}+\mathcal{D}\right)}{l_{\mathsmaller{B}}^2}\left(\nu_{\pm}^\uparrow+\frac{1}{2}\right)+\frac{g_{\mathsmaller{\mathrm{e}}}}{2}\mu_{\mathsmaller{B}}B.
\end{equation}
However, there is a second set of---in general---independent solutions to Eq.~(\ref{effective_1D_SE_up}) that can be obtained from the ansatz 
\begin{equation}\label{ansatz_-xi}
f_\uparrow(\xi)=u_1D_\nu(-\xi)\quad\text{and}\quad g_\uparrow(\xi)=u_2D_{\nu-1}(-\xi),
\end{equation}
where $u_1$ and $u_2$ are complex numbers as before. With this ansatz yielding two further solutions,
\begin{equation}\label{spin_up_solution_-xi}
\eta_{\pm}^\uparrow(\xi)=\left(\sqrt{2}\mathcal{A}D_{\nu_{\pm}^\uparrow}(-\xi)/l_{\mathsmaller{B}},-c_{\pm}^\uparrow D_{\nu_{\pm}^\uparrow-1}(-\xi)\right)^T,
\end{equation}
the general solution to Eq.~(\ref{effective_1D_SE_up})---if $\mathcal{A}\neq0$---is given by
\begin{equation}\label{general_solution_up}
\left(\begin{array}{c}f_\uparrow(\xi)\\ g_\uparrow(\xi)\end{array}\right)=\alpha\,\chi_{+}^\uparrow(\xi)+\beta\,\chi_{-}^\uparrow(\xi)+\gamma\,\eta_{+}^\uparrow(\xi)+\delta\,\eta_{-}^\uparrow(\xi),
\end{equation}
where the coefficients $\alpha$, $\beta$, $\gamma$, and $\delta$ are complex numbers to be determined by the boundary conditions of the problem.

A procedure similar to the one above can also be applied for the spin-down electrons in Eq.~(\ref{general_ansatz}). Then, we find \begin{equation}\label{general_solution_down}
\left(\begin{array}{c}f_\downarrow(\xi)\\ g_\downarrow(\xi)\end{array}\right)=\tilde{\alpha}\,\chi_{+}^\downarrow(\xi)+\tilde{\beta}\,\chi_{-}^\downarrow(\xi)+\tilde{\gamma}\,\eta_{+}^\downarrow(\xi)+\tilde{\delta}\,\eta_{-}^\downarrow(\xi),
\end{equation}
where we have introduced the vectors
\begin{equation}\label{spin_down_solution_xi}
\chi_{\pm}^\downarrow(\xi)=\left(c_{\pm}^\downarrow D_{\nu_{\pm}^\downarrow-1}(\xi),\sqrt{2}\mathcal{A}D_{\nu_{\pm}^\downarrow}(\xi)/l_{\mathsmaller{B}}\right)^T
\end{equation}
and
\begin{equation}\label{spin_down_solution_-xi}
\eta_{\pm}^\downarrow(\xi)=\left(-c_{\pm}^\downarrow D_{\nu_{\pm}^\downarrow-1}(-\xi),\sqrt{2}\mathcal{A}D_{\nu_{\pm}^\downarrow}(-\xi)/l_{\mathsmaller{B}}\right)^T,
\end{equation}
with
\begin{equation}\label{index_down}
\nu_{\pm}^\downarrow=\frac{l_{\mathsmaller{B}}^2}{2}\left[F(-1)\pm\sqrt{F^2(-1)+\frac{G_{\mathsmaller{\mathrm{e}}}(-1)G_{\mathsmaller{\mathrm{h}}}(-1)}{\mathcal{B}^2-\mathcal{D}^2}}\right]
\end{equation}
and
\begin{equation}\label{coefficient_down}
c_{\pm}^\downarrow=\mathcal{M}+\left(E-\mathcal{C}\right)-\frac{2\left(\mathcal{B}-\mathcal{D}\right)}{l_{\mathsmaller{B}}^2}\left(\nu_{\pm}^\downarrow+\frac{1}{2}\right)+\frac{g_{\mathsmaller{\mathrm{h}}}}{2}\mu_{\mathsmaller{B}}B.
\end{equation}
As in the case of spin-up electrons, the coefficients $\tilde{\alpha}$, $\tilde{\beta}$, $\tilde{\gamma}$, and $\tilde{\delta}$ need to be fixed by boundary conditions. In the following, we will use the general solutions given by Eqs.~(\ref{general_solution_up}) and~(\ref{general_solution_down}) to determine the energy spectrum and wave functions for several different geometries.

\noindent\emph{(i) Bulk.}

If there is no confining potential $V(y)$, that is, if we consider an infinite system, where Eq.~(\ref{effective_1D_SE_up}) holds for any $\xi\in\mathbb{R}$, we only have to require the wave function to be normalizable and accordingly we impose the boundary conditions $\lim\limits_{\xi\to\pm\infty}f_\uparrow(\xi)=\lim\limits_{\xi\to\pm\infty}g_\uparrow(\xi)=0$. These requirements can only be satisfied if $\nu$ is a non-negative integer $n$ in Eq.~(\ref{ansatz_xi}). In this case, $D_n(\xi)=2^{-n/2}\e^{-\xi^2/4}H_n(\xi/\sqrt{2})$ can be expressed by Hermite polynomials $H_n(\xi)$,\cite{Olver2010} and both Eqs.~(\ref{ansatz_xi}) and~(\ref{ansatz_-xi}) lead to the same solution. If $n\geq1$, the ansatz from Eq.~(\ref{ansatz_xi}) leads to an eigenvalue problem for $E$ from which the following Landau levels for spin-up electrons can be determined:
\begin{equation}\label{Landau_levels_up}
\begin{aligned}
E_{\pm}^\uparrow(n)&=\mathcal{C}-\frac{2\mathcal{D}n+\mathcal{B}}{l_{\mathsmaller{B}}^2}+\frac{g_{\mathsmaller{\mathrm{e}}}+g_{\mathsmaller{\mathrm{h}}}}{4}\mu_{\mathsmaller{B}}B\\
&\pm\sqrt{\frac{2n\mathcal{A}^2}{l_{\mathsmaller{B}}^2}+\left(\mathcal{M}-\frac{2\mathcal{B}n+\mathcal{D}}{l_{\mathsmaller{B}}^2}+\frac{g_{\mathsmaller{\mathrm{e}}}-g_{\mathsmaller{\mathrm{h}}}}{4}\mu_{\mathsmaller{B}}B\right)^2}.
\end{aligned}
\end{equation}
For $n=0$, on the other hand, Eqs.~(\ref{ansatz_xi}) and~(\ref{ansatz_-xi}) reduce to the ansatz $f_\uparrow(\xi)=v_1D_0(\xi)$ and $g_\uparrow(\xi)=0$ and we obtain the Landau level
\begin{equation}\label{zero_Landau_level_up}
E^\uparrow(0)=\mathcal{C}+\mathcal{M}-\frac{\mathcal{D}+\mathcal{B}}{l_{\mathsmaller{B}}^2}+\frac{g_{\mathsmaller{\mathrm{e}}}}{2}\mu_{\mathsmaller{B}}B.
\end{equation}

By requiring $\lim\limits_{\xi\to\pm\infty}f_\downarrow(\xi)=\lim\limits_{\xi\to\pm\infty}g_\downarrow(\xi)=0$, the Landau levels for spin-down electrons can be calculated similarly as
\begin{equation}\label{Landau_levels_down}
\begin{aligned}
E_{\pm}^\downarrow(n)&=\mathcal{C}-\frac{2\mathcal{D}n-\mathcal{B}}{l_{\mathsmaller{B}}^2}-\frac{g_{\mathsmaller{\mathrm{e}}}+g_{\mathsmaller{\mathrm{h}}}}{4}\mu_{\mathsmaller{B}}B\\
&\pm\sqrt{\frac{2n\mathcal{A}^2}{l_{\mathsmaller{B}}^2}+\left(\mathcal{M}-\frac{2\mathcal{B}n-\mathcal{D}}{l_{\mathsmaller{B}}^2}-\frac{g_{\mathsmaller{\mathrm{e}}}-g_{\mathsmaller{\mathrm{h}}}}{4}\mu_{\mathsmaller{B}}B\right)^2}
\end{aligned}
\end{equation}
and
\begin{equation}\label{zero_Landau_level_down}
E^\downarrow(0)=\mathcal{C}-\mathcal{M}-\frac{\mathcal{D}-\mathcal{B}}{l_{\mathsmaller{B}}^2}-\frac{g_{\mathsmaller{\mathrm{h}}}}{2}\mu_{\mathsmaller{B}}B.
\end{equation}
With Eqs.~(\ref{Landau_levels_up})-(\ref{zero_Landau_level_down}), we have recovered the Landau levels found in Ref.~\onlinecite{Buettner2011:NatPhys}. The corresponding eigenstates are given in the Appendix~\ref{Sec:AppendixLLstates}.

In writing down Eqs.~(\ref{Landau_levels_up})-(\ref{zero_Landau_level_down}), we have adopted the convention that $B>0$, that is, the magnetic field points in the $z$-direction. The formulas of the Landau levels for $B<0$ can be obtained from Eqs.~(\ref{Landau_levels_up})-(\ref{zero_Landau_level_down}) via the relations $E^{s}(0,B)=E^{-s}(0,-B)$ and $E_{\pm}^{s}(n,B)=E_{\pm}^{-s}(n,-B)$ [note that the magnetic length in Eqs.~(\ref{Landau_levels_up})-(\ref{zero_Landau_level_down}) is given by $l_{\mathsmaller{B}}=\sqrt{\hbar/e|B|}$].

\noindent\emph{(ii) Semi-infinite system.}

In the presence of the confining potential given by Eq.~(\ref{confining_potential_semiinfinite}), the wave function is required to vanish at the boundary $y=0$ as well as at $y\to\infty$. Thus, we invoke the boundary conditions $\lim\limits_{\xi\to\infty}f_{\uparrow/\downarrow}(\xi)=\lim\limits_{\xi\to\infty}g_{\uparrow/\downarrow}(\xi)=0$ and $f_{\uparrow/\downarrow}(\xi_0)=g_{\uparrow/\downarrow}(\xi_0)=0$ for spin-up as well as spin-down electrons, where $\xi_0=-\sqrt{2}l_{\mathsmaller{B}}k$. The condition for $\xi\to\infty$ can only be satisfied for $\gamma=\delta=0$ and $\tilde{\gamma}=\tilde{\delta}=0$, respectively. Then, each remaining pair of coefficients, $\alpha$ and $\beta$ as well as $\tilde{\alpha}$ and $\tilde{\beta}$, from Eqs.~(\ref{general_solution_up}) and~(\ref{general_solution_down}) has to be calculated from the condition at $y=0$, that is, at $\xi_0$. The resulting linear systems of equations have non-trivial solutions if
\begin{equation}\label{si_trans}
\begin{aligned} 
c_{-}^{\uparrow/\downarrow}D_{\nu_{-}^{\uparrow/\downarrow}-1}(\xi_0)D_{\nu_{+}^{\uparrow/\downarrow}}(\xi_0)&\\
-c_{+}^{\uparrow/\downarrow}D_{\nu_{+}^{\uparrow/\downarrow}-1}(\xi_0)D_{\nu_{-}^{\uparrow/\downarrow}}(\xi_0)&=0.
\end{aligned}
\end{equation}
This transcendental equation enables us to calculate the electron dispersion for spin-up [$s=\uparrow$ in Eq.~(\ref{si_trans})] as well as for spin-down electrons [$s=\downarrow$ in Eq.~(\ref{si_trans})]. The corresponding eigenstates can be determined by explicitly calculating the coefficients $\alpha$, $\beta$ and $\tilde{\alpha},\tilde{\beta}$, respectively.

\noindent\emph{(iii) Finite-strip geometry.}

In the finite-strip geometry described by Eq.~(\ref{confining_potential_finite_strip}), the wave function has to vanish at the potential boundaries, that is, Eqs.~(\ref{general_solution_up}) and~(\ref{general_solution_down}) have to vanish at $\xi_{1/2}=\sqrt{2}\left(\mp w/2-l_{\mathsmaller{B}}^2k\right)/l_{\mathsmaller{B}}$. The corresponding linear systems of equations defined by this condition have non-trivial solutions if
\begin{equation}\label{det}
\det\left(\begin{array}{ccccc}\chi_{+}^{\uparrow/\downarrow}(\xi_1) & \chi_{-}^{\uparrow/\downarrow}(\xi_1) & & \eta_{+}^{\uparrow/\downarrow}(\xi_1) & \eta_{-}^{\uparrow/\downarrow}(\xi_1)\\
\\
\chi_{+}^{\uparrow/\downarrow}(\xi_2) & \chi_{-}^{\uparrow/\downarrow}(\xi_2) & & \eta_{+}^{\uparrow/\downarrow}(\xi_2) & \eta_{-}^{\uparrow/\downarrow}(\xi_2)
\end{array}\right)=0
\end{equation}
for spin-up ($s=\uparrow$) and spin-down ($s=\downarrow$) electrons, respectively. Similarly to (ii), the transcendental Eq.~(\ref{det}) represents exact expressions from which the dispersion of the electrons can be calculated. The corresponding eigenstates can be determined by explicitly calculating the coefficients $\alpha$, $\beta$, $\gamma$, and $\delta$ for spin-up electrons and $\tilde{\alpha}$, $\tilde{\beta}$, $\tilde{\gamma}$, and $\tilde{\delta}$ for spin-down electrons, respectively.

Having derived transcendental equations from which the electronic dispersion (and indirectly the eigenstates) can be determined for semi-infinite as well as finite-strip systems, we will also introduce an alternative method to calculate the spectrum and eigenstates of a finite strip.

\subsection{Numerical finite-difference solution}\label{Sec:MES_FD}
In addition to solving the exact expression~(\ref{det}), we calculate the eigenspectrum and eigenstates also by using a finite-difference scheme to express Eq.~(\ref{effective_Hamiltonian}).\cite{Datta2007} We discretize Eq.~(\ref{effective_Hamiltonian}) for $B=0$ and account for the magnetic field by introducing the Peierls' phase\cite{Peierls1933:ZfP} to describe the vector potential given by Eq.~(\ref{gauge}) and an additional on-site term to describe the Zeeman term. If only nearest neighbors are considered and there is no magnetic field, this procedure leads to the Hamiltonian introduced in Ref.~\onlinecite{Koenig2008:JPSJ}.

For reasons of improving the convergence of our calculation, we go beyond the nearest-neighbor approximation and include the next-nearest neighbors. Due to translational invariance along the $x$-direction, the $x$-coordinate can be Fourier transformed to the reciprocal space and we obtain the Hamiltonian
\begin{equation}\label{FD_Hamiltonian}
\hat{H}_{\mathrm{FD}}=\sum\limits_{k,n,n'}\sum\limits_{\alpha\beta}\mathcal{H}_{\alpha\beta}(k;n,n')\hat{c}_{kn\alpha}^\dagger\hat{c}_{kn'\beta},
\end{equation}
where $k$ is the momentum along the $x$-direction, $n$ and $n'\in\mathbb{Z}$ are discrete $y$-coordinates, $\alpha$ and $\beta$ denote the basis states $\ket{E\uparrow}$, $\ket{H\uparrow}$, $\ket{E\downarrow}$, $\ket{H\downarrow}$, and $\hat{c}_{kn\alpha}^\dagger$ ($\hat{c}_{kn\alpha}$) is the creation (annihilation) operator of those states. Furthermore, we have introduced the matrix
\begin{widetext}
\begin{equation}\label{FD_Hamiltonian_matrix}
\begin{aligned}
\mathcal{H}_{\alpha\beta}(k;n,n')=&\left[\mathcal{C}\left(\mathbf{1}\right)_{\alpha\beta}+\mathcal{M}\left(\Gamma_5\right)_{\alpha\beta}-\frac{\mathcal{D}\left(\mathbf{1}\right)_{\alpha\beta}+\mathcal{B}\left(\Gamma_5\right)_{\alpha\beta}}{a^2}\mathcal{F}(k,B,n)+\frac{\mathcal{A}}{a}\left(\Gamma_1\right)_{\alpha\beta}\mathcal{G}(k,B,n)+\frac{\mu_{\mathsmaller{B}}B}{2}\left(\Gamma_g\right)_{\alpha\beta}\right]\delta_{nn'}\\
&+\left\{\frac{4\left[\mathcal{D}\left(\mathbf{1}\right)_{\alpha\beta}+\mathcal{B}\left(\Gamma_5\right)_{\alpha\beta}\right]}{3a^2}+\frac{2\i\mathcal{A}\left(n-n'\right)}{3a}\left(\Gamma_2\right)_{\alpha\beta}\right\}\left(\delta_{n,n'+1}+\delta_{n,n'-1}\right)\\
&-\left[\frac{\mathcal{D}\left(\mathbf{1}\right)_{\alpha\beta}+\mathcal{B}\left(\Gamma_5\right)_{\alpha\beta}}{12a^2}+\frac{\i\mathcal{A}\left(n-n'\right)}{24a}\left(\Gamma_2\right)_{\alpha\beta}\right]\left(\delta_{n,n'+2}+\delta_{n,n'-2}\right),
\end{aligned}
\end{equation}
\end{widetext}
where
\begin{equation}
\begin{aligned}
\mathcal{F}&(k,B,n)=\\
&5-\frac{8\cos\left(ka-a^2n/l_{\mathsmaller{B}}^2\right)}{3}+\frac{\cos\left(2ka-2a^2n/l_{\mathsmaller{B}}^2\right)}{6},
\end{aligned}
\end{equation}
\begin{equation}
\begin{aligned}
\mathcal{G}&(k,B,n)=\\
&\frac{4\sin\left(ka-a^2n/l_{\mathsmaller{B}}^2\right)}{3}-\frac{\sin\left(2ka-2a^2n/l_{\mathsmaller{B}}^2\right)}{6},
\end{aligned}
\end{equation}
and $a$ denotes the distance between two lattice points in $y$-direction. However, in the finite-strip geometry considered here, the matrix given by Eq.~(\ref{FD_Hamiltonian_matrix}) has to be modified at the edges along the $y$-direction, where only nearest neighbors can be used for the approximation of the derivatives with respect to $y$. Following these modifications, the eigenspectrum and the eigenstates of the system in a finite-strip geometry can be determined numerically.

\subsection{Comparison between the analytical and numerical solutions}\label{Sec:MES_Comparison}

\begin{figure}[t]
\centering
\includegraphics*[width=8cm]{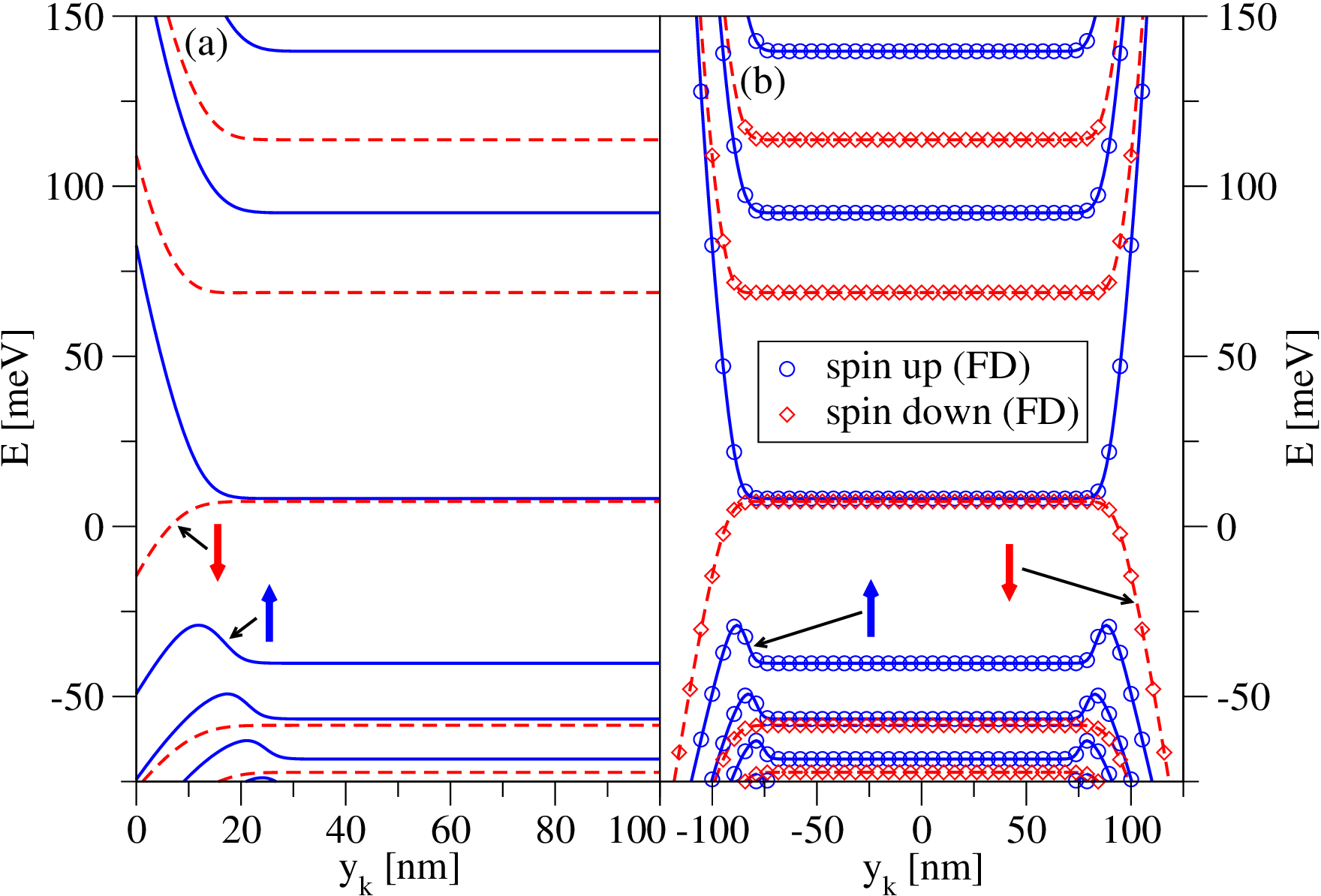}
\caption{(Color online) Calculated energy spectra of (a) a semi-infinite system and (b) a finite strip of width $w=200$ nm for $B=10$ T, $\mathcal{A}=364.5$ meV nm, $\mathcal{B}=-686.0$ meV nm$^2$, $\mathcal{C}=0$, $\mathcal{D}=-512.0$ meV nm$^2$, $\mathcal{M}=-10.0$ meV, and $g_\mathsmaller{\mathrm{e}}=g_\mathsmaller{\mathrm{h}}=0$. Here, the energy spectra are plotted versus $y_k=l^2_\mathsmaller{B}k$. The solid and dashed lines represent $s=\uparrow$ and $s=\downarrow$ states, respectively, which have been calculated using the analytical methods from Sec.~\ref{Sec:MES_analytical} [case (ii) for Fig.~(a) and case (iii) for Fig.~(b)]. Results obtained by the finite-difference method from Sec.~\ref{Sec:MES_FD} are represented by circles (spin up) and diamonds (spin down) in Fig.~(b).}
\label{fig:spectrum_comparison_10T}
\end{figure}

We compare the results obtained by the analytical procedures described in Sec.~\ref{Sec:MES_analytical} with those of the finite-difference method introduced in Sec.~\ref{Sec:MES_FD}. For illustration, Fig.~\ref{fig:spectrum_comparison_10T} shows the energy spectra of a semi-infinite system [Fig.~\ref{fig:spectrum_comparison_10T}~(a)] and a finite strip of width $w=200$ nm [Fig.~\ref{fig:spectrum_comparison_10T}~(b)]. Here, we have chosen the magnetic field $B=10$ T and the parameters $\mathcal{A}=364.5$ meV nm, $\mathcal{B}=-686.0$ meV nm$^2$, $\mathcal{C}=0$, $\mathcal{D}=-512.0$ meV nm$^2$, $\mathcal{M}=-10.0$ meV, and $g_\mathsmaller{\mathrm{e}}=g_\mathsmaller{\mathrm{h}}=0$, which (apart from the vanishing $\mathrm{g}$-factors) correspond to the thickness of $d=7.0$ nm.\cite{Koenig2008:JPSJ,Qi2011:RMP} Whereas the energy spectrum of a semi-infinite system is calculated using the transcendental Eq.~(\ref{si_trans}), both procedures described above, solving the transcendental Eq.~(\ref{det}) or diagonalizing the finite-difference Hamiltonian~(\ref{FD_Hamiltonian}), can be used to calculate the eigenspectrum of the Hamiltonian~(\ref{effective_Hamiltonian}) in a finite-strip geometry. The finite-difference calculations for Fig.~\ref{fig:spectrum_comparison_10T}~(b) have been conducted for 201 lattice sites along the $y$-direction, for which we get a relative error of $10^{-6}$-$10^{-5}$. Figure~\ref{fig:spectrum_comparison_10T}~(b) also clearly illustrates the nearly perfect agreement between the analytical and numerical solutions. As can be expected if the magnetic length $l_\mathsmaller{B}$ is small compared to the width of the sample $w$, the energy spectra near the edge as well as the energy spectra in the bulk are almost identical for the semi-infinite and finite systems as shown in Figs.~\ref{fig:spectrum_comparison_10T}~(a) and~\ref{fig:spectrum_comparison_10T}~(b). The bulk Landau levels are perfectly characterized by Eqs.~(\ref{Landau_levels_up})-(\ref{zero_Landau_level_down}).

\subsection{Results}\label{Sec:MES_Results}

\begin{figure}[t]
\centering
\includegraphics*[width=8cm]{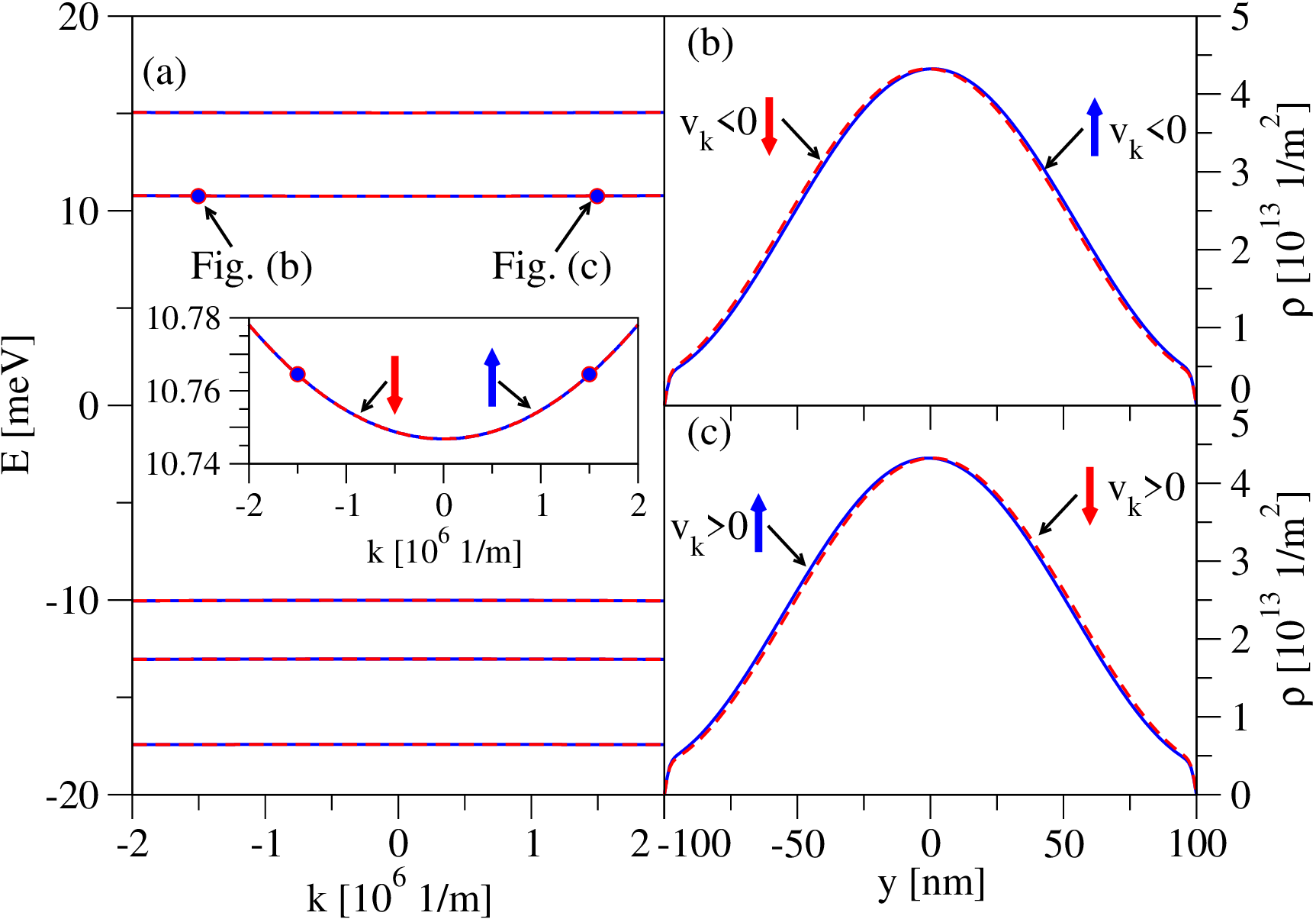}
\caption{(Color online) (a) Calculated energy spectrum and (b), (c) probability densities, $\rho(x,y)=\left|\Psi(x,y)\right|^2$, of selected states for $d=5.5$ nm, $w=200$ nm, and $B=0$ T, where solid and dashed lines represent $s=\uparrow$ and $s=\downarrow$ states, respectively. Here, the states shown in Figs.~(b) and~(c) are marked in the energy spectrum, Fig.~(a), by dots. The velocity with which the states propagate along the $x$-direction is given by $v_k=[\partial E(k)/\partial k]/\hbar$.}
\label{fig:spectrum0T_OI}
\end{figure}

\begin{figure}[t]
\centering
\includegraphics*[width=8cm]{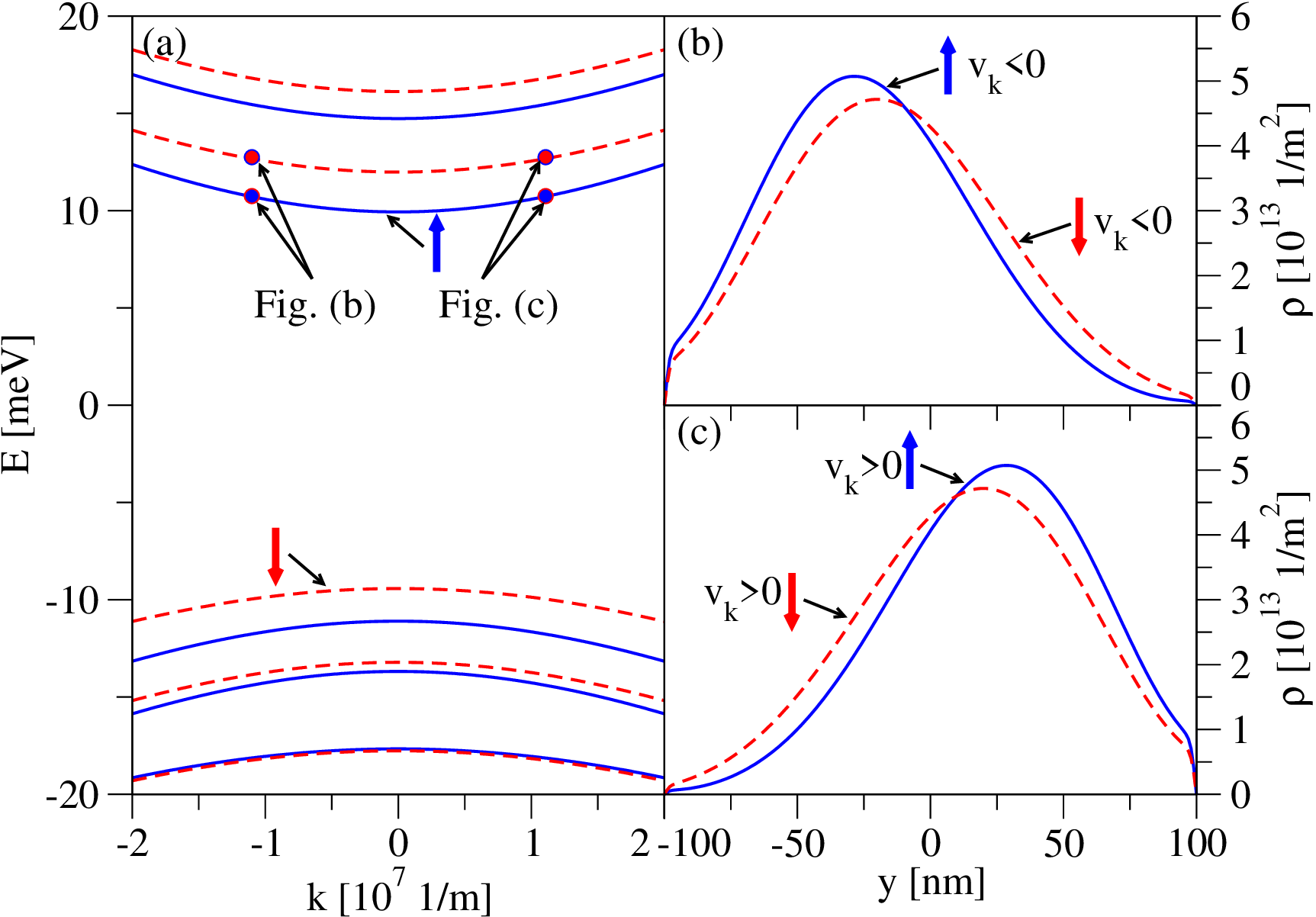}
\caption{(Color online) (a) Calculated energy spectrum and (b), (c) probability densities, $\rho(x,y)=\left|\Psi(x,y)\right|^2$, of selected states for $d=5.5$ nm, $w=200$ nm, and $B=0.1$ T, where solid and dashed lines represent $s=\uparrow$ and $s=\downarrow$ states, respectively. Here, the states shown in Figs.~(b) and~(c) are marked in the energy spectrum, Fig.~(a), by dots. The velocity with which the states propagate along the $x$-direction is given by $v_k=[\partial E(k)/\partial k]/\hbar$.}
\label{fig:spectrum01T_OI}
\end{figure}

\begin{figure}[t]
\centering
\includegraphics*[width=8cm]{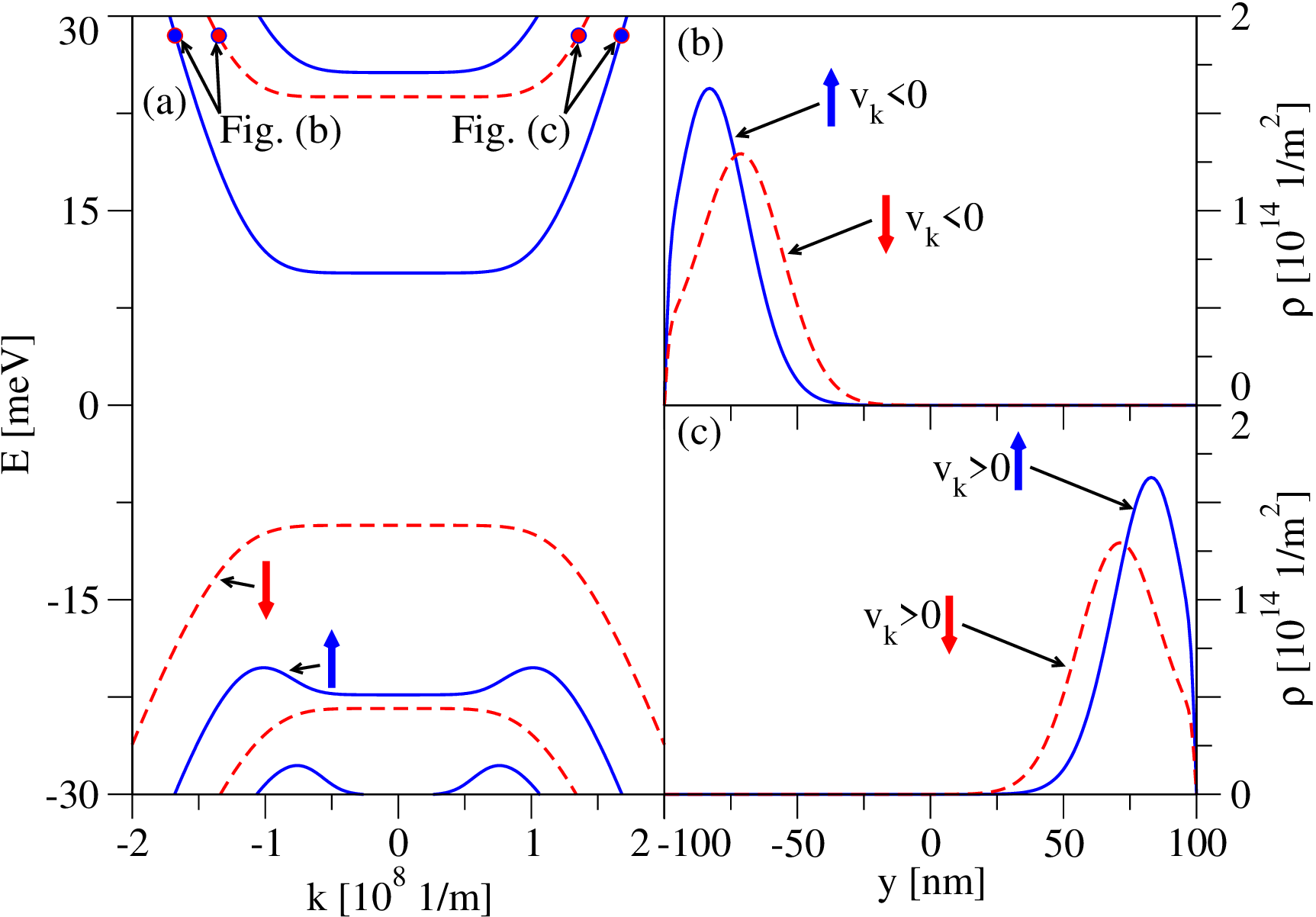}
\caption{(Color online) (a) Calculated energy spectrum and (b), (c) probability densities, $\rho(x,y)=\left|\Psi(x,y)\right|^2$, of selected states for $d=5.5$ nm, $w=200$ nm, and $B=1$ T, where solid and dashed lines represent $s=\uparrow$ and $s=\downarrow$ states, respectively. Here, the states shown in Figs.~(b) and~(c) are marked in the energy spectrum, Fig.~(a), by dots. The velocity with which the states propagate along the $x$-direction is given by $v_k=[\partial E(k)/\partial k]/\hbar$.}
\label{fig:spectrum1T_OI}
\end{figure}

\begin{figure}[t]
\centering
\includegraphics*[width=8cm]{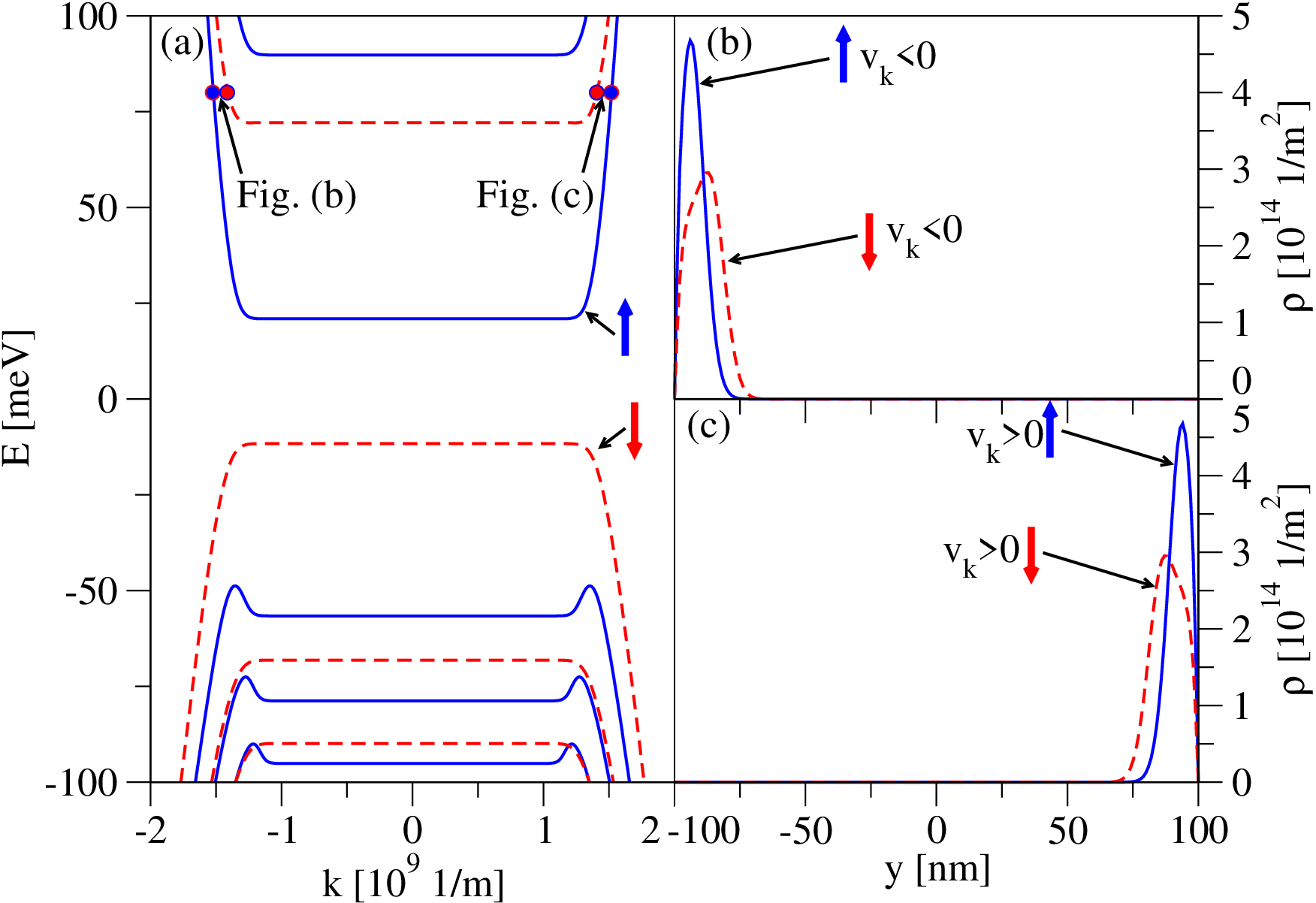}
\caption{(Color online) (a) Calculated energy spectrum and (b), (c) probability densities, $\rho(x,y)=\left|\Psi(x,y)\right|^2$, of selected states for $d=5.5$ nm, $w=200$ nm, and $B=10$ T, where solid and dashed lines represent $s=\uparrow$ and $s=\downarrow$ states, respectively. Here, the states shown in Figs.~(b) and~(c) are marked in the energy spectrum, Fig.~(a), by dots. The velocity with which the states propagate along the $x$-direction is given by $v_k=[\partial E(k)/\partial k]/\hbar$.}
\label{fig:spectrum10T_OI}
\end{figure}

In this section, we investigate the magnetic field dependence of the energy spectrum and its corresponding eigenstates in a finite-strip geometry with the width $w=200$ nm. The graphs shown in this section have been calculated using the finite-difference scheme from Sec.~\ref{Sec:MES_FD} with 201 lattice sites along the $y$-direction (see also Sec.~\ref{Sec:MES_Comparison}).
\subsubsection{Ordinary insulator regime}
First, we examine the quantum-well spectrum in the ordinary insulator regime, that is, for a thickness $d<d_c$, where the band structure is normal and there are no QSH states (at zero magnetic field). Figures~\ref{fig:spectrum0T_OI}-\ref{fig:spectrum10T_OI} show the energy spectrum and (selected) eigenstates at different magnetic fields for the material parameters $\mathcal{A}=387$ meV nm, $\mathcal{B}=-480.0$ meV nm$^2$, $\mathcal{C}=0$, $\mathcal{D}=-306.0$ meV nm$^2$, and $\mathcal{M}=9.0$ meV, which correspond to a quantum-well thickness of $d=5.5$ nm.\cite{Qi2011:RMP} As illustrated by Fig.~\ref{fig:spectrum0T_OI}~(a), which shows the spectrum for $B=0$, only bulk states, but no edge states can be found [see Figs.~\ref{fig:spectrum0T_OI}~(b) and~(c)], a situation which changes little if small magnetic fields are applied (see Fig.~\ref{fig:spectrum01T_OI}). Only if the magnetic field is increased further, do Landau levels [given by Eqs.~(\ref{Landau_levels_up})-(\ref{zero_Landau_level_down})] and corresponding QH edge states begin to form as can be seen in Figs.~\ref{fig:spectrum1T_OI} and~\ref{fig:spectrum10T_OI}. Comparing Figs.~\ref{fig:spectrum1T_OI} and~\ref{fig:spectrum10T_OI}, one can also discern that with increasing magnetic field the QH edge states become more localized.

\subsubsection{QSH regime}

\begin{figure}[t]
\centering
\includegraphics*[width=8cm]{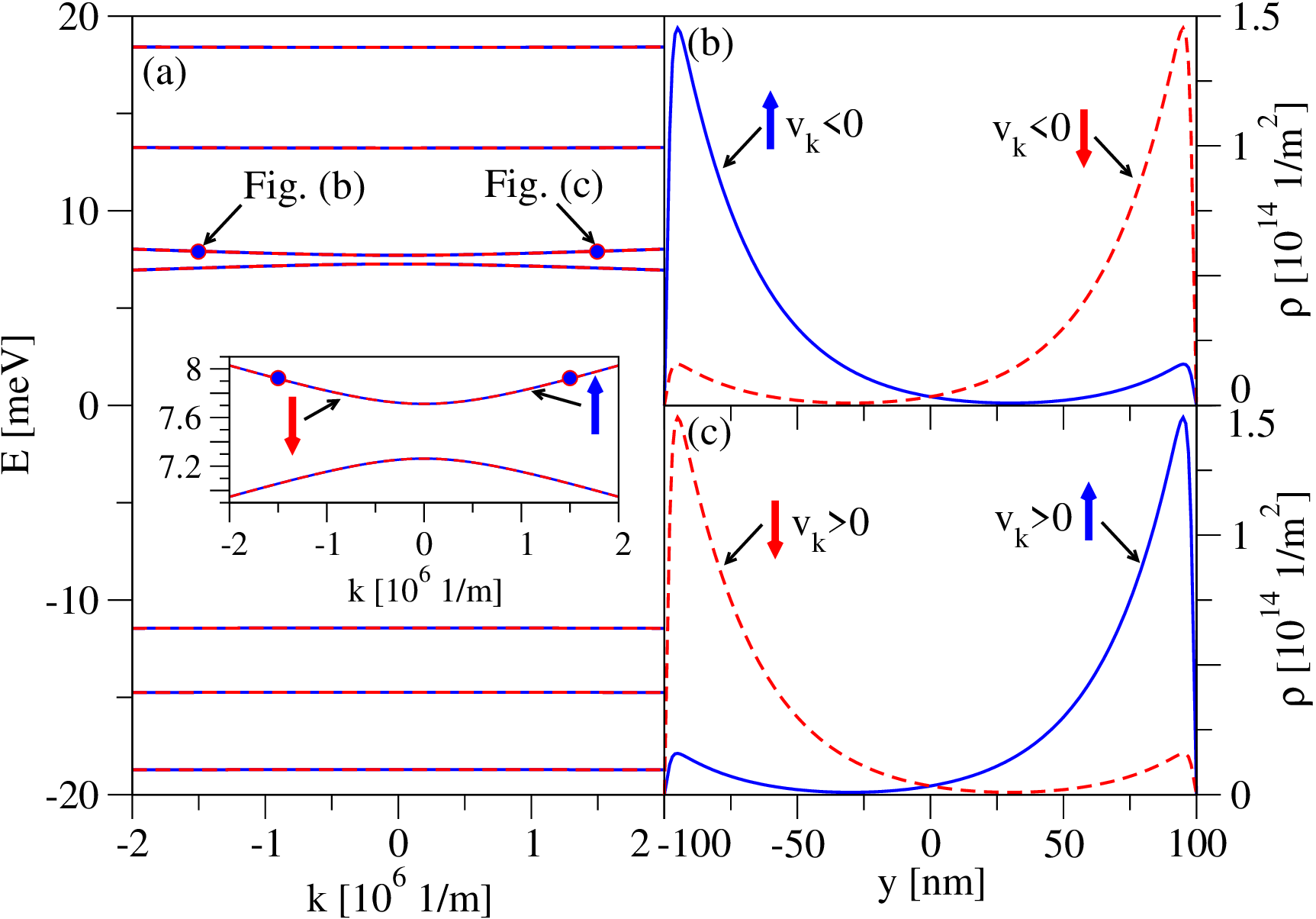}
\caption{(Color online) (a) Calculated energy spectrum and (b), (c) probability densities, $\rho(x,y)=\left|\Psi(x,y)\right|^2$, of selected states for $d=7.0$ nm, $w=200$ nm, and $B=0$ T, where solid and dashed lines represent $s=\uparrow$ and $s=\downarrow$ states, respectively. Here, the states shown in Figs.~(b) and~(c) are marked in the energy spectrum, Fig.~(a), by dots. The velocity with which the states propagate along the $x$-direction is given by $v_k=[\partial E(k)/\partial k]/\hbar$.}
\label{fig:spectrum0T}
\end{figure}

\begin{figure}[t]
\centering
\includegraphics*[width=8cm]{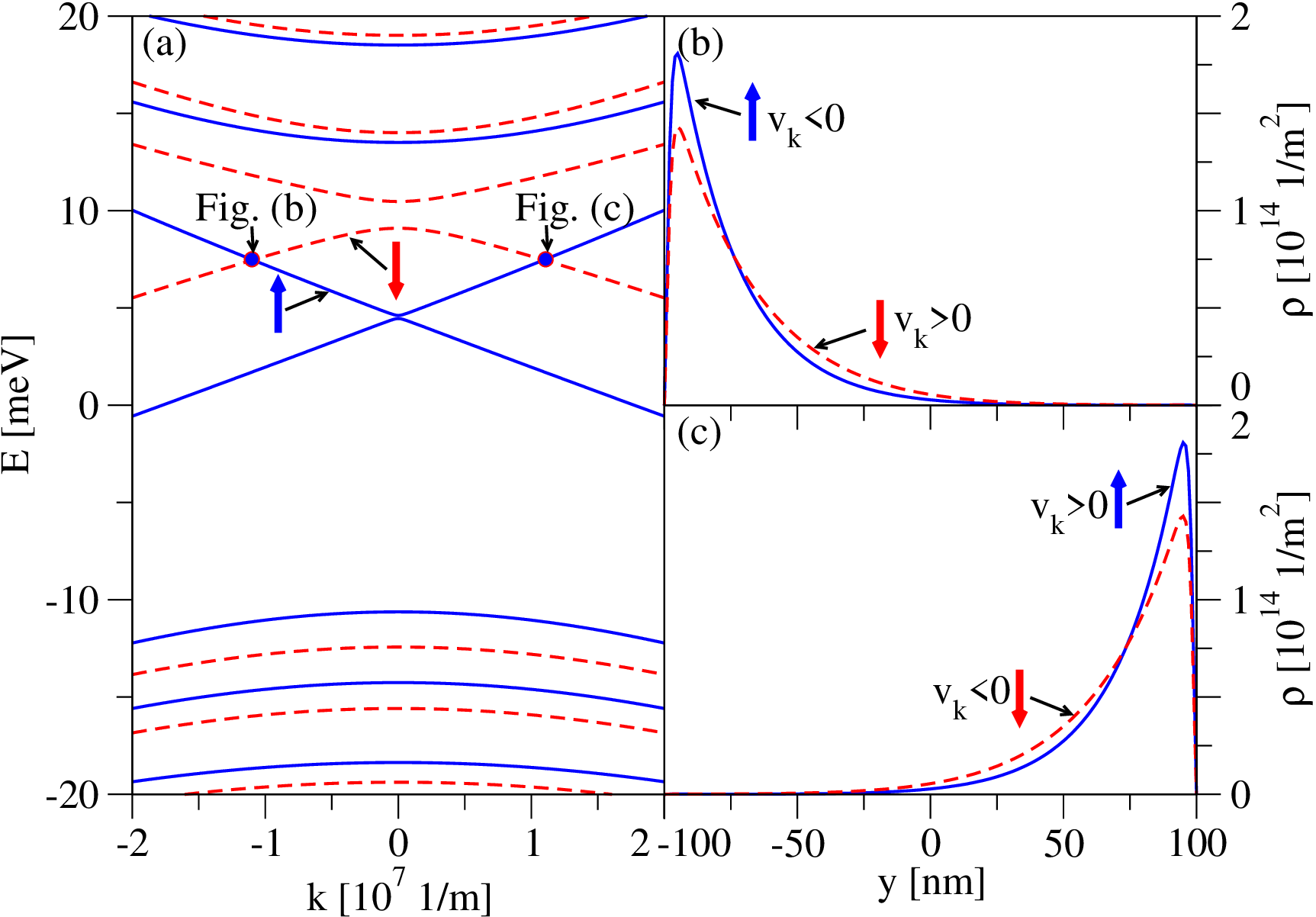}
\caption{(Color online) (a) Calculated energy spectrum and (b), (c) probability densities, $\rho(x,y)=\left|\Psi(x,y)\right|^2$, of selected states for $d=7.0$ nm, $w=200$ nm, and $B=0.1$ T, where solid and dashed lines represent $s=\uparrow$ and $s=\downarrow$ states, respectively. Here, the states shown in Figs.~(b) and~(c) are marked in the energy spectrum, Fig.~(a), by dots. The velocity with which the states propagate along the $x$-direction is given by $v_k=[\partial E(k)/\partial k]/\hbar$.}
\label{fig:spectrum01T}
\end{figure}

\begin{figure}[t]
\centering
\includegraphics*[width=8cm]{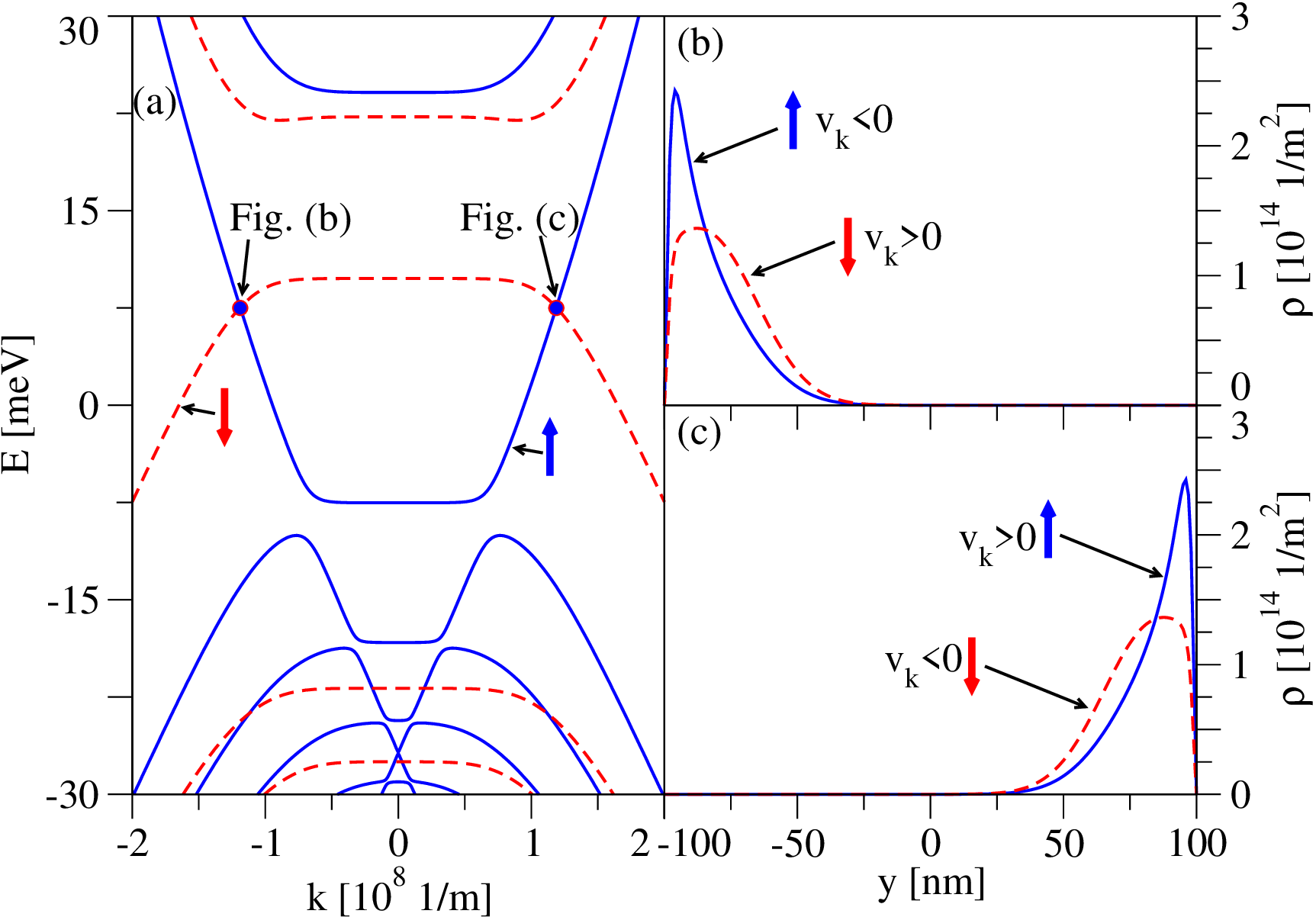}
\caption{(Color online) (a) Calculated energy spectrum and (b), (c) probability densities, $\rho(x,y)=\left|\Psi(x,y)\right|^2$, of selected states for $d=7.0$ nm, $w=200$ nm, and $B=1$ T, where solid and dashed lines represent $s=\uparrow$ and $s=\downarrow$ states, respectively. Here, the states shown in Figs.~(b) and~(c) are marked in the energy spectrum, Fig.~(a), by dots. The velocity with which the states propagate along the $x$-direction is given by $v_k=[\partial E(k)/\partial k]/\hbar$.}
\label{fig:spectrum1T}
\end{figure}

\begin{figure}[t]
\centering
\includegraphics*[width=8cm]{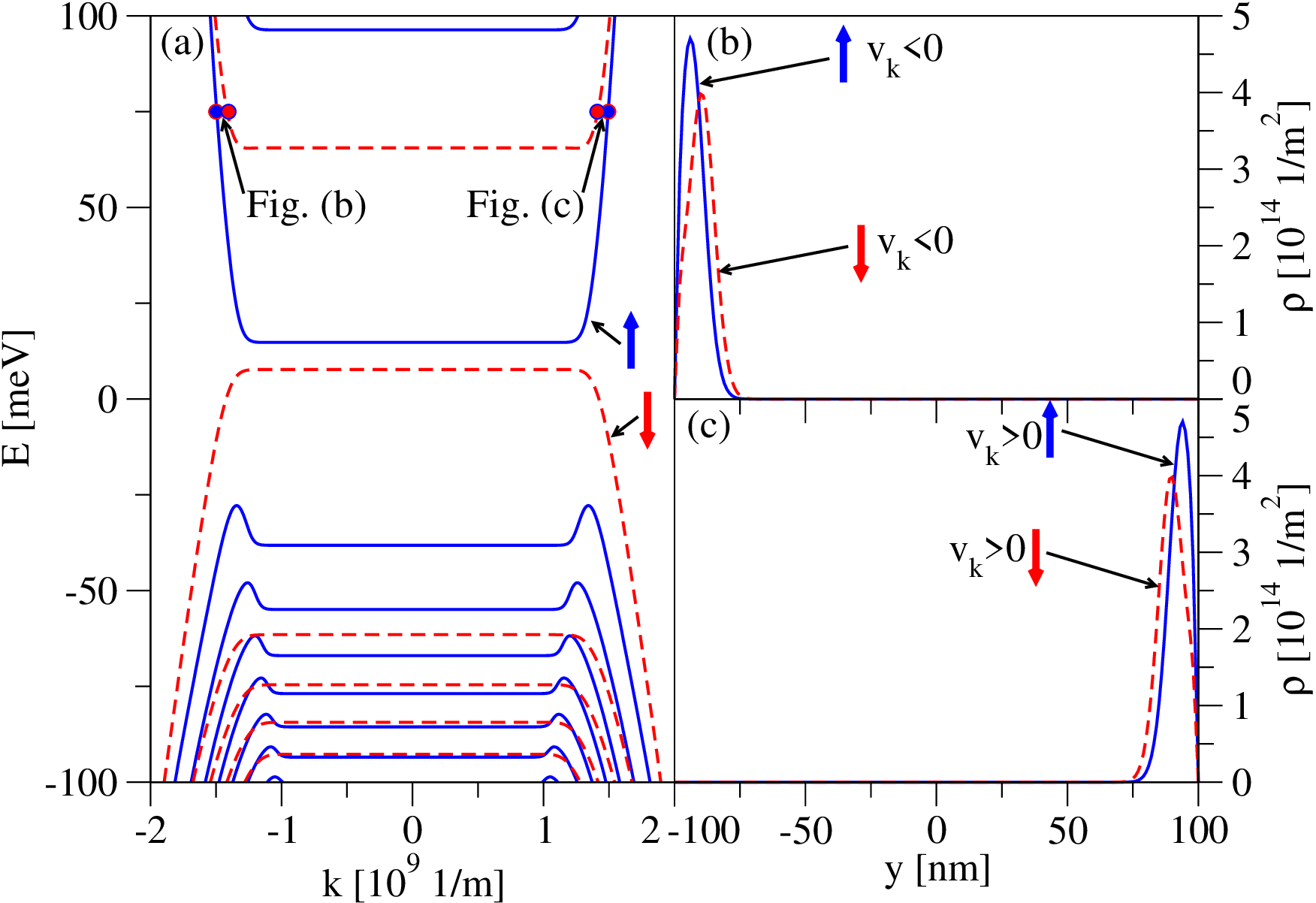}
\caption{(Color online) (a) Calculated energy spectrum and (b), (c) probability densities, $\rho(x,y)=\left|\Psi(x,y)\right|^2$, of selected states for $d=7.0$ nm, $w=200$ nm, and $B=10$ T, where solid and dashed lines represent $s=\uparrow$ and $s=\downarrow$ states, respectively. Here, the states shown in Figs.~(b) and~(c) are marked in the energy spectrum, Fig.~(a), by dots. The velocity with which the states propagate along the $x$-direction is given by $v_k=[\partial E(k)/\partial k]/\hbar$.}
\label{fig:spectrum10T}
\end{figure}

In Figs.~\ref{fig:spectrum0T}-\ref{fig:spectrum10T}, by contrast, the energy spectrum and (selected) eigenstates of a strip with the width $w=200$ nm are presented for the material parameters $\mathcal{A}=364.5$ meV nm, $\mathcal{B}=-686.0$ meV nm$^2$, $\mathcal{C}=0$, $\mathcal{D}=-512.0$ meV nm$^2$, $\mathcal{M}=-10.0$ meV, $g_\mathsmaller{\mathrm{e}}=22.7$, and $g_\mathsmaller{\mathrm{h}}=-1.21$, corresponding to a quantum-well thickness $d=7.0$ nm,\cite{Koenig2008:JPSJ,Qi2011:RMP} that is, for parameters in the QSH regime (at $B=0$), and several strengths of the perpendicular magnetic field. The spectra and states in Figs.~\ref{fig:spectrum0T}-\ref{fig:spectrum10T} illustrate the evolution of QSH and QH states in HgTe.

Figure~\ref{fig:spectrum0T}~(a) shows the spectrum at zero magnetic field. At this magnetic field, one can observe the QSH state inside the bulk gap, that is, two degenerate pairs of counterpropagating, spin-polarized edge states, one pair at each edge [see Figs.~\ref{fig:spectrum0T}~(b) and~(c)]. As found in Ref.~\onlinecite{Zhou2008:PRL}, at $k=0$ the wave functions of QSH edge states with the same spin, but at opposite edges overlap thereby opening up a gap [see the inset in Fig.~\ref{fig:spectrum0T}~(a)]. By increasing the width of the strip, the overlap of the edge-state wave functions with the same spin is diminished and one can remove this finite-size effect.

For small magnetic fields (Fig.~\ref{fig:spectrum01T}), apart from the splitting of spin-up and down states, the situation is at first glance quite comparable to the one in Fig.~\ref{fig:spectrum0T}. Most importantly, one can still find pairs of counterpropagating, spin-polarized states in the vicinity of each neutrality point [for example, the states shown in Figs.~\ref{fig:spectrum01T}~(b) and~(c)], that is, the crossovers between the lowest (hole-like) conduction band and uppermost (electron-like) valence band [marked by dots in Fig.~\ref{fig:spectrum01T}~(a)]. However, we stress that these counterpropagating, spin-polarized states which can be found (at a given edge) if the Fermi level is close to the neutrality points, are not connected with each other by time-reversal symmetry and are therefore not topologically protected (for example, against spin-orbit coupling).

\begin{figure}[t]
\centering
\includegraphics*[width=8cm]{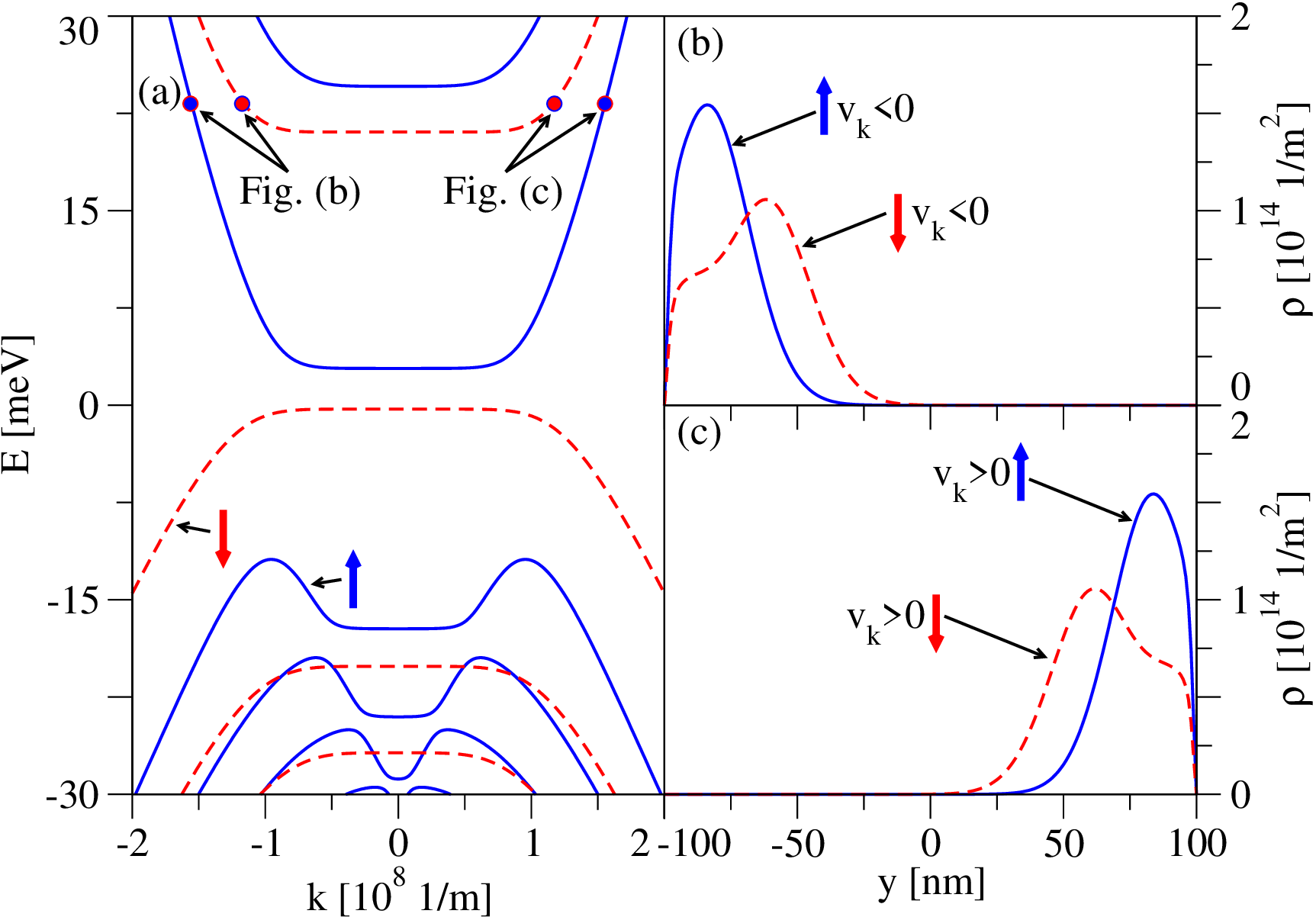}
\caption{(Color online) (a) Calculated energy spectrum and (b), (c) probability densities, $\rho(x,y)=\left|\Psi(x,y)\right|^2$, of selected states for $d=6.3$ nm, $w=200$ nm, and $B=1$ T, where solid and dashed lines represent $s=\uparrow$ and $s=\downarrow$ states, respectively. Here, the states shown in Figs.~(b) and~(c) are marked in the energy spectrum, Fig.~(a), by dots. The velocity with which the states propagate along the $x$-direction is given by $v_k=[\partial E(k)/\partial k]/\hbar$.}
\label{fig:spectrum1T_critical}
\end{figure}

\begin{figure}[t]
\centering
\includegraphics*[width=8cm]{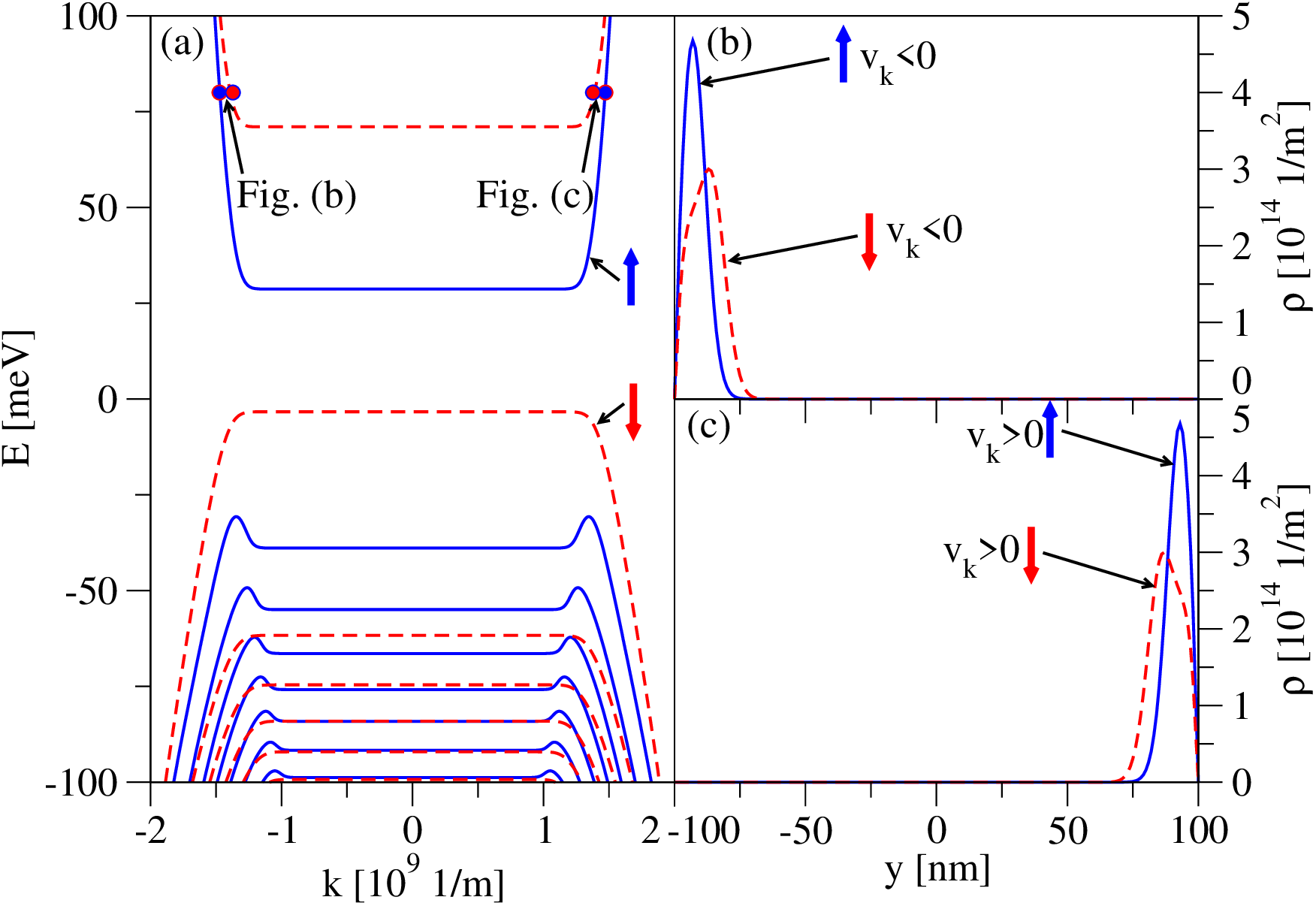}
\caption{(Color online) (a) Calculated energy spectrum and (b), (c) probability densities, $\rho(x,y)=\left|\Psi(x,y)\right|^2$, of selected states for $d=6.3$ nm, $w=200$ nm, and $B=10$ T, where solid and dashed lines represent $s=\uparrow$ and $s=\downarrow$ states, respectively. Here, the states shown in Figs.~(b) and~(c) are marked in the energy spectrum, Fig.~(a), by dots. The velocity with which the states propagate along the $x$-direction is given by $v_k=[\partial E(k)/\partial k]/\hbar$.}
\label{fig:spectrum10T_critical}
\end{figure}

Going to $B=1$ T (Fig.~\ref{fig:spectrum1T}), we can still find counterpropagating, spin-polarized states near and at the crossovers between the lowest (hole-like) conduction and uppermost (electron-like) valence bands, which (in the bulk) have evolved into the $E^\uparrow(0)$ and $E^\downarrow(0)$ Landau levels. As the center of the orbital motion is given by $\sqrt{2}l_{\mathsmaller{B}}k$, one can see that those states are now no longer as localized as before at the edges [see Figs.~\ref{fig:spectrum1T}~(b) and~(c)]. Meanwhile, the bulk states from Fig.~\ref{fig:spectrum0T} have also evolved into Landau levels given by Eqs.~(\ref{Landau_levels_up}) and~(\ref{Landau_levels_down}) with localized QH edge as well as bulk states. From Fig.~\ref{fig:spectrum1T}, one can also discern another feature of the energy spectrum and eigenstates that develops with an increasing magnetic field, namely the appearance of 'bumps' [see the spin-up valence bands in Fig.~\ref{fig:spectrum1T}~(a)]. If the Fermi level crosses those 'bumps', one finds states which are localized near the same edge and carry the same spin, but counterpropagate. This has also been observed in Ref.~\onlinecite{Chen2012:PRB}, where those states gave rise to exotic plateaus in the longitudinal and Hall resistances. As can be seen in Figs.~\ref{fig:spectrum1T_OI} and~\ref{fig:spectrum10T_OI} (as well as later in Figs.~\ref{fig:spectrum10T}, \ref{fig:spectrum1T_critical}, and~\ref{fig:spectrum10T_critical}), this behavior can also be found for other quantum well parameters.

The situation described so far changes for high magnetic fields (Fig.~\ref{fig:spectrum10T}), when the electron-like band described by $E^\uparrow(0)$ (in the bulk) is above the hole-like $E^\downarrow(0)$ band. Then, there is no longer any crossover between the dispersions of electron- and hole-like bands and one consequently cannot find counterpropagating, spin-polarized states anymore, just QH edge states propagating in the same direction [for example, the states shown in Figs.~\ref{fig:spectrum10T}~(b) and~(c)].

As has been known for a long time, the uppermost (electron-like) valence and the lowest (hole-like) conduction Landau levels cross at a finite magnetic field $B_\mathrm{c}$ in inverted HgTe/CdTe quantum wells.\cite{Meyer1990:PRB,Truchsess1997:HMFPS,Schultz1998:PRB} The transition between the two situations, the one where counterpropagating, spin-polarized states exist and the one where they do not, happens exactly at this crossover point: As long as the hole-like band is above the electron-like band, that is, as long as the band structure remains inverted, one can find counterpropagating, spin-polarized states in addition to the QH states. Otherwise, there are only QH states.

\begin{figure}[t]
\centering
\includegraphics*[width=8cm]{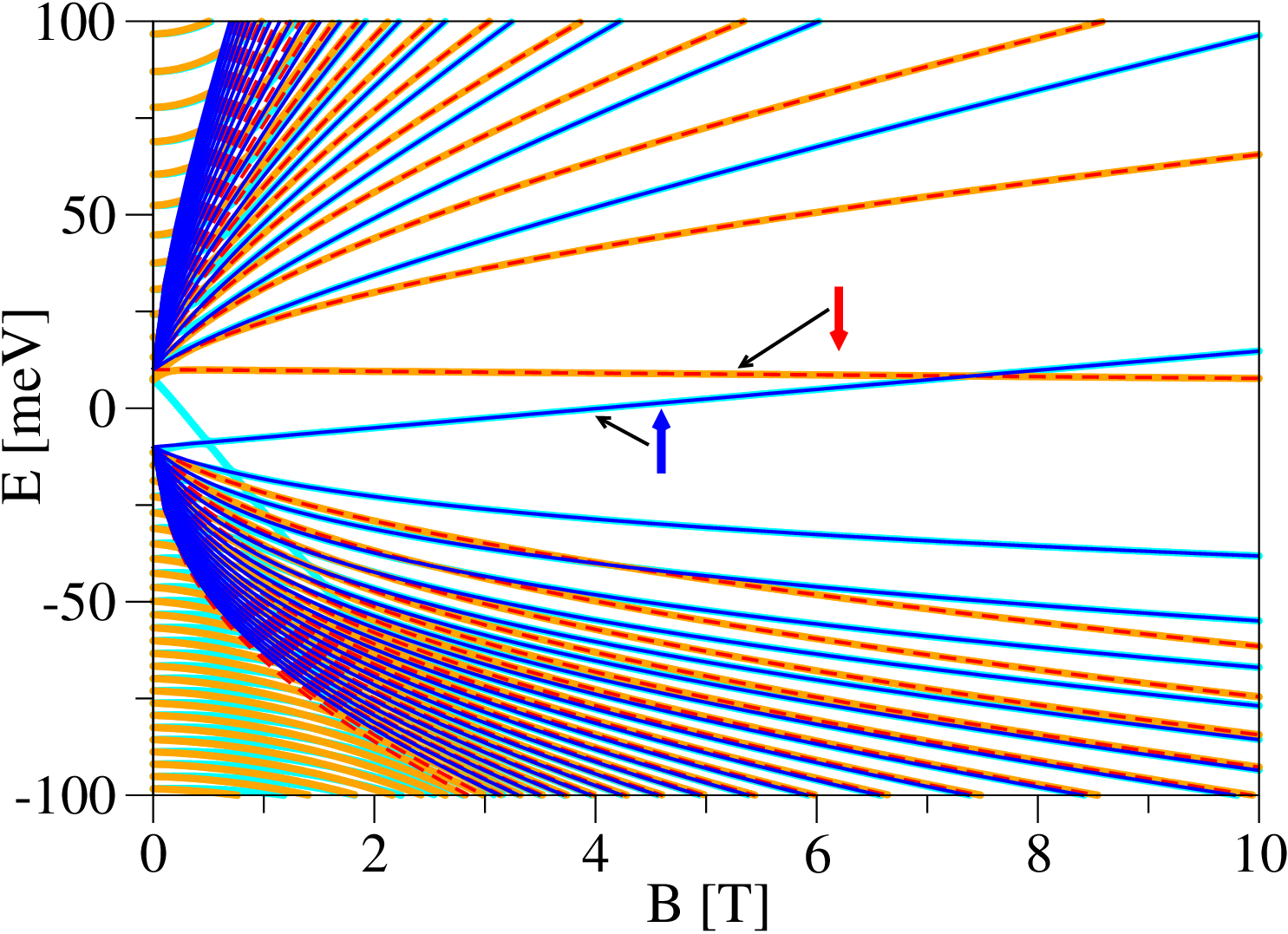}
\caption{(Color online) Magnetic field dependence of the states at $k=0$ in a finite strip of width $w=200$ nm compared to the bulk Landau levels given by Eqs.~(\ref{Landau_levels_up})-(\ref{zero_Landau_level_down}). The thinner solid and dashed lines represent bulk Landau levels for $s=\uparrow$ and $s=\downarrow$, respectively. The levels of the finite-strip geometry are displayed by thick lines. All levels displayed here have been calculated for band parameters corresponding to $d=7.0$ nm.}\label{fig:Landau_levels}
\end{figure}

This crossover point can be easily calculated from the Landau levels via the condition $E^\uparrow(0)=E^\downarrow(0)$, from which we get
\begin{equation}\label{B_crossing}
B_\mathrm{c}=\frac{\mathcal{M}}{2\pi\mathcal{B}/\Phi_0-\left(g_{\mathsmaller{\mathrm{e}}}+g_{\mathsmaller{\mathrm{h}}}\right)\mu_\mathsmaller{B}/4}
\end{equation}
for the magnetic field at which the transition happens (valid only for $B_\mathrm{c}>0$). Here, $\Phi_0=2\pi\hbar/e$ denotes the magnetic flux quantum. The validity of the result given by Eq.~(\ref{B_crossing}) is also illustrated by Fig.~\ref{fig:Landau_levels}, which shows the magnetic field dependence of the energies of the finite strip with width $w=200$ nm at $k=0$ and of the bulk Landau levels for the same band parameters as above. As can be expected, the energies at $k=0$ are given by the Landau levels~(\ref{Landau_levels_up})-(\ref{zero_Landau_level_down}) at high magnetic fields. Most importantly, the crossover between the electron-like $E^\uparrow(0)$ and the hole-like $E^\downarrow(0)$ bands happens in the region, where the $B$-dependence of the energy levels at $k=0$ is already described extremely well by those Landau levels and from Eq.~(\ref{B_crossing}) we find $B_\mathrm{c}\approx7.4$ T, consistent with the numerical result that can be extracted from Fig.~\ref{fig:Landau_levels}. Furthermore, one can see how the $E^\uparrow(0)$ band is below the $E^\downarrow(0)$ band for $B<B_\mathrm{c}$, and how the situation is reversed for $B>B_\mathrm{c}$.

\begin{figure}[t]
\centering
\includegraphics*[width=8cm]{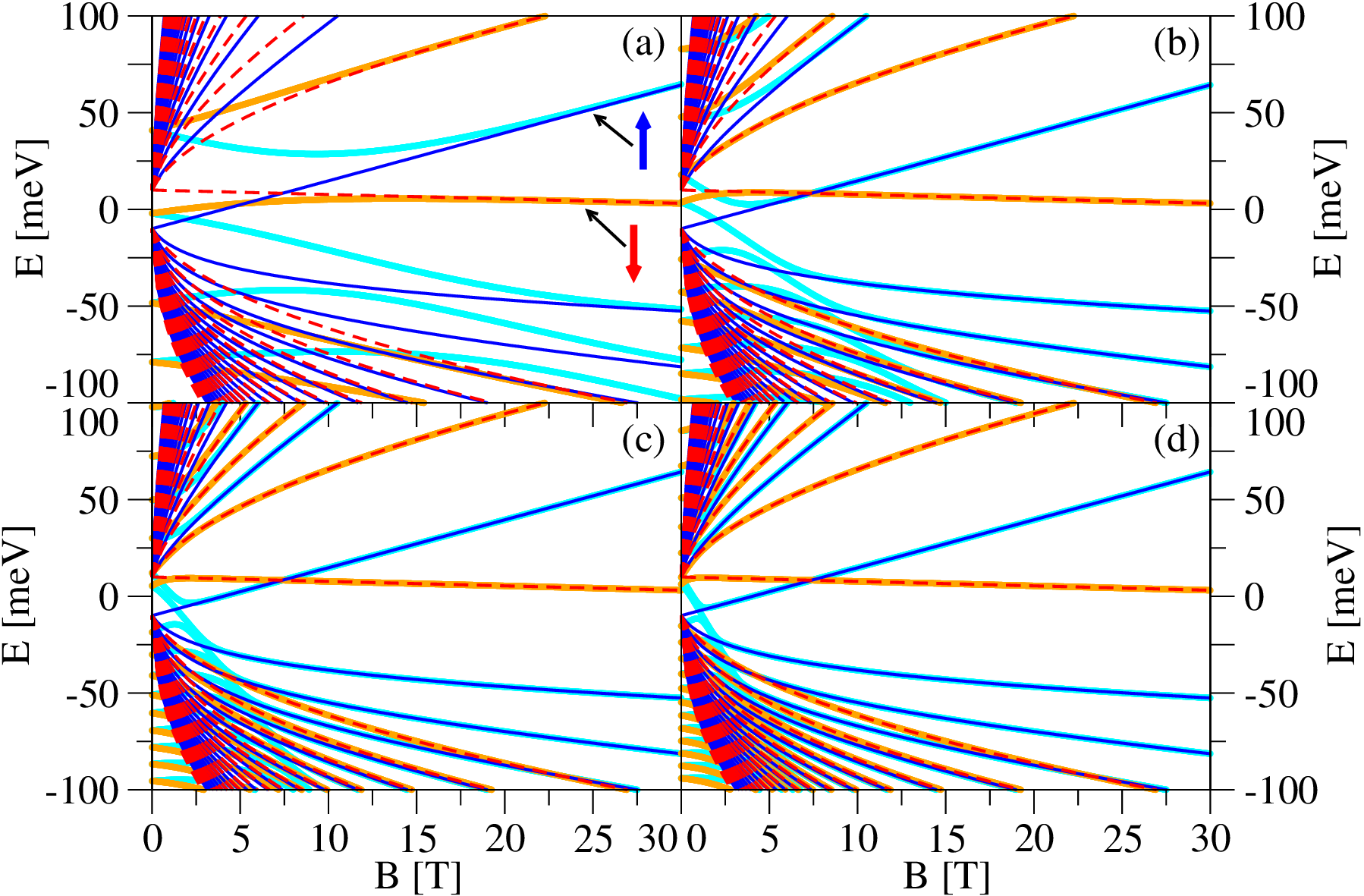}
\caption{(Color online) Magnetic field dependence of the states at $k=0$ in finite strips with the widths (a) $w=25$ nm, (b) $w=50$ nm, (c) $w=75$ nm, and (d) $w=100$ nm compared to the bulk Landau levels given by Eqs.~(\ref{Landau_levels_up})-(\ref{zero_Landau_level_down}). The thinner solid and dashed lines represent bulk Landau levels for $s=\uparrow$ and $s=\downarrow$, respectively. The levels of the finite-strip geometry are displayed by thick lines. All levels displayed here have been calculated for band parameters corresponding to $d=7.0$ nm.}\label{fig:Landau_level_evolution}
\end{figure}

Therefore, we find that if the magnetic field is not too high, the counterpropagating, spin-polarized states persist at finite magnetic fields, consistent with the conclusions in Refs.~\onlinecite{Tkachov2010:PRL,Tkachov2012:PhysicaE}, where the reduced model (mentioned in Sec.~\ref{Sec:Model}) for HgTe has been used, and Ref.~\onlinecite{Chen2012:PRB}. Only for high magnetic fields, the band structure becomes normal and one enters the ordinary insulator regime, in which no counterpropagating, spin-polarized states can be found (see also Ref.~\onlinecite{Chen2012:PRB}). We remark that the description presented in this section also bears out if other widths $w\gtrsim100$ nm of the finite strip are investigated. For larger widths, the formation of Landau levels sets in already at lower magnetic fields, whereas higher fields are needed to observe Landau levels in more narrow strips. If very small samples ($w\lesssim50$ nm) are investigated, however, we find that there is no crossover between the electron-like $E^\uparrow(0)$ and the hole-like $E^\downarrow(0)$ bands, as illustrated by Fig.~\ref{fig:Landau_level_evolution}, which shows a comparison between the bulk Landau levels and the states calculated at $k=0$ for band parameters corresponding to $d=7.0$ nm and several small widths $w$. Only if $w\gtrsim50$ nm, the gap due to the finite size of the sample at $B=0$ is reduced far enough and one can observe a crossover of the $E^\uparrow(0)$ and $E^\downarrow(0)$ bands at $B=B_\mathrm{c}$ which is then give by Eq.~(\ref{B_crossing}).

\begin{figure}[t]
\centering
\includegraphics*[width=8cm]{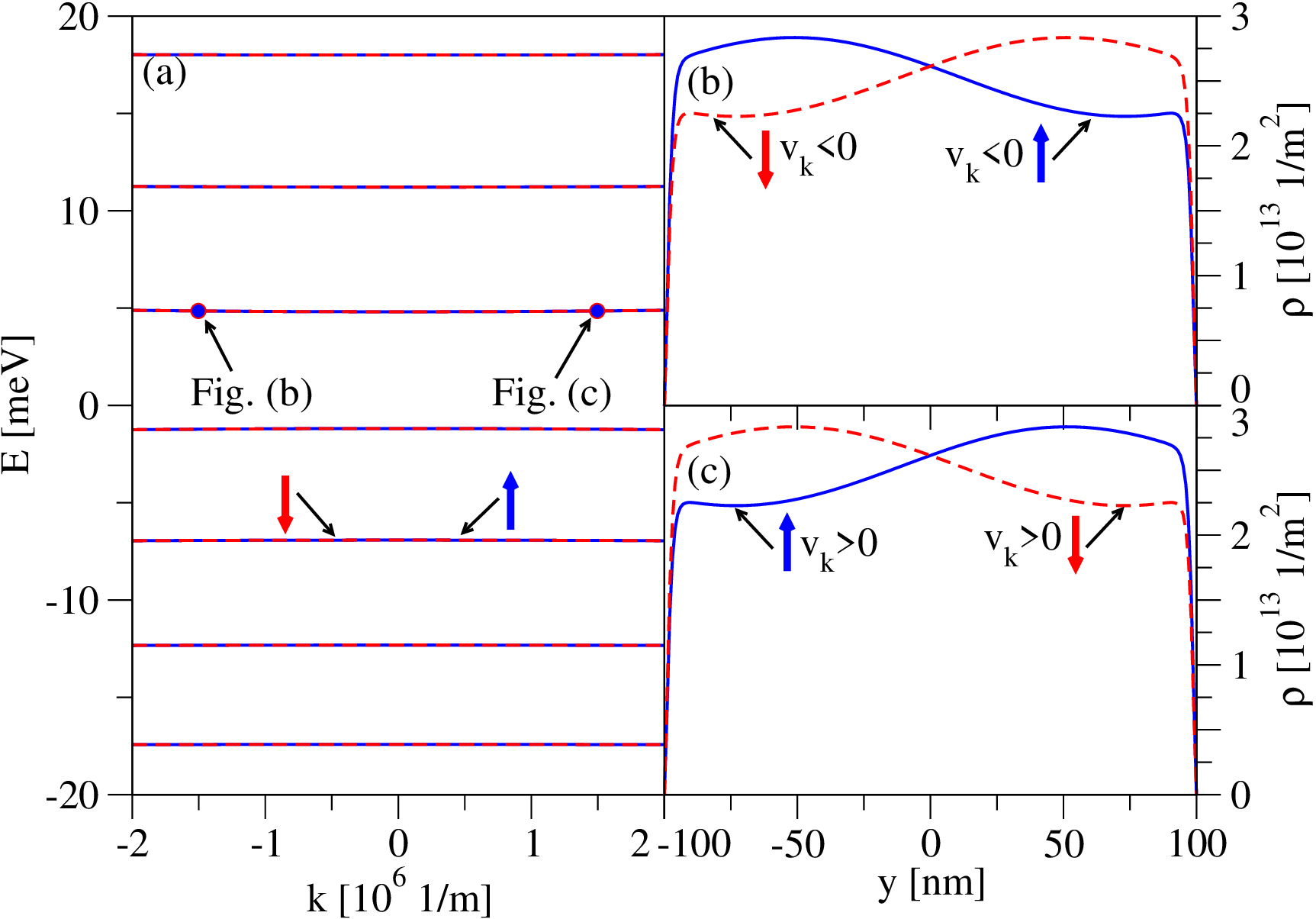}
\caption{(Color online) (a) Calculated energy spectrum and (b), (c) probability densities, $\rho(x,y)=\left|\Psi(x,y)\right|^2$, of selected states for $d=6.3$ nm, $w=200$ nm, and $B=0$ T, where solid and dashed lines represent $s=\uparrow$ and $s=\downarrow$ states, respectively. Here, the states shown in Figs.~(b) and~(c) are marked in the energy spectrum, Fig.~(a), by dots. The velocity with which the states propagate along the $x$-direction is given by $v_k=[\partial E(k)/\partial k]/\hbar$.}
\label{fig:spectrum0T_critical}
\end{figure}

\begin{figure}[t]
\centering
\includegraphics*[width=8cm]{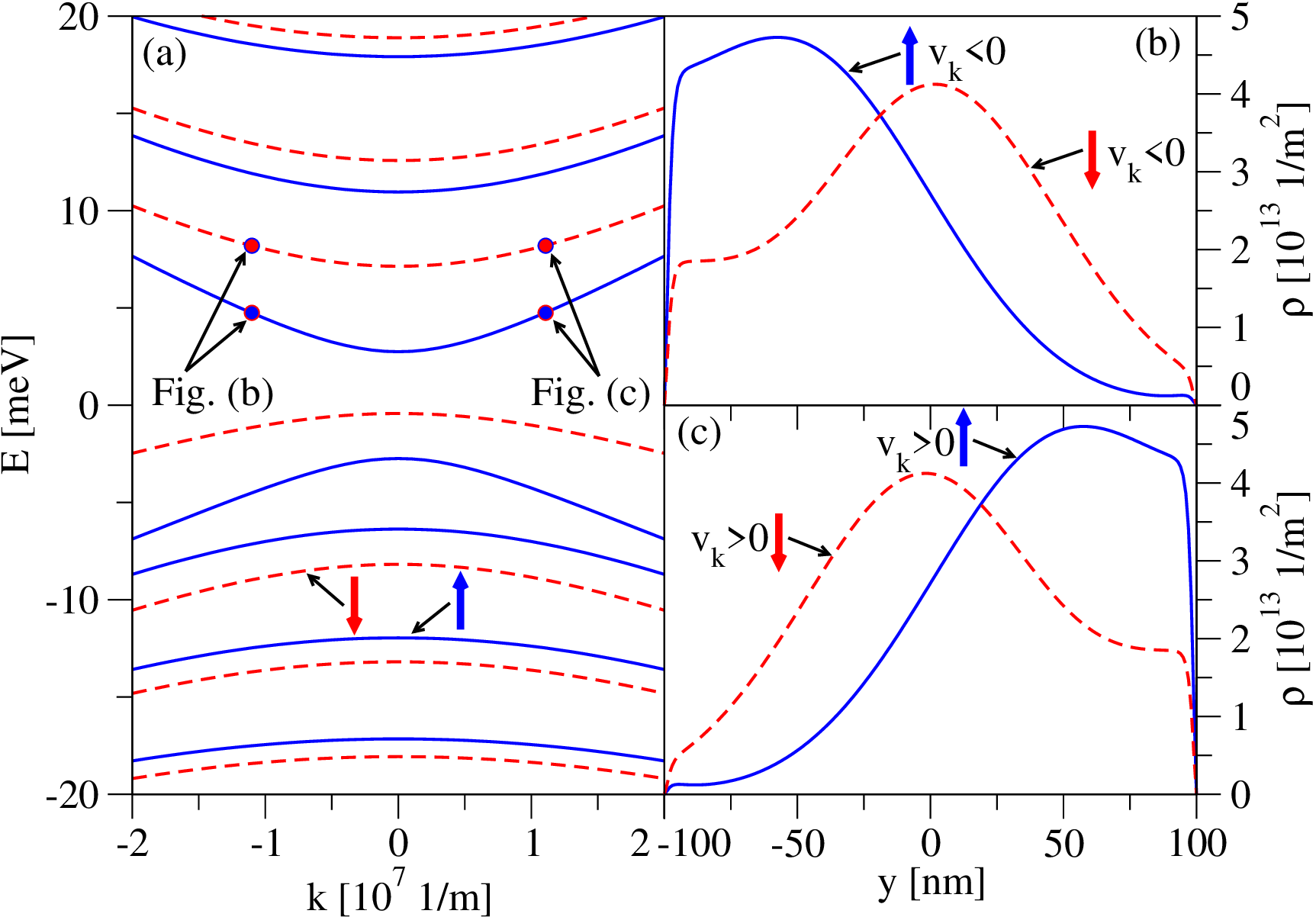}
\caption{(Color online) (a) Calculated energy spectrum and (b), (c) probability densities, $\rho(x,y)=\left|\Psi(x,y)\right|^2$, of selected states for $d=6.3$ nm, $w=200$ nm, and $B=0.1$ T, where solid and dashed lines represent $s=\uparrow$ and $s=\downarrow$ states, respectively. Here, the states shown in Figs.~(b) and~(c) are marked in the energy spectrum, Fig.~(a), by dots. The velocity with which the states propagate along the $x$-direction is given by $v_k=[\partial E(k)/\partial k]/\hbar$.}
\label{fig:spectrum01T_critical}
\end{figure}

\subsubsection{Critical regime}

Finally, for the purpose of comparison to the discussion above, Figs.~\ref{fig:spectrum0T_critical}-\ref{fig:spectrum10T_critical} show the energy spectrum and (selected) eigenstates at different magnetic fields for a strip with the width $w=200$ nm and the material parameters $\mathcal{A}=373.5$ meV nm, $\mathcal{B}=-857.0$ meV nm$^2$, $\mathcal{C}=0$, $\mathcal{D}=-682.0$ meV nm$^2$, $\mathcal{M}=-0.035$ meV, $g_\mathsmaller{\mathrm{e}}=18.5$, and $g_\mathsmaller{\mathrm{h}}=2.4$, which correspond to the critical regime at a quantum-well thickness of $d=d_c=6.3$ nm.\cite{Buettner2011:NatPhys,Qi2011:RMP} For $B=0$, instead of edge states, we find states whose probability densities are spread over the entire width of the strip with a slight preponderance near one of the edges [see Figs.~\ref{fig:spectrum0T_critical}~(b) and~(c)]. with increasing magnetic field the states become more localized (see Figs.~\ref{fig:spectrum01T_critical} and~\ref{fig:spectrum1T_critical}) and, finally, one can find QH edge states (see Fig.~\ref{fig:spectrum10T_critical}).

\section{Magnetic oscillations}\label{Sec:Mag}
\subsection{General formalism}\label{Sec:Mag_Formalism}
In this section, we discuss the magnetization and magnetic oscillations in HgTe quantum wells. Our starting point is the grand potential
\begin{equation}\label{grand_potential_general}
\Omega\left(T,\mu,B\right)=-\frac{S}{\beta}\int\d\epsilon\,\rho(\epsilon)\ln\left\{1+\exp\left[-\beta\left(\epsilon-\mu\right)\right]\right\},
\end{equation}
where $\beta=1/(k_\mathsmaller{B}T)$ and $T$ denotes the temperature, $k_\mathsmaller{B}$ the Boltzmann constant, $\mu$ the chemical potential, $\rho(\epsilon)$ the density of states per unit area, and $S$ is the surface area.

We make the electron-hole transformation and divide the spectrum in the electron and hole contributions, $\rho_\mathsmaller{\mathrm{e}}(\epsilon)=\rho(\epsilon)\Theta(\epsilon-E_\mathrm{n})$ and $\rho_\mathsmaller{\mathrm{h}}(\epsilon)=\rho(\epsilon)\Theta(E_\mathrm{n}-\epsilon)$, where $E_\mathrm{n}=E_\mathrm{n}(B)$ denotes the neutrality point. Then, we can rewrite $\Omega\left(T,\mu,B\right)$ as
\begin{equation}\label{grand_potential_eh_t0}
\begin{aligned}
\Omega\left(T,\mu,B\right)=&\Omega_\mathsmaller{\mathrm{e}}\left(T,\mu,B\right)+\Omega_\mathsmaller{\mathrm{h}}\left(T,\mu,B\right)\\
&+S\int\d\epsilon\,\rho_h(\epsilon)\left(\epsilon-\mu\right),
\end{aligned}
\end{equation}
where
\begin{equation}\label{grand_potential_e}
\Omega_\mathsmaller{\mathrm{e}}\left(T,\mu,B\right)=-\frac{S}{\beta}\int\d\epsilon\,\rho_\mathsmaller{\mathrm{e}}(\epsilon)\ln\left\{1+\exp\left[-\beta\left(\epsilon-\mu\right)\right]\right\}
\end{equation}
and
\begin{equation}\label{grand_potential_h}
\Omega_\mathsmaller{\mathrm{h}}\left(T,\mu,B\right)=-\frac{S}{\beta}\int\d\epsilon\,\rho_\mathsmaller{\mathrm{h}}(\epsilon)\ln\left\{1+\exp\left[\beta\left(\epsilon-\mu\right)\right]\right\}
\end{equation}
denote the grand potentials of electrons and holes, respectively. The total particle number in the system is given by $N_\mathsmaller{\mathrm{tot}}=-\left[\partial\Omega\left(T,\mu,B\right)/\partial\mu\right]$. However, it is more convenient to distinguish between electrons and holes and to work with the carrier imbalance $N=N_\mathsmaller{\mathrm{e}}-N_\mathsmaller{\mathrm{h}}$ (with $N_\mathsmaller{\mathrm{e/h}}$ denoting the number of electrons and holes, respectively). Following Ref.~\onlinecite{Sharapov2004:PRB}, we redefine the grand potential and use
\begin{equation}\label{grand_potential_eh_t}
\begin{aligned}
\Omega'\left(T,\mu,B\right)&=\Omega\left(T,\mu,B\right)+S\mu\int\d\epsilon\,\rho_h(\epsilon)\\
&=\Omega_\mathsmaller{\mathrm{e}}\left(T,\mu,B\right)+\Omega_\mathsmaller{\mathrm{h}}\left(T,\mu,B\right)+\Omega_0(B),
\end{aligned}
\end{equation}
where
\begin{equation}
\Omega_0(B)=S\int\d\epsilon\,\rho_h(\epsilon)\epsilon
\end{equation}
is the ground-state/vacuum energy. The carrier imbalance is then given by $N=-\left[\partial\Omega'\left(T,\mu,B\right)/\partial\mu\right]$.

The magnetization (as a function of the chemical potential, the temperature, and the magnetic field) can be extracted from $\Omega'\left(T,\mu,B\right)$ via
\begin{equation}\label{Mag_gen}
\begin{aligned}
M_\mathsmaller{\mathrm{tot}}\left(T,\mu,B\right)&=-\frac{1}{S}\frac{\partial\Omega'\left(T,\mu,B\right)}{\partial B}\\
&=M_0(B)+M\left(T,\mu,B\right),
\end{aligned}
\end{equation}
where we have split the magnetization in the vacuum part
\begin{equation}\label{Mag_gs}
M_0(B)=-\frac{1}{S}\frac{\partial\Omega_0(B)}{\partial B}
\end{equation}
and the non-vacuum part
\begin{equation}\label{Mag_nv}
M\left(T,\mu,B\right)=-\frac{1}{S}\left[\frac{\partial\Omega_\mathsmaller{\mathrm{e}}\left(T,\mu,B\right)}{\partial B}+\frac{\partial\Omega_\mathsmaller{\mathrm{h}}\left(T,\mu,B\right)}{\partial B}\right].
\end{equation}
At zero temperature, the magnetization of an undoped system is given by $M_0(B)$, whereas at finite temperatures or in doped systems the additional contribution $M\left(T,\mu,B\right)$ arises. The magnetization as a function of the carrier imbalance density $n_d=N/S$ ($n_d>0:$ $n$-doped, $n_d<0:$ $p$-doped) is given by $M\left[T,\mu\left(T,n_d,B\right),B\right]$, where the chemical potential is determined by
\begin{equation}\label{chempot_calc}
n_d=-\frac{1}{S}\left.\left[\frac{\partial\Omega'\left(T,\mu,B\right)}{\partial\mu}\right]\right|_{\mu=\mu\left(T,n_d,B\right)}.
\end{equation}
Finally, we remark that the magnetic susceptibility $\chi_\mathsmaller{\mathrm{tot}}\left(T,\mu,B\right)=\chi_0(B)+\chi\left(T,\mu,B\right)$ can also be split in the vacuum part
\begin{equation}\label{chi_gs}
\chi_0(B)=\frac{\partial M_0(B)}{\partial B}=-\frac{1}{S}\frac{\partial^2\Omega_0(B)}{\partial B^2}
\end{equation}
and the non-vacuum part
\begin{equation}\label{chi_nv}
\begin{aligned}
\chi\left(T,\mu,B\right)&=\frac{\partial M_\mathsmaller{\mathrm{e}}\left(T,\mu,B\right)}{\partial B}+\frac{\partial M_\mathsmaller{\mathrm{h}}\left(T,\mu,B\right)}{\partial B}\\
&=-\frac{1}{S}\left[\frac{\partial^2\Omega_\mathsmaller{\mathrm{e}}\left(T,\mu,B\right)}{\partial B^2}+\frac{\partial^2\Omega_\mathsmaller{\mathrm{h}}\left(T,\mu,B\right)}{\partial B^2}\right].
\end{aligned}
\end{equation}

For the (bulk) Landau levels (and typical parameters of HgTe quantum wells), the different contributions to the grand potential read as
\begin{equation}\label{grand_potential_bulk_e}
\begin{aligned}
\Omega_\mathsmaller{\mathrm{e}}\left(T,\mu,B\right)&=\\
-\frac{SB}{\beta\Phi_0}\Biggl\{&\ln\left[1+\e^{-\beta\left(E^\uparrow(0)-\mu\right)}\right]\Theta\left[E^\uparrow(0)-E^\downarrow(0)\right]\\
&+\ln\left[1+\e^{-\beta\left(E^\downarrow(0)-\mu\right)}\right]\Theta\left[E^\downarrow(0)-E^\uparrow(0)\right]\\
&+\sum\limits_{n=1}^{\infty}\sum\limits_{s=\uparrow,\downarrow}\ln\left[1+\e^{-\beta\left(E^s_\mathsmaller{+}(n)-\mu\right)}\right]\Biggl\},
\end{aligned}
\end{equation}
\begin{equation}\label{grand_potential_bulk_h}
\begin{aligned}
\Omega_\mathsmaller{\mathrm{h}}\left(T,\mu,B\right)&=\\
-\frac{SB}{\beta\Phi_0}\Biggl\{&\ln\left[1+\e^{\beta\left(E^\downarrow(0)-\mu\right)}\right]\Theta\left[E^\uparrow(0)-E^\downarrow(0)\right]\\
&+\ln\left[1+\e^{\beta\left(E^\uparrow(0)-\mu\right)}\right]\Theta\left[E^\downarrow(0)-E^\uparrow(0)\right]\\
&+\sum\limits_{n=1}^{\infty}\sum\limits_{s=\uparrow,\downarrow}\ln\left[1+\e^{\beta\left(E^s_\mathsmaller{-}(n)-\mu\right)}\right]\Biggl\},
\end{aligned}
\end{equation}
and
\begin{equation}\label{grand_potential_bulk_gs}
\Omega_0(B)=\Omega_\mathsmaller{\mathrm{dis}}(B)+\tilde{\Omega}_0(B),
\end{equation}
where the energies are given by Eqs.~(\ref{Landau_levels_up})-(\ref{zero_Landau_level_down}) and $\Phi_0=2\pi\hbar/e$ is the magnetic flux quantum. In Eq.~(\ref{grand_potential_bulk_gs}), we have split the ground-state potential into a contribution from the uppermost valence band [which may not be continuously differentiable if there is a crossover between the hole-like $E^\downarrow(0)$ and the electron-like $E^\uparrow(0)$ bands like at the transition point in Fig.~\ref{fig:Landau_levels}],
\begin{equation}\label{uppervalence}
\begin{aligned}
\Omega_\mathsmaller{\mathrm{dis}}(B)=&E^\downarrow(0)\Theta\left[E^\uparrow(0)-E^\downarrow(0)\right]\\
&+E^\uparrow(0)\Theta\left[E^\downarrow(0)-E^\uparrow(0)\right],
\end{aligned}
\end{equation}
and a contribution from the remaining valence bands,
\begin{equation}\label{remainingvalence}
\begin{aligned}
\tilde{\Omega}_0(B)=\sum\limits_{n=1}^{\infty}\sum\limits_{s=\uparrow,\downarrow}E^s_\mathsmaller{-}(n).
\end{aligned}
\end{equation}

Since the energies in Eq.~(\ref{remainingvalence}) are not bounded from below (for typical parameters of HgTe quantum wells), the sum is divergent; following Refs.~\onlinecite{Koshino2007:PRB,Koshino2010:PRB,Ominato2012:PRB}, we introduce a smooth cutoff function which results in a smooth $\tilde{\Omega}_0(B)$ (we refer to the Appendix~\ref{Sec:AppendixGroundState} for more details). If there is no crossover between the electron-like $E^\uparrow(0)$ band and the hole-like $E^\downarrow(0)$ band, that is, if one deals with an ordinary insulator, then the total ground-state magnetization $M_0(B)$ is continuous. Due to $\Omega_\mathsmaller{\mathrm{dis}}(B)$, which is not continuously differentiable if the $E^\uparrow(0)$ and $E^\downarrow(0)$ bands cross (see Fig.~\ref{fig:Landau_levels}), the ground-state magnetization is not continuous at the crossover point in this case. For bulk Landau levels, we find the jumps
\begin{equation}\label{jump}
\Delta M_0=\lim\limits_{\delta B\to0}\left[M_0(B_\mathrm{c}+\delta B)-M_0(B_\mathrm{c}-\delta B)\right]=-\frac{2\mathcal{M}}{\Phi_0}
\end{equation}
at the crossover point $B_\mathrm{c}$, where there is a transition from the inverted [$E^\uparrow(0)<E^\downarrow(0)$] to the normal regime [$E^\downarrow(0)<E^\uparrow(0)$].

However, at finite temperatures or doping, the total magnetization is given by the sum of the ground-state magnetization $M_0(B)$ and the contribution from the electrons and holes, $M\left(T,\mu,B\right)$. Analyzing this contribution for the case of a transition from the inverted to the normal band structure, one finds that $M\left(T,\mu,B\right)$ vanishes for zero temperature and zero doping, but otherwise always contains a discontinuity at $B_\mathrm{c}$ which exactly cancels the discontinuity of the intrinsic magnetization. Thus, the total magnetization is a continuous function. If there is no transition between the normal and inverted band structures,  the non-vacuum contribution and therefore the total magnetization are also continuous. For a given quantum-well thickness $d$, the vacuum contribution $M_0(B)$ constitutes the same background for every set of thermodynamic variables ($\mu$, $T$) or ($n_d$, $T$) of the system. Thus, the quantity of interest which allows one to compare different doping levels, chemical potentials or temperatures of the system is the non-vacuum contribution $M\left(T,\mu,B\right)$.

Equations~(\ref{Mag_gs})-(\ref{grand_potential_bulk_gs}) allow us to calculate the (bulk) magnetization and susceptibility of HgTe quantum wells, the results of which are discussed in the following section.

\subsection{Results}\label{Sec:Mag_Results}

\begin{figure}[t]
\centering
\includegraphics*[width=8cm]{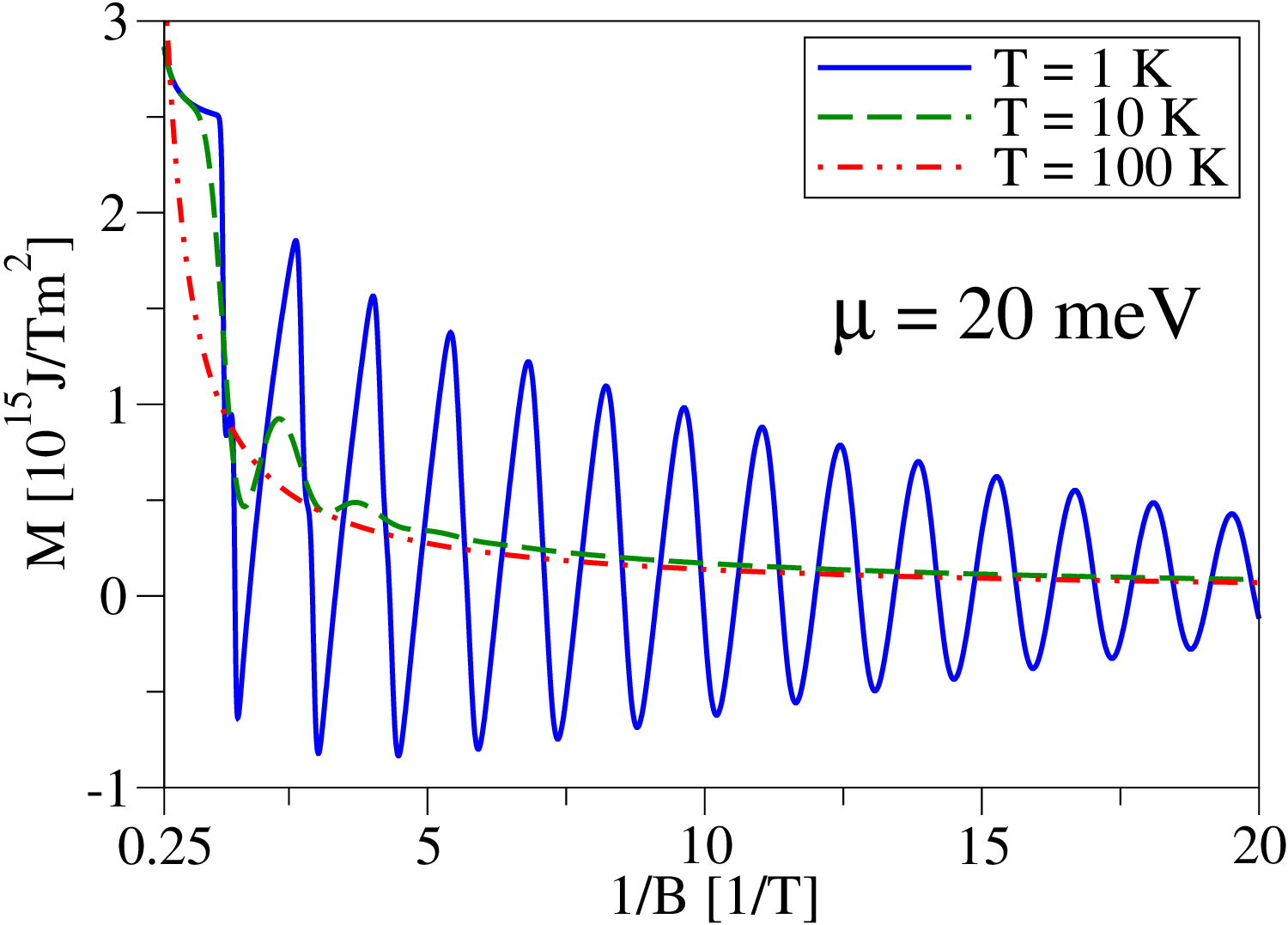}
\caption{(Color online) The non-vacuum magnetization $M\left(T,\mu,B\right)$ (corresponding to a quantum-well thickness of $d=7.0$ nm) plotted versus $1/B$ for a fixed chemical potential, $\mu=20$ meV, and different temperatures ($T=1,\,10,\,100$ K).}\label{fig:IMagNV_cp20}
\end{figure}

\begin{figure}[t]
\centering
\includegraphics*[width=8cm]{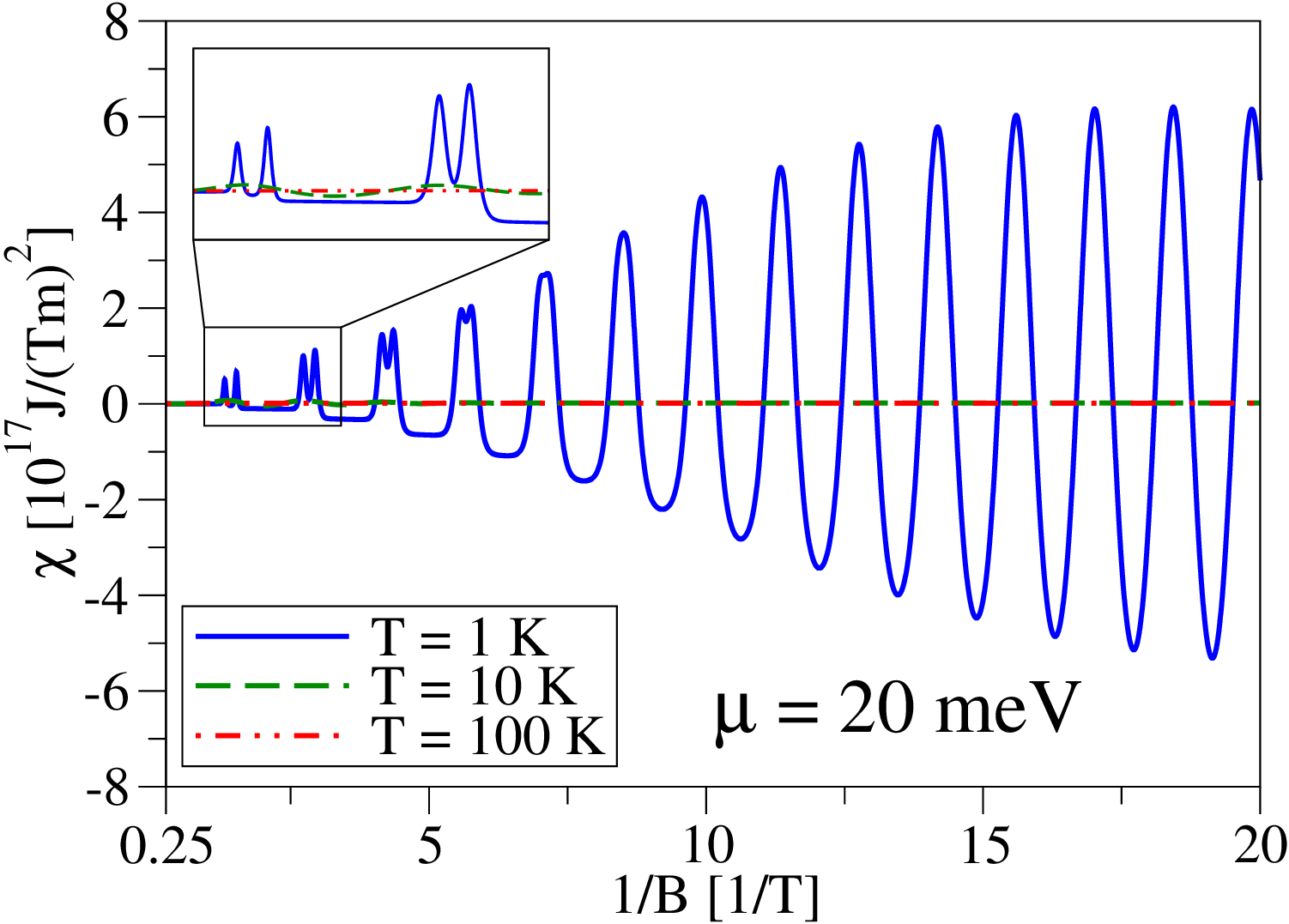}
\caption{(Color online) The non-vacuum susceptibility $\chi\left(T,\mu,B\right)$ (corresponding to a quantum-well thickness of $d=7.0$ nm) plotted versus $1/B$ for a fixed chemical potential, $\mu=20$ meV, and different temperatures ($T=1,\,10,\,100$ K).}\label{fig:ISusNV_cp20}
\end{figure}

In this section, we apply the formalism introduced above to calculate the bulk magnetization of HgTe for the parameter set corresponding to a quantum-well thickness of $d=7.0$ nm (nominally the QSH regime; see above), that is, a situation where there is a crossover between the $E^\uparrow(0)$ and $E^\downarrow(0)$ bands. Figures~\ref{fig:IMagNV_cp20} and~\ref{fig:ISusNV_cp20} show the magnetic field  dependence of the non-vacuum contributions, that is, the contribution arising from electrons and holes, to the magnetization and the susceptibility for a fixed chemical potential, several different temperatures, and magnetic fields well below the crossover point $B_\mathrm{c}\approx7.4$ T (compare to Sec.~\ref{Sec:MES_Results}). As different Landau levels cross the Fermi level with increasing magnetic field, one can observe the de Haas-van Alphen oscillations in the magnetization as well as in the susceptibility whose amplitude decreases with increasing temperature. For high magnetic fields (see the inset in Fig.~\ref{fig:ISusNV_cp20}), the spacing between the energies of spin-up and spin-down Landau levels (with the same quantum number $n$) is large enough compared to thermal broadening to observe spin-resolved peaks in the susceptibility. Fitting the oscillations of the magnetization to a periodic function, we find that the periodicity of those oscillations is given by $\Delta(1/B)\approx1.43$ 1/T [see also the Appendix~\ref{Sec:AppendixMag_Simple}, where Eq.~(\ref{gp_osc_e_final}) yields a period of $\Delta(1/B)\approx1.35$ 1/T for the main contribution to the oscillations in the reduced model].

\begin{figure}[t]
\centering
\includegraphics*[width=8cm]{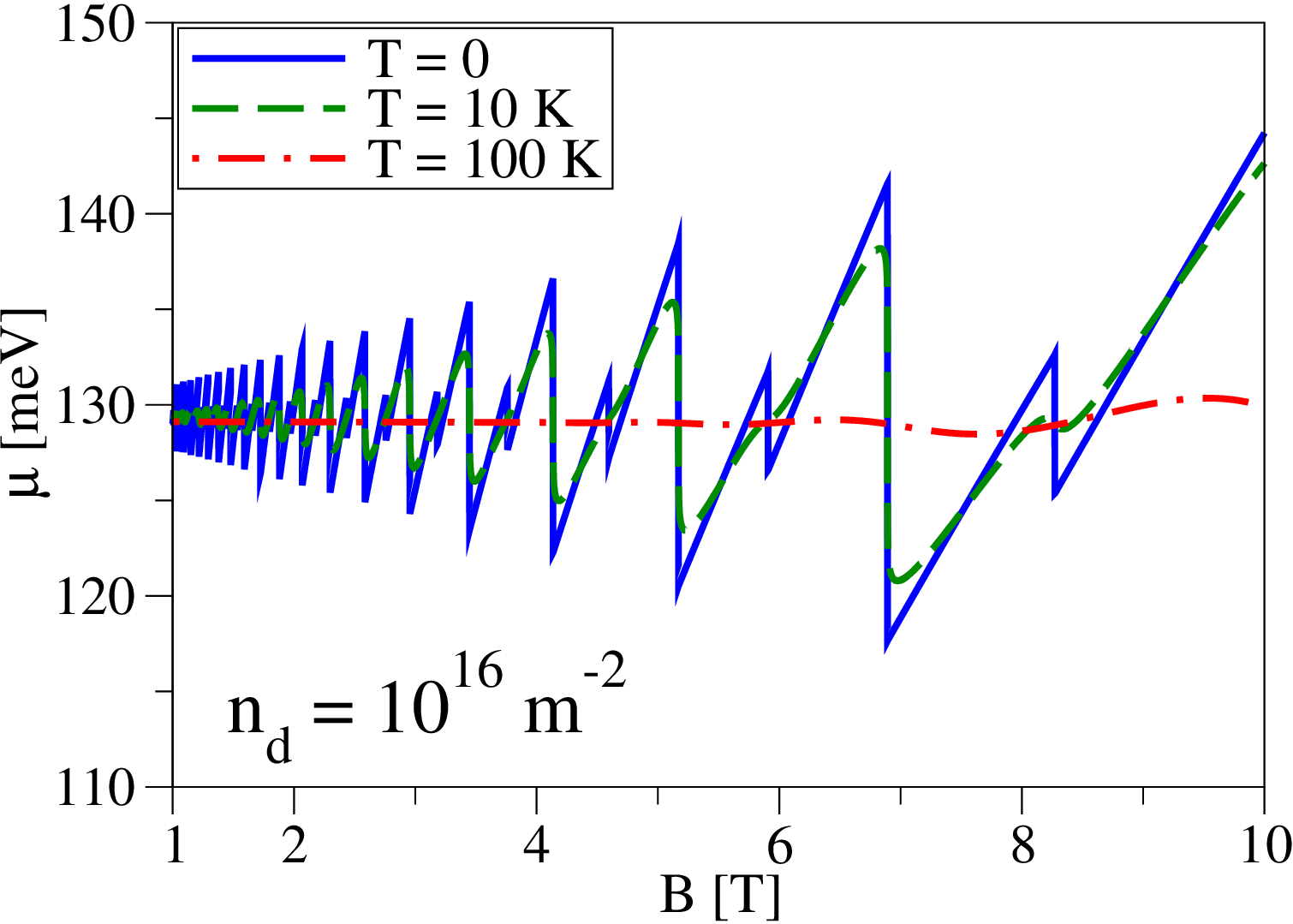}
\caption{(Color online) Magnetic field dependence of the chemical potential $\mu\left(T,n_d,B\right)$ (corresponding to a quantum-well thickness of $d=7.0$ nm) for $n_d=10^{16}\;1/\textrm{m}^2$ and different temperatures ($T=0,\,10,\,100\;K$).}\label{fig:cp_bd1e16}
\end{figure}

Next, we consider a fixed carrier density $n_d>0$. The corresponding chemical potential as a function of the magnetic field is calculated via Eq.~(\ref{chempot_calc}) and is displayed in Fig.~\ref{fig:cp_bd1e16} for the density $n_d=10^{16}\;1/\textrm{m}^2$ and different temperatures. With varying magnetic field, the Fermi energy $\mu\left(0,n_d,B\right)$ shows oscillations consisting of a pair of spin-resolved peaks, where each of those oscillations corresponds to a crossing of a Landau level with the Fermi level. Higher temperatures result in a smoothening of the oscillations and a diminution of their amplitudes. Moreover, thermal broadening leads to a removal of the spin-resolution at small magnetic fields.

\begin{figure}[t]
\centering
\includegraphics*[width=8cm]{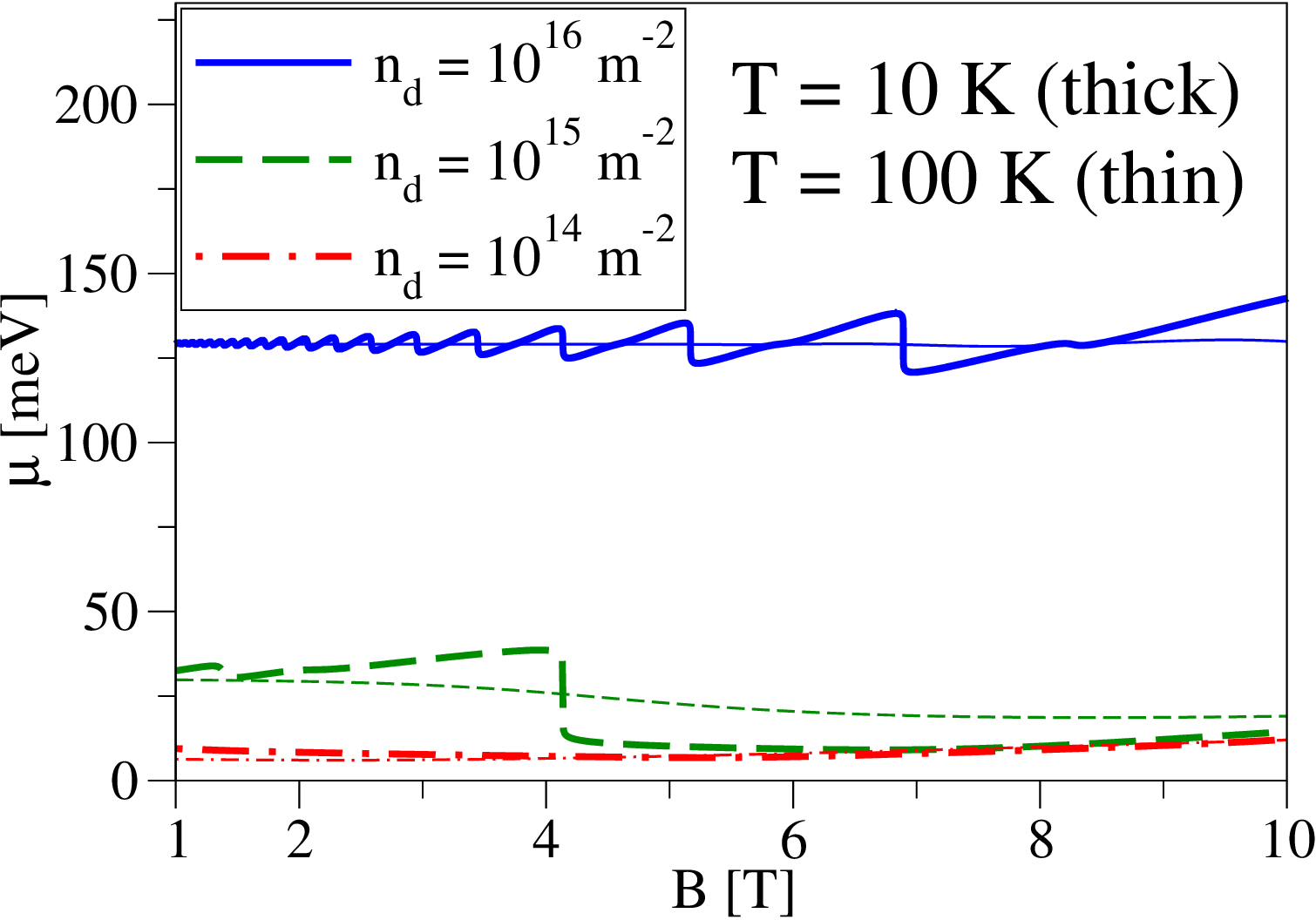}
\caption{(Color online) Magnetic field dependence of the chemical potential $\mu\left(T,n_d,B\right)$ (corresponding to a quantum-well thickness of $d=7.0$ nm) for $T=10$ K and $T=100$ K and different densities ($n_d=10^{14},\,10^{15},\,10^{16}\;1/\textrm{m}^2$).}\label{fig:cp}
\end{figure}

\begin{figure}[t]
\centering
\includegraphics*[width=8cm]{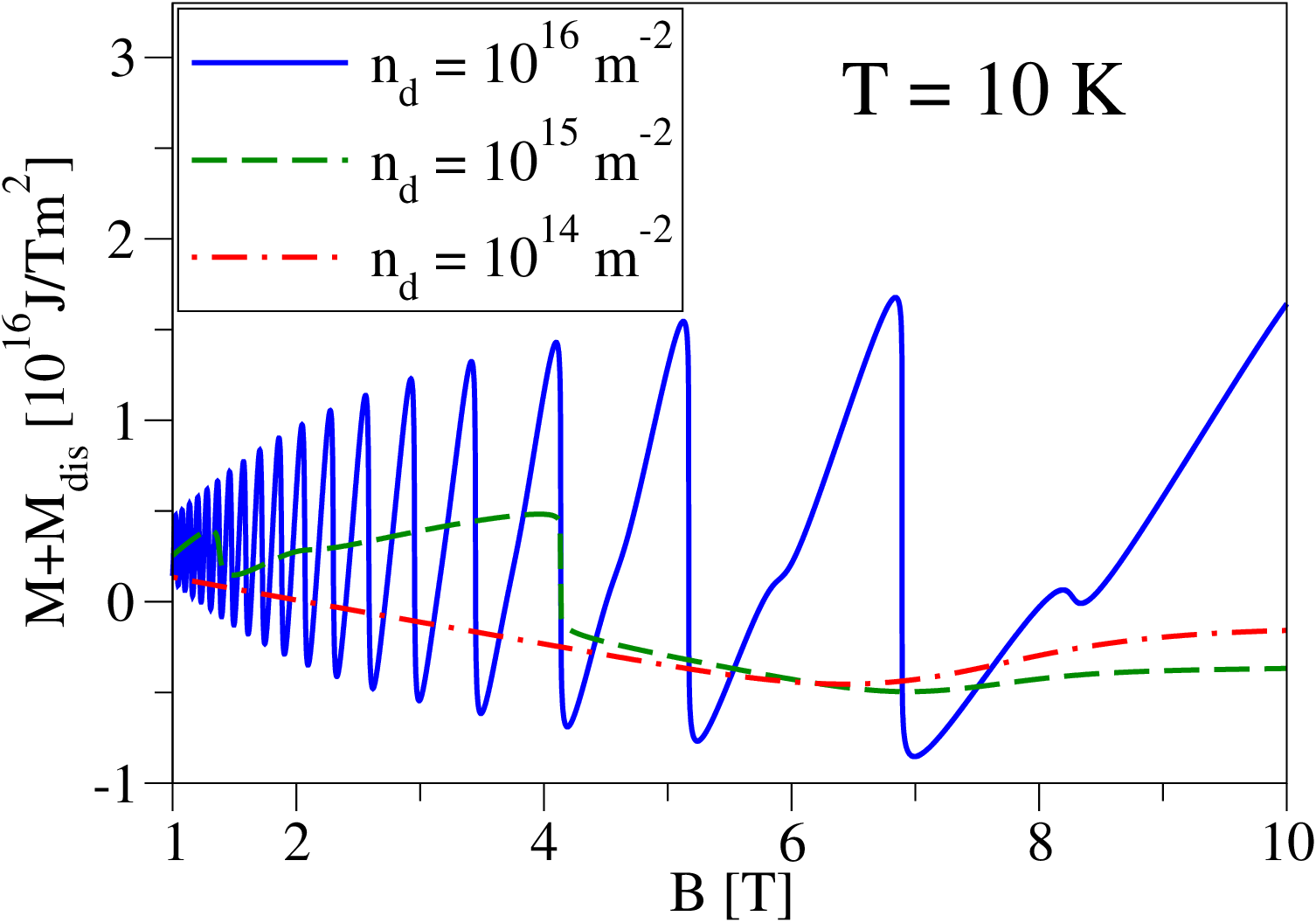}
\caption{(Color online) Magnetic field dependence of the contribution $M_\mathsmaller{\mathrm{dis}}(B)+M\left(T,\mu,B\right)$ (corresponding to a quantum-well thickness of $d=7.0$ nm) for $T=10$ K and different densities ($n_d=10^{14},\,10^{15},\,10^{16}\;1/\textrm{m}^2$).}\label{fig:MagehDis_T10}
\end{figure}

Figures~\ref{fig:cp} and~\ref{fig:MagehDis_T10} show the chemical potential and the combined contribution $M_\mathsmaller{\mathrm{dis}}(B)+M\left(T,\mu,B\right)$ to magnetization as functions of the magnetic field for $T=10$ K and different carrier densities $n_d$. \{Here, we have added the discontinuous contribution from the ground-state magnetization, $M_\mathsmaller{\mathrm{dis}}(B)=-(1/S)[\partial\Omega_\mathsmaller{\mathrm{dis}}(B)/\partial B]$, to the non-vacuum magnetization in order that the discontinuity at $B=B_\mathrm{c}$ be canceled.\} As above, one can see the de Haas-van Alphen oscillations in the magnetization (see Fig.~\ref{fig:MagehDis_T10}), which---for fixed carrier densities---follow the oscillations in the chemical potential (see Fig.~\ref{fig:cp}). At low densities, on the other hand, only the lowest conduction Landau level is occupied and the chemical potential roughly follows this level and there are consequently no oscillations.

\begin{figure}[t]
\centering
\includegraphics*[width=8cm]{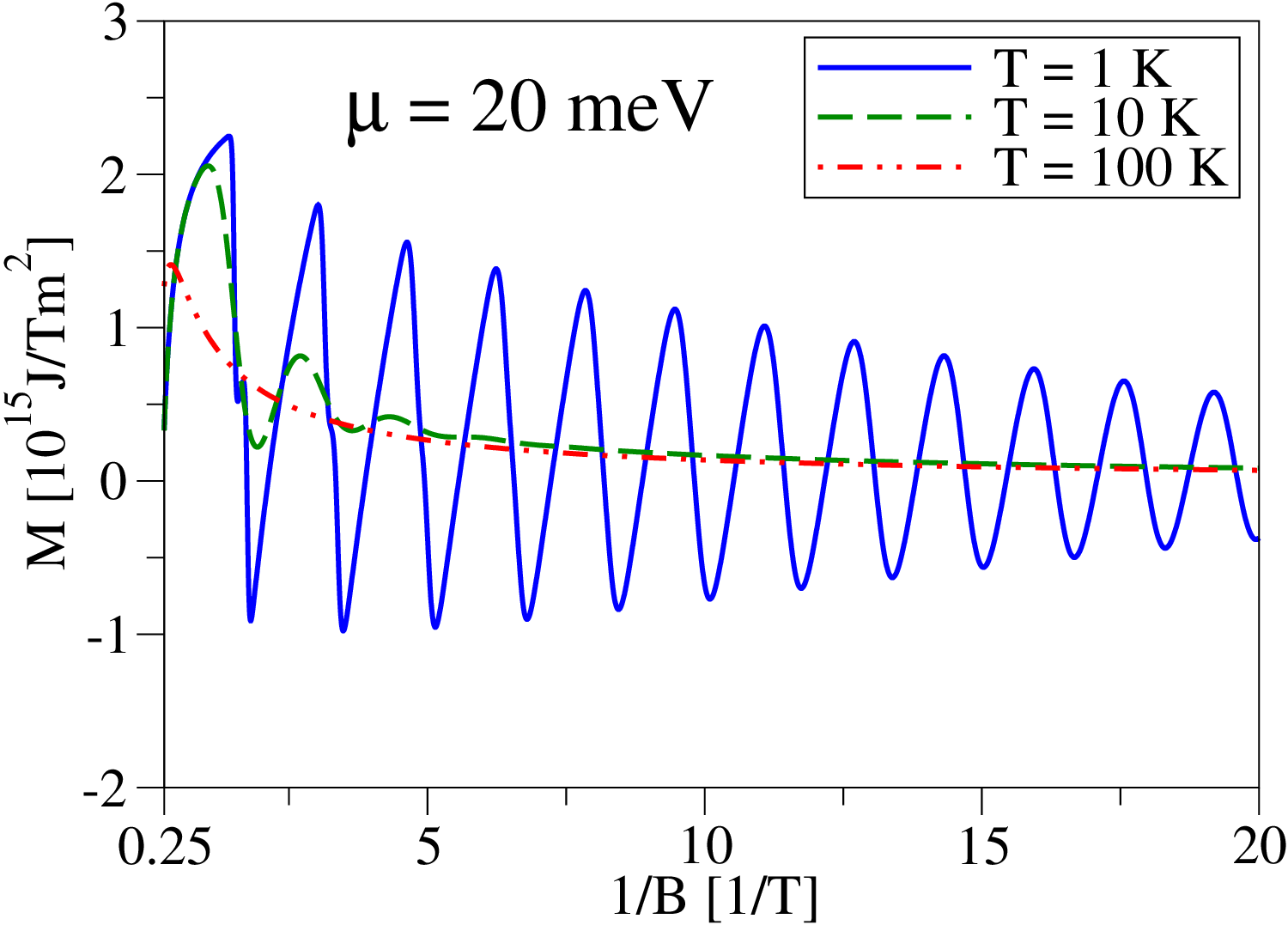}
\caption{(Color online) The non-vacuum magnetization $M\left(T,\mu,B\right)$ (corresponding to a quantum-well thickness of $d=5.5$ nm) plotted versus $1/B$ for a fixed chemical potential, $\mu=20$ meV, and different temperatures ($T=1,\,10,\,100$ K).}\label{fig:IMagNV_cp20_OI}
\end{figure}

\begin{figure}[t]
\centering
\includegraphics*[width=8cm]{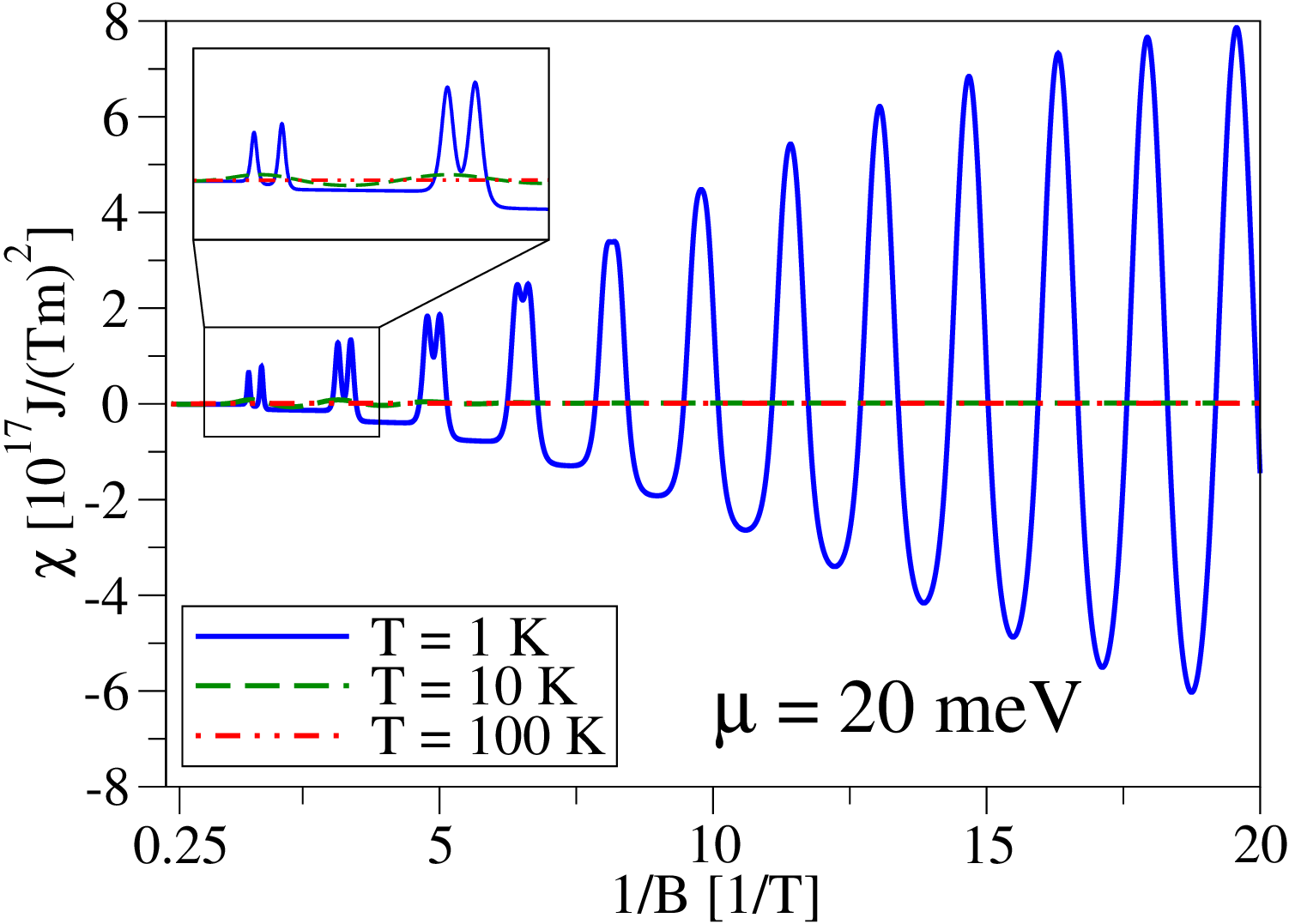}
\caption{(Color online) The non-vacuum susceptibility $\chi\left(T,\mu,B\right)$ (corresponding to a quantum-well thickness of $d=5.5$ nm) plotted versus $1/B$ for a fixed chemical potential, $\mu=20$ meV, and different temperatures ($T=1,\,10,\,100$ K).}\label{fig:ISusNV_cp20_OI}
\end{figure}

For the sake of comparison to the situation in the inverted regime discussed so far, Figs.~\ref{fig:IMagNV_cp20_OI} and~\ref{fig:ISusNV_cp20_OI} show the magnetic field  dependence of the non-vacuum contributions to the magnetization and the susceptibility in the normal regime (corresponding to the parameters for a quantum-well thickness of $d=5.5$ nm as in Sec.~\ref{Sec:MES_Results}) for a fixed chemical potential and several different temperatures. As in Figs.~\ref{fig:IMagNV_cp20} and~\ref{fig:ISusNV_cp20}, one can observe the de Haas-van Alphen oscillations. No discernible features are seen when comparing the inverted and normal regimes in the bulk.

In limiting cases, compact analytical formulas to describe some of the main features of the magnetization and the susceptibility shown above can be given for the reduced model and are presented in the Appendix~\ref{Sec:AppendixMag_Simple}.

\section{Conclusions}\label{Sec:Conclusions}
We have derived analytical formulas to calculate the energy spectra of HgTe quantum wells in infinite, semi-infinite, and finite-strip systems in the presence of perpendicular magnetic fields and hard walls. Complementary to the analytical formulas, we have also used a finite-difference scheme to investigate the magnetic field dependence of the energy spectra and their respective eigenstates in a finite-strip geometry for parameters corresponding to the normal ($d<d_c$), inverted ($d>d_c$), and critical regimes ($d\approx d_c$). In the inverted regime ($d>d_c$), we found that for magnetic fields below the crossover point between the uppermost (electron-like) valence and lowest (hole-like) conduction Landau levels, one can still observe counterpropagating, spin-polarized states at finite magnetic fields, although these states are no longer protected by time-reversal symmetry. Above the crossover point, the band structure becomes normal and one can no longer find those states. This situation is similar for parameters corresponding to the normal regime ($d<d_c$), where one cannot find counterpropagating, spin-polarized states even for zero or weak magnetic fields. Finally, we have studied the bulk magnetization and susceptibility in HgTe quantum wells and have investigated their dependence on the magnetic field, chemical potential, and carrier density. In the case of fixed chemical potentials as well as in the case of fixed densities, the magnetization (for both, the normal as well as the inverted regime) exhibits characteristic de Haas-van Alphen oscillations, which in the case of fixed carrier densities follow the oscillations in the chemical potential. Corresponding to those oscillations of the magnetization, on can also observe oscillations in the magnetic susceptibility. With increasing temperature, the amplitude of these oscillations decreases. Furthermore, we found that, if the band structure is inverted, the ground-state magnetization (and consequently also the ground-state susceptibility) is discontinuous at the crossover point between the uppermost valence and lowest conduction Landau levels.
At finite temperatures and/or doping, however, this discontinuity is canceled by the contribution from electrons and holes and the total magnetization and susceptibility are continuous.

\acknowledgments
This work was supported by the DFG SFB 689 and GRK 1570.

\appendix

\section{Landau levels}
\label{Sec:AppendixLLstates}
In the absence of any confining potential, we require the wave functions given by Eqs.~(\ref{general_solution_up}) and~(\ref{general_solution_down}) to vanish for $\xi\to\pm\infty$, which can only be satisfied if the indices of the parabolic cylindrical functions are non-negative integers $n$. As above, we first consider spin-up electrons. Then, Eqs.~(\ref{ansatz_xi}) and~(\ref{ansatz_-xi}) reduce to the ansatz
\begin{equation}\label{ansatz_LLup}
f_\uparrow(\xi)=v_1\phi_n(\xi/\sqrt{2})\quad\text{and}\quad g_\uparrow(\xi)=v_2\phi_{n-1}(\xi/\sqrt{2}),
\end{equation}
valid for $n\geq1$. For convenience, we have expressed the parabolic cylindrical functions $D_n(\xi)$ by the eigenfunctions of the one-dimensional harmonic oscillator,
\begin{equation}\label{HO_eigenfunctions}
\phi_n(\xi')=D_n(\sqrt{2}\xi')/\sqrt{n!\sqrt{\pi}}=\e^{-\xi'^2/2}H_n(\xi')/\sqrt{2^nn!\sqrt{\pi}},
\end{equation}
where $H_n(\xi')$ is the $n$-th Hermite polynomial. Inserting Eq.~(\ref{ansatz_LLup}) into Eq.~(\ref{effective_1D_SE_up}) and using the recurrence relations for the parabolic cylindrical functions~(\ref{reccurence_1}) and~(\ref{reccurence_2}) leads to the eigenvalue problem
\begin{widetext}
\begin{equation}\label{EVP_up}
\left(\begin{array}{cc}\left[\mathcal{C}+\mathcal{M}-\frac{\left(\mathcal{D}+\mathcal{B}\right)\left(2n+1\right)}{l_{\mathsmaller{B}}^2}+\frac{g_{\mathsmaller{\mathrm{e}}}\mu_{\mathsmaller{B}}B}{2}\right] & -\frac{\sqrt{2n}\mathcal{A}}{l_{\mathsmaller{B}}} \\
-\frac{\sqrt{2n}\mathcal{A}}{l_{\mathsmaller{B}}} & \left[\mathcal{C}-\mathcal{M}-\frac{\left(\mathcal{D}-\mathcal{B}\right)\left(2n-1\right)}{l_{\mathsmaller{B}}^2}+\frac{g_{\mathsmaller{\mathrm{h}}}\mu_{\mathsmaller{B}}B}{2}\right]
\end{array}\right)\left(\begin{array}{c}v_1\\v_2\end{array}\right)=E\left(\begin{array}{c}v_1\\v_2\end{array}\right).
\end{equation}
\end{widetext}
By determining the eigenvalues of Eq.~(\ref{EVP_up}) and their corresponding eigenvectors, we find the Landau levels~(\ref{Landau_levels_up}) and their respective (normalized) eigenstates
\begin{widetext}
\begin{equation}\label{LL_states_up}
\Psi^{\uparrow,\pm}_{n,k}(x,y)=\frac{\e^{\i kx}}{\sqrt{L}}\left(\begin{array}{c} \frac{\left(\sqrt{2n}\mathcal{A}/l_{\mathsmaller{B}}\mp\Delta_{\uparrow,n}/2\right)-\left[\mathcal{M}-\left(2\mathcal{B}n+\mathcal{D}\right)/l_{\mathsmaller{B}}^2+\left(g_{\mathsmaller{\mathrm{e}}}-g_{\mathsmaller{\mathrm{h}}}\right)\mu_{\mathsmaller{B}}B/4\right]}{\sqrt{\Delta_{\uparrow,n}\left(\Delta_{\uparrow,n}\mp2\sqrt{2n}\mathcal{A}/l_{\mathsmaller{B}}\right)}\sqrt{l_{\mathsmaller{B}}}}\phi_n\left(\frac{y-kl_{\mathsmaller{B}}^2}{l_{\mathsmaller{B}}}\right)\\ \frac{\left(\sqrt{2n}\mathcal{A}/l_{\mathsmaller{B}}\mp\Delta_{\uparrow,n}/2\right)+\left[\mathcal{M}-\left(2\mathcal{B}n+\mathcal{D}\right)/l_{\mathsmaller{B}}^2+\left(g_{\mathsmaller{\mathrm{e}}}-g_{\mathsmaller{\mathrm{h}}}\right)\mu_{\mathsmaller{B}}B/4\right]}{\sqrt{\Delta_{\uparrow,n}\left(\Delta_{\uparrow,n}\mp2\sqrt{2n}\mathcal{A}/l_{\mathsmaller{B}}\right)}\sqrt{l_{\mathsmaller{B}}}}\phi_{n-1}\left(\frac{y-kl_{\mathsmaller{B}}^2}{l_{\mathsmaller{B}}}\right)\\ 0\\ 0\\ \end{array}\right),
\end{equation}
\end{widetext}
where
\begin{equation}
\Delta_{\uparrow,n}=2\sqrt{\frac{2n\mathcal{A}^2}{l_{\mathsmaller{B}}^2}+\left(\mathcal{M}-\frac{2\mathcal{B}n+\mathcal{D}}{l_{\mathsmaller{B}}^2}+\frac{g_{\mathsmaller{\mathrm{e}}}-g_{\mathsmaller{\mathrm{h}}}}{4}\mu_{\mathsmaller{B}}B\right)^2}.
\end{equation}
Whereas Eqs.~(\ref{ansatz_LLup}) and~(\ref{LL_states_up}) are valid for $n\geq1$, one can also choose $n=0$ to satisfy the boundary conditions. Instead of Eq.~(\ref{ansatz_LLup}), one then has the ansatz
\begin{equation}\label{ansatz_0LLup}
f_\uparrow(\xi)=v_1\phi_0(\xi/\sqrt{2})\quad\text{and}\quad g_\uparrow(\xi)=0,
\end{equation}
which yields the single Landau level given by Eq.~(\ref{zero_Landau_level_up}) and its corresponding (normalized) eigenstates
\begin{equation}\label{LL_state_zero_up}
\Psi^{\uparrow}_{0,k}(x,y)=\frac{\e^{\i kx}}{\sqrt{L}}\frac{1}{\sqrt{l_{\mathsmaller{B}}}}\phi_0\left(\frac{y-kl_{\mathsmaller{B}}^2}{l_{\mathsmaller{B}}}\right)\left(\begin{array}{c} 1\\ 0\\ 0\\ 0\\ \end{array}\right).
\end{equation}

If a similar procedure is applied for the spin-down states, one finds the Landau levels given by Eq.~(\ref{Landau_levels_down}) with the eigenstates
\begin{widetext}
\begin{equation}\label{LL_states_down}
\Psi^{\downarrow,\pm}_{n,k}(x,y)=\frac{\e^{\i kx}}{\sqrt{L}}\left(\begin{array}{c}0\\ 0\\ \frac{-\left(\sqrt{2n}\mathcal{A}/l_{\mathsmaller{B}}\pm\Delta_{\downarrow,n}/2\right)-\left[\mathcal{M}-\left(2\mathcal{B}n-\mathcal{D}\right)/l_{\mathsmaller{B}}^2-\left(g_{\mathsmaller{\mathrm{e}}}-g_{\mathsmaller{\mathrm{h}}}\right)\mu_{\mathsmaller{B}}B/4\right]}{\sqrt{\Delta_{\downarrow,n}\left(\Delta_{\downarrow,n}\pm2\sqrt{2n}\mathcal{A}/l_{\mathsmaller{B}}\right)}\sqrt{l_{\mathsmaller{B}}}}\phi_{n-1}\left(\frac{y-kl_{\mathsmaller{B}}^2}{l_{\mathsmaller{B}}}\right)\\ \frac{-\left(\sqrt{2n}\mathcal{A}/l_{\mathsmaller{B}}\pm\Delta_{\downarrow,n}/2\right)+\left[\mathcal{M}-\left(2\mathcal{B}n-\mathcal{D}\right)/l_{\mathsmaller{B}}^2-\left(g_{\mathsmaller{\mathrm{e}}}-g_{\mathsmaller{\mathrm{h}}}\right)\mu_{\mathsmaller{B}}B/4\right]}{\sqrt{\Delta_{\downarrow,n}\left(\Delta_{\downarrow,n}\pm2\sqrt{2n}\mathcal{A}/l_{\mathsmaller{B}}\right)}\sqrt{l_{\mathsmaller{B}}}}\phi_n\left(\frac{y-kl_{\mathsmaller{B}}^2}{l_{\mathsmaller{B}}}\right)\\ \end{array}\right),
\end{equation}
\end{widetext}
where
\begin{equation}
\Delta_{\downarrow,n}=2\sqrt{\frac{2n\mathcal{A}^2}{l_{\mathsmaller{B}}^2}+\left(\mathcal{M}-\frac{2\mathcal{B}n-\mathcal{D}}{l_{\mathsmaller{B}}^2}-\frac{g_{\mathsmaller{\mathrm{e}}}-g_{\mathsmaller{\mathrm{h}}}}{4}\mu_{\mathsmaller{B}}B\right)^2},
\end{equation}
and the single Landau level given by Eq.~(\ref{zero_Landau_level_down}) with the eigenstate
\begin{equation}\label{LL_state_zero_down}
\Psi^{\downarrow}_{0,k}(x,y)=\frac{\e^{\i kx}}{\sqrt{L}}\frac{1}{\sqrt{l_{\mathsmaller{B}}}}\phi_0\left(\frac{y-kl_{\mathsmaller{B}}^2}{l_{\mathsmaller{B}}}\right)\left(\begin{array}{c} 0\\ 0\\ 0\\ 1\\ \end{array}\right).
\end{equation}

\section{Ground-state magnetization}\label{Sec:AppendixGroundState}
As mentioned in Sec.~\ref{Sec:Mag_Formalism}, the ground-state energy~(\ref{grand_potential_bulk_gs}) can be split in a---possibly not continuously differentiable---contribution from the uppermost valence band, $\Omega_\mathsmaller{\mathrm{dis}}(B)$ given by Eq.~(\ref{uppervalence}), and a contribution from the remaining valence bands, $\tilde{\Omega}_0(B)$ given by Eq.~(\ref{remainingvalence}). Likewise, one can divide the magnetization of the ground state into
\begin{equation}\label{gsMag_uppervalence}
M_\mathsmaller{\mathrm{dis}}(B)=-\frac{1}{S}\frac{\partial\Omega_\mathsmaller{\mathrm{dis}}(B)}{\partial B}
\end{equation}
and
\begin{equation}\label{gsMag_remainingvalence}
\tilde{M}_0(B)=-\frac{1}{S}\frac{\partial\tilde{\Omega}_0(B)}{\partial B}.
\end{equation}

\begin{figure}[t]
\centering
\includegraphics*[width=8cm]{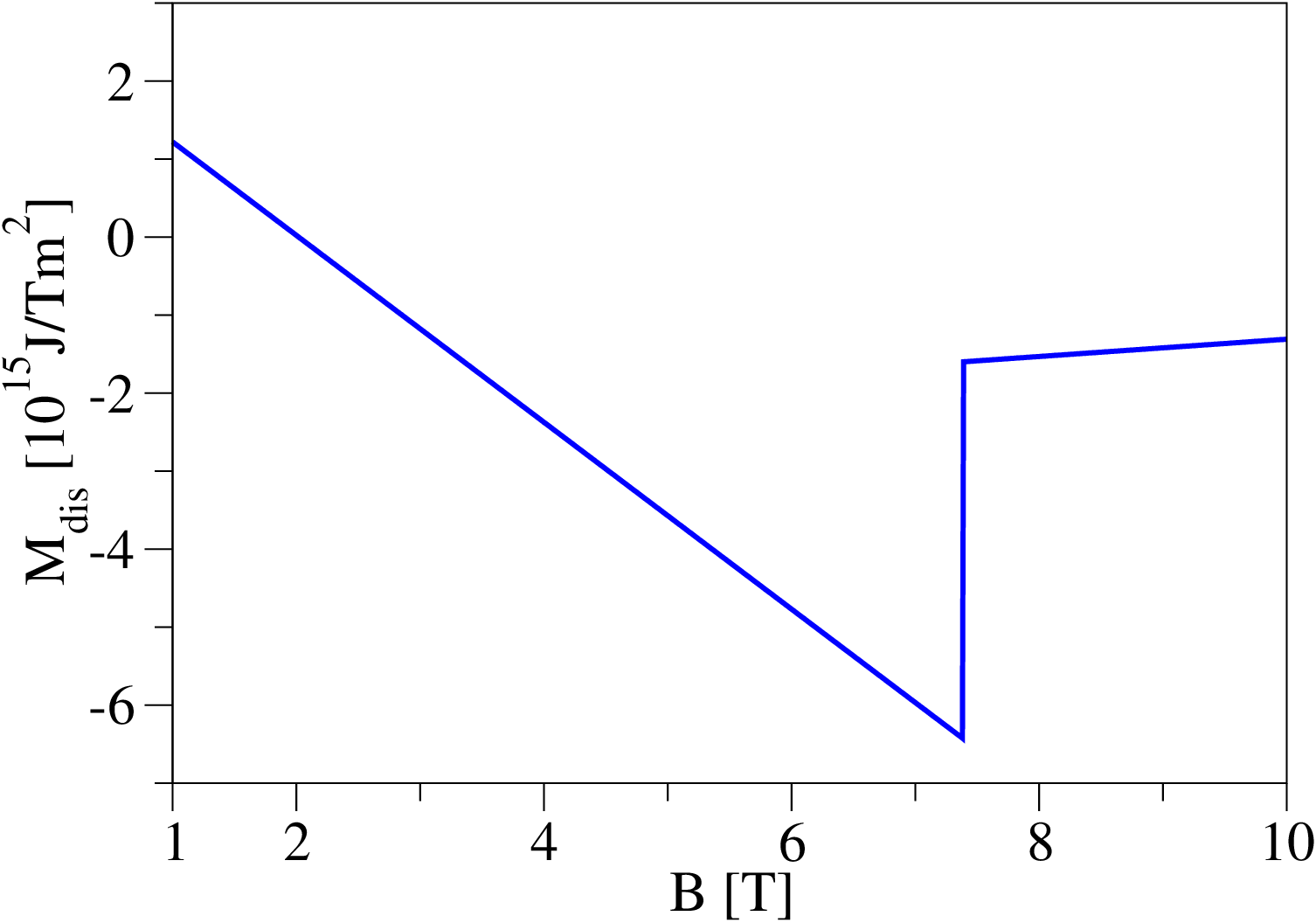}
\caption{(Color online) Magnetic field dependence of the magnetization $M_\mathrm{dis}\left(B\right)$ (corresponding to a quantum-well thickness of $d=7.0$ nm).}\label{fig:MagGS_Dis}
\end{figure}

\begin{figure}[t]
\centering
\includegraphics*[width=8cm]{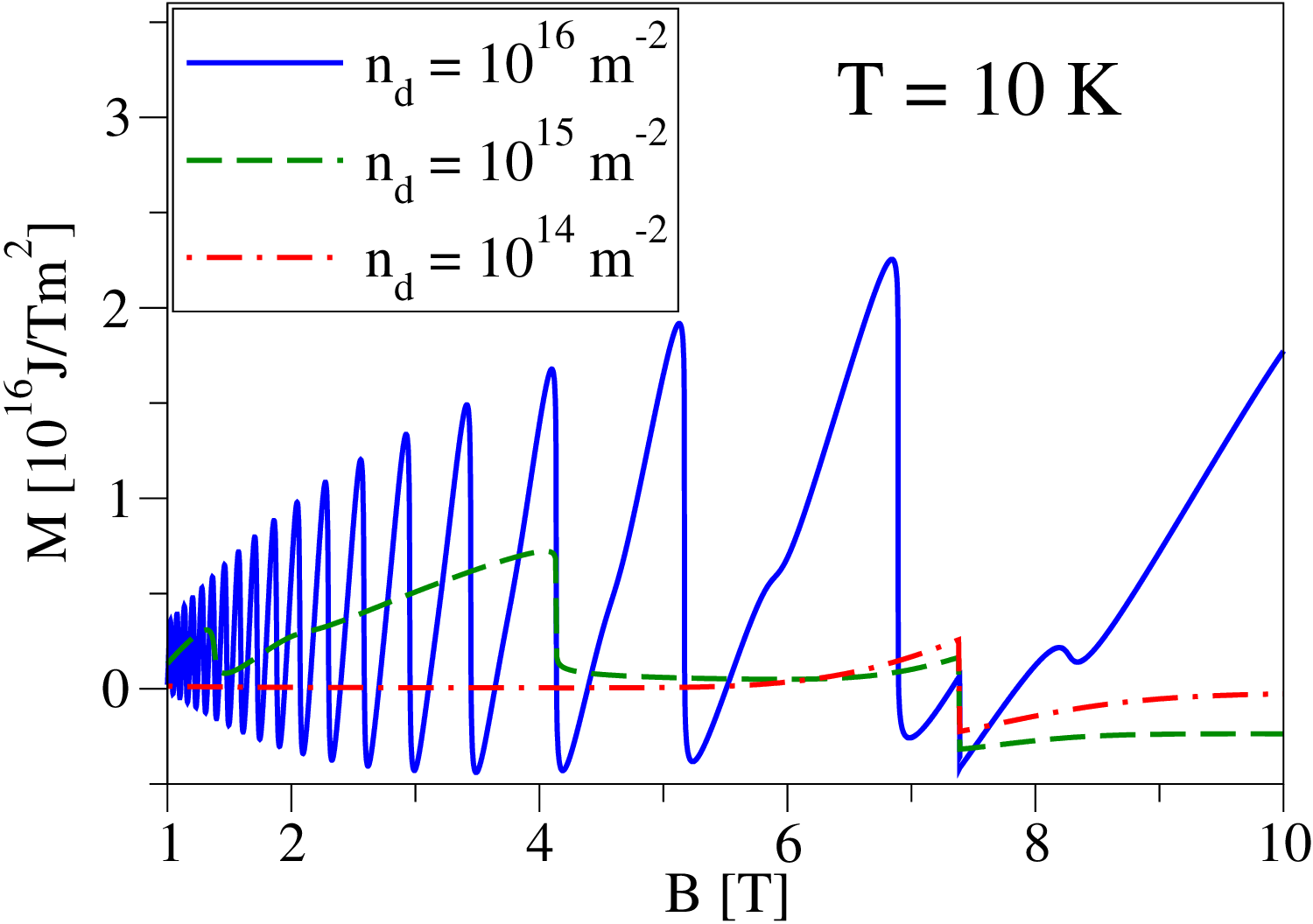}
\caption{(Color online) Magnetic field dependence of the non-vacuum magnetization $M\left(T,\mu,B\right)$ (corresponding to a quantum-well thickness of $d=7.0$ nm) for $T=10$ K and different densities ($n_d=10^{14},\,10^{15},\,10^{16}\;1/\textrm{m}^2$).}\label{fig:MagNV_T10}
\end{figure}

Figure~\ref{fig:MagGS_Dis} shows the contribution to the magnetization from the uppermost valence band, $M_\mathrm{dis}\left(B\right)$, for parameters corresponding to the quantum-well thickness of $d=7.0$ nm, that is, the inverted regime. Here, one can clearly see the discontinuity of $M_\mathrm{dis}\left(B\right)$ at $B=B_\mathrm{c}$. Comparing $M_\mathrm{dis}\left(B\right)$ to the non-vacuum contribution $M\left(T,\mu,B\right)$ which is shown in Fig.~\ref{fig:MagNV_T10} for $T=10$ K and different densities illustrates how the discontinuity of $M\left(T,\mu,B\right)$ is canceled by the discontinuity of $M_\mathrm{dis}\left(B\right)$. The resulting magnetization can be seen in Fig.~\ref{fig:MagehDis_T10} in Sec.~\ref{Sec:Mag_Results}.

\begin{figure}[t]
\centering
\includegraphics*[width=8cm]{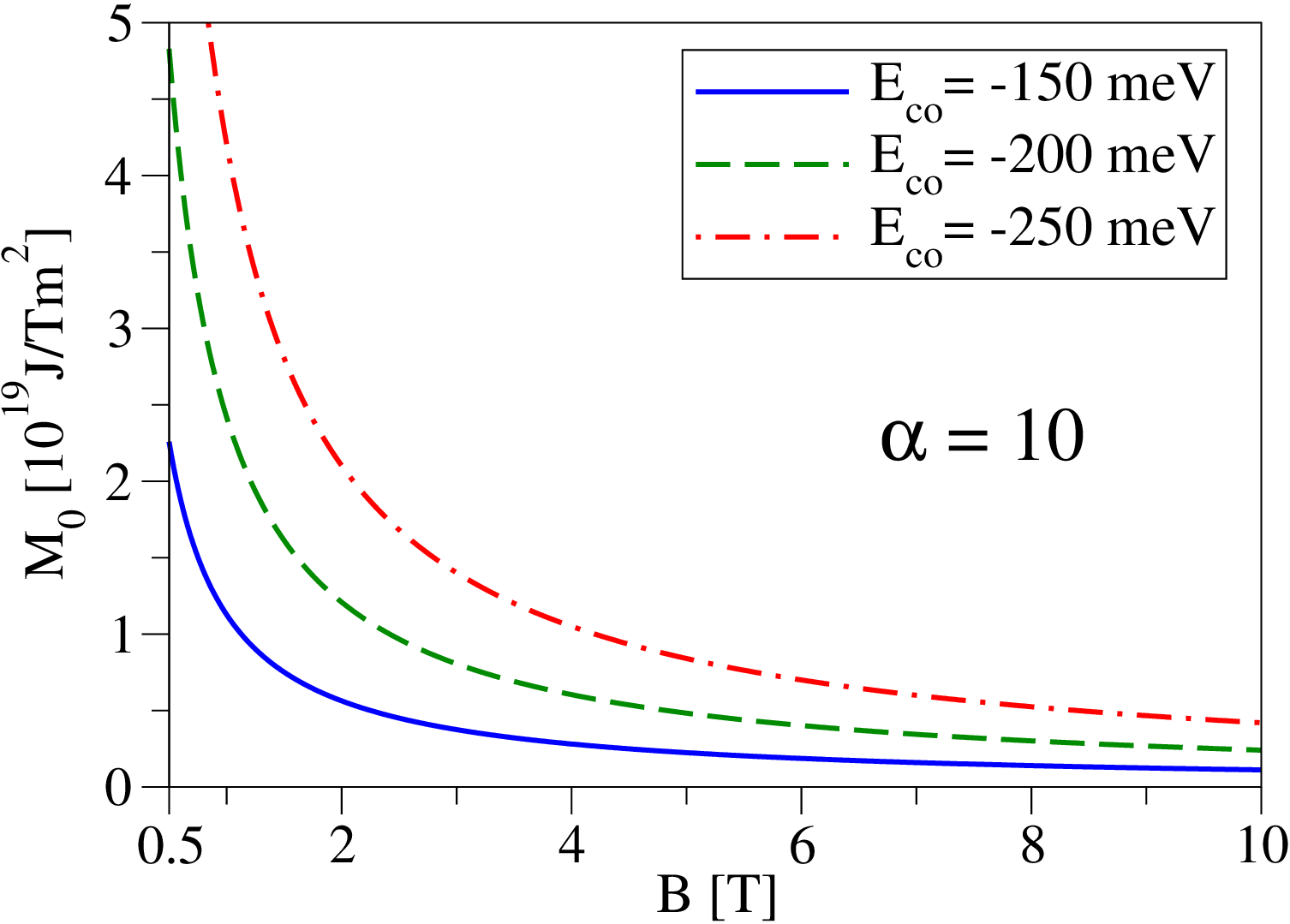}
\caption{(Color online) Magnetic field dependence of the vacuum magnetization $\tilde{M}_0\left(B\right)$ corresponding to a quantum-well thickness of $d=7.0$ nm and $\alpha=10$.}\label{fig:MagGS_QSH}
\end{figure}

\begin{figure}[t]
\centering
\includegraphics*[width=8cm]{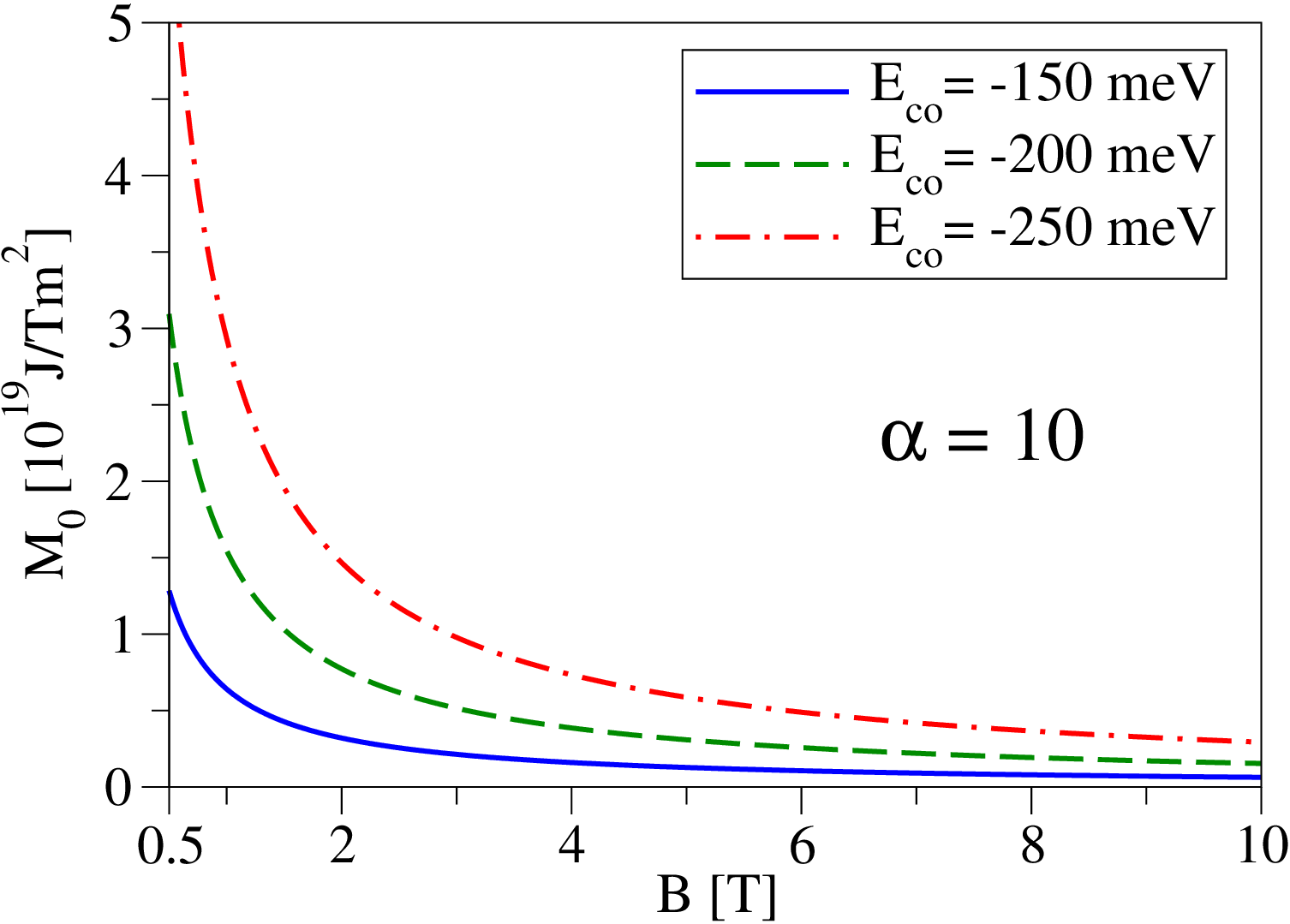}
\caption{(Color online) Magnetic field dependence of the vacuum magnetization $\tilde{M}_0\left(B\right)$ corresponding to a quantum-well thickness of $d=5.5$ nm and $\alpha=10$.}\label{fig:MagGS_OI}
\end{figure}

Apart from the contribution of $M_\mathrm{dis}\left(B\right)+M\left(T,\mu,B\right)$, there is also a contribution arising from the remaining valence bands, $\tilde{M}_0(B)$. When using the effective model for HgTe quantum wells given by Eq.~(\ref{effective_Hamiltonian}), the valence band Landau levels are not bounded from below and, thus, the sum over them is divergent. However, the effective model used in this manuscript is only valid for low energies and there should be a lower bound for the valence band Landau levels of the real band structure. To remedy this, we adopt the approach from Refs.~\onlinecite{Koshino2007:PRB,Koshino2010:PRB,Ominato2012:PRB} and introduce a smooth cutoff function $g_\mathrm{co}(\epsilon)=E_\mathrm{co}^\alpha/(\epsilon^\alpha+E_\mathrm{co}^\alpha)$ which we include in the thermodynamical quantities to smoothly cut off the respective summation over the Landau levels. Here, $E_\mathrm{co}$ and $\alpha$ denote the energy cutoff for the valence band Landau levels and a positive integer, respectively. Figures~\ref{fig:MagGS_QSH} and~\ref{fig:MagGS_OI} show the contribution from $\tilde{M}_0(B)$ for $\alpha=10$, several different energy cutoffs $E_\mathrm{co}$, and band parameters in the inverted ($d=7.0$ nm) and normal ($d=5.5$ nm) regimes, respectively. The main feature in these graphs is the decay of the magnetization with increasing magnetic field, indicating a negative susceptibility and therefore diamagnetism.

\section{Magnetization: Simplified model}\label{Sec:AppendixMag_Simple}
In the following, we briefly discuss the magnetization for the special case of the reduced model for Eq.~(\ref{effective_Hamiltonian}) mentioned in Sec.~\ref{Sec:Model}. If one chooses $\mathcal{C}=0$, the bulk Landau levels~(\ref{Landau_levels_up})-(\ref{zero_Landau_level_down}) reduce to
\begin{equation}\label{zero_Landau_levels_simple}
E^{\uparrow/\downarrow}(0)=\pm\mathcal{M}
\end{equation}
and the degenerate levels
\begin{equation}\label{Landau_levels_simple}
\begin{aligned}
E_{\pm}^{\uparrow/\downarrow}(n)=\pm\sqrt{\frac{2n\mathcal{A}^2}{l_{\mathsmaller{B}}^2}+\mathcal{M}^2}
\end{aligned}
\end{equation}
in this case.

If the simplified expressions~(\ref{zero_Landau_levels_simple}) and~(\ref{Landau_levels_simple}) are used, the different contributions to the grand potential, $\Omega'\left(T,\mu,B\right)$, Eqs.~(\ref{grand_potential_bulk_e}), (\ref{grand_potential_bulk_h}), and~(\ref{grand_potential_bulk_gs}), read as
\begin{equation}\label{grand_potential_e_simple}
\Omega_\mathsmaller{\mathrm{e}}\left(T,\mu,B\right)=\frac{1}{2}f(0)+\sum\limits_{n=1}^{\infty}f(n),
\end{equation}
\begin{equation}\label{grand_potential_h_simple}
\Omega_\mathsmaller{\mathrm{h}}\left(T,\mu,B\right)=\Omega_\mathsmaller{\mathrm{e}}\left(T,-\mu,B\right),
\end{equation}
and
\begin{equation}\label{grand_potential_gs_simple}
\Omega_0(B)=\frac{1}{2}g(0)+\sum\limits_{n=1}^{\infty}g(n),
\end{equation}
where
\begin{equation}\label{def_grand_potential_e_simple}
f(x)=-\frac{2SB}{\beta\Phi_0}\ln\left[1+\e^{-\beta\left(\sqrt{2x\mathcal{A}^2/l_{\mathsmaller{B}}^2+\mathcal{M}^2}-\mu\right)}\right]
\end{equation}
and
\begin{equation}\label{def_grand_potential_gs_simple}
g(x)=-\frac{2SB}{\Phi_0}\sqrt{\frac{2x\mathcal{A}^2}{l_{\mathsmaller{B}}^2}+\mathcal{M}^2}.
\end{equation}

In the following, we will look at the behavior of the magnetization in the regime of $2(\mathcal{A}\beta/l_{\mathsmaller{B}})^2\ll1$ as well as the de Haas-van Alphen oscillations within the model given by Eqs.~(\ref{zero_Landau_levels_simple}) and~(\ref{Landau_levels_simple}). For both cases, we assume to be in the degenerate limit, that is, $\beta|\mu|\gg1$. Since the Landau levels of this reduced model correspond to those of two-dimensional Dirac fermions, most notably those of (monolayer) graphene, one can apply the same procedures as in these cases.

\subsubsection{'Weak' magnetic fields}
For magnetic fields with $2(\mathcal{A}\beta/l_{\mathsmaller{B}})^2\ll1$, we follow the classic Landau approach\cite{LandauLifshitz1999} and use the Euler-Maclaurin formula to express $\Omega_\mathsmaller{\mathrm{e}}\left(T,\mu,B\right)$ as
\begin{equation}\label{expansion_e_simple}
\Omega_\mathsmaller{\mathrm{e}}\left(T,\mu,B\right)\approx\int\limits_0^\infty\d x\,f(x)-\frac{1}{12}\left.\frac{\d f(x)}{\d x}\right|_{x=0}.
\end{equation}
When conducting the transformation $x/l^2_{\mathsmaller{B}}\to x$, one can see that the integral in Eq.~(\ref{expansion_e_simple}) [denoted as $F(T,\mu)$ in the following] does not depend on the magnetic field and one arrives at
\begin{equation}\label{weakfields_grandpotential_e_simple}
\Omega_\mathsmaller{\mathrm{e}}\left(T,\mu,B\right)\approx F(T,\mu)-\frac{S\mathcal{A}^2}{12\pi l^4_{\mathsmaller{B}}\left|\mathcal{M}\right|}\frac{1}{\e^{\beta\left(\left|\mathcal{M}\right|-\mu\right)}+1}.
\end{equation}

By the same procedure [and assuming a cutoff for $g(x)$], we obtain
\begin{equation}\label{weakfields_grandpotential_gs_simple}
\Omega_0\left(B\right)\approx c_0+\frac{S\mathcal{A}^2}{12\pi l^4_{\mathsmaller{B}}\left|\mathcal{M}\right|},
\end{equation}
where $c_0$ does not depend on the magnetic field. Then, the grand canonical potential can be written as
\begin{equation}\label{weakfields_grandpotential_tot_simple}
\begin{aligned}
\Omega'&\left(T,\mu,B\right)=\Omega_0(B)+\Omega_\mathsmaller{\mathrm{e}}\left(T,\mu,B\right)+\Omega_\mathsmaller{\mathrm{e}}\left(T,-\mu,B\right)\\
&\approx \tilde{F}(T,\mu)+\frac{S\pi\mathcal{A}^2B^2}{3\Phi_0^2\left|\mathcal{M}\right|}\frac{\sinh\left(\beta\left|\mathcal{M}\right|\right)}{\cosh\left(\beta\left|\mathcal{M}\right|\right)+\cosh\left(\beta\mu\right)},
\end{aligned}
\end{equation}
where the different $B$-independent contributions have been combined in the function $\tilde{F}(T,\mu)$. Note, that the expansion used to arrive at Eq.~(\ref{weakfields_grandpotential_tot_simple}) is valid for $2(\mathcal{A}\beta/l_{\mathsmaller{B}})^2\ll1$.

Consequently, we find for the magnetic susceptibility
\begin{equation}
\chi_\mathsmaller{\mathrm{tot}}\left(T,\mu\right)=-\frac{2\pi\mathcal{A}^2}{3\Phi_0^2\left|\mathcal{M}\right|}\frac{\sinh\left(\beta\left|\mathcal{M}\right|\right)}{\cosh\left(\beta\left|\mathcal{M}\right|\right)+\cosh\left(\beta\mu\right)},
\end{equation}
implying that the system is diamagnetic. This result generalizes the zero-temperature formula of graphene found in Ref.~\onlinecite{Koshino2010:PRB}, but also the $\mathcal{M}=0$ model of Pb$_{1-x}$Sn$_x$Te interface states found in Ref.~\onlinecite{Volkov1985:JETPL}.

\subsubsection{De Haas-van Alphen oscillations}
To calculate the de Haas-van Alphen oscillations for $|\mu|>|\mathcal{M}|$, we only need to look at the non-vacuum contributions $\Omega_\mathsmaller{\mathrm{e}}\left(T,\mu,B\right)$ and $\Omega_\mathsmaller{\mathrm{h}}\left(T,\mu,B\right)$. We again follow Ref.~\onlinecite{LandauLifshitz1999} as well as Ref.~\onlinecite{Cheremisin2011:arXiv} and use Poisson's summation formula to write
\begin{equation}\label{Poisson_e_simple}
\Omega_\mathsmaller{\mathrm{e}}\left(T,\mu,B\right)\approx\int\limits_0^\infty\d x\,f(x)+2\mathrm{Re}\left[\sum\limits_{k=1}^{\infty}\int\limits_0^\infty\d x\,f(x)\e^{2\pi\i kx}\right],
\end{equation}
where the first and second terms describe the non-oscillating and oscillating parts of the grand potential, respectively. Here, we are interested in the oscillating part [denoted by $\Omega^{\mathsmaller{\mathrm{e}}}_\mathsmaller{\mathrm{osc}}\left(T,\mu,B\right)$ in the following]. This part can be rewritten as
\begin{equation}\label{int_osc_e_simple}
\begin{aligned}
\Omega^{\mathsmaller{\mathrm{e}}}_\mathsmaller{\mathrm{osc}}&\left(T,\mu,B\right)=\\
&-\frac{4SB}{\beta\Phi_0}\mathrm{Re}\left\{\sum\limits_{k=1}^{\infty}\frac{1}{2\pi\i k\xi}\int\limits_{\left|\mathcal{M}/\mu\right|}^\infty\d y\frac{\e^{2\pi\i kx(y)}}{\e^{[y-\mathrm{sgn}(\mu)]/\xi}+1}\right\},
\end{aligned}
\end{equation}
where
\begin{equation}
x(y)=\frac{1}{2}\left(\frac{\mu l_{\mathsmaller{B}}}{\mathcal{A}}\right)^2\left(y^2-\frac{\mathcal{M}^2}{\mu^2}\right)
\end{equation}
and $\xi=1/(\beta|\mu|)$.

We first consider the case $\mu>|\mathcal{M}|$. In this case, a major contribution to the integral originates from the vicinity of the Fermi level, that is, from $y\sim1$, whereas the integrand is damped for values $y\gtrsim1$. Therefore, we expand $x(y)$ around $y=1$ and replace the lower boundary of the integral by $y\to-\infty$. Changing the integration variable to $x=(y-1)/\xi$, we find that the oscillating part of the grand potential is given by
\begin{equation}\label{int2_osc_e_simple}
\begin{aligned}
\Omega^{\mathsmaller{\mathrm{e}}}_\mathsmaller{\mathrm{osc}}&\left(T,\mu,B\right)=\\
&\frac{2SB}{\pi\Phi_0\beta}\mathrm{Re}\Biggl\{\sum\limits_{k=1}^{\infty}\frac{\i\e^{\i\pi k(l_{\mathsmaller{B}}/\mathcal{A})^2\left(\mu^2-\mathcal{M}^2\right)}}{k}\\
&\times\int\limits_{-\infty}^\infty\d x\frac{\e^{2\pi\i(\mu l_{\mathsmaller{B}}/\mathcal{A})^2\xi kx}}{\e^x+1}\Biggr\}.
\end{aligned}
\end{equation}
Computing the above integral, we can write the oscillating part of the electronic contribution to the grand potential as
\begin{equation}\label{gp_osc_e_final}
\begin{aligned}
\Omega^{\mathsmaller{\mathrm{e}}}_\mathsmaller{\mathrm{osc}}\left(T,\mu,B\right)=\frac{2SB}{\Phi_0\beta}\sum\limits_{k=1}^{\infty}\frac{\cos\left[\pi k(l_{\mathsmaller{B}}/\mathcal{A})^2\left(\mu^2-\mathcal{M}^2\right)\right]}{k\sinh\left[2\pi^2k\xi(\mu l_{\mathsmaller{B}}/\mathcal{A})^2\right]},
\end{aligned}
\end{equation}
with $\mu>|\mathcal{M}|$. For $\mu<|\mathcal{M}|$, the contribution from the oscillating part of the electrons is much smaller than Eq.~(\ref{gp_osc_e_final}) and in the case of $\mu<-|\mathcal{M}|$, the main contribution arises from the hole contribution given by $\Omega^{\mathsmaller{\mathrm{h}}}_\mathsmaller{\mathrm{osc}}\left(T,\mu,B\right)=\Omega^{\mathsmaller{\mathrm{e}}}_\mathsmaller{\mathrm{osc}}\left(T,-\mu,B\right)$. Thus, the total oscillating part of the grand potential is given by Eq.~(\ref{gp_osc_e_final}) for any $|\mu|>|\mathcal{M}|$. By taking the derivative, one obtains the oscillating part of the total magnetization, which is periodic in $1/B$.

Finally, we emphasize that this reduced model discussed here cannot describe a transition between inverted and normal band structures and can thus only be used for magnetic fields well below the crossover point (or for situations where there is no crossover at all).

\bibliographystyle{apsrev}
\bibliography{BibTopIns}

\begin{thebibliography}{49}
\expandafter\ifx\csname natexlab\endcsname\relax\def\natexlab#1{#1}\fi
\expandafter\ifx\csname bibnamefont\endcsname\relax
  \def\bibnamefont#1{#1}\fi
\expandafter\ifx\csname bibfnamefont\endcsname\relax
  \def\bibfnamefont#1{#1}\fi
\expandafter\ifx\csname citenamefont\endcsname\relax
  \def\citenamefont#1{#1}\fi
\expandafter\ifx\csname url\endcsname\relax
  \def\url#1{\texttt{#1}}\fi
\expandafter\ifx\csname urlprefix\endcsname\relax\def\urlprefix{URL }\fi
\providecommand{\bibinfo}[2]{#2}
\providecommand{\eprint}[2][]{\url{#2}}

\bibitem[{\citenamefont{Hasan and Kane}(2010)}]{Hasan2010:RMP}
\bibinfo{author}{\bibfnamefont{M.~Z.} \bibnamefont{Hasan}} \bibnamefont{and}
  \bibinfo{author}{\bibfnamefont{C.~L.} \bibnamefont{Kane}},
  \bibinfo{journal}{Rev. Mod. Phys.} \textbf{\bibinfo{volume}{82}},
  \bibinfo{pages}{3045} (\bibinfo{year}{2010}).

\bibitem[{\citenamefont{Qi and Zhang}(2011)}]{Qi2011:RMP}
\bibinfo{author}{\bibfnamefont{X.-L.} \bibnamefont{Qi}} \bibnamefont{and}
  \bibinfo{author}{\bibfnamefont{S.-C.} \bibnamefont{Zhang}},
  \bibinfo{journal}{Rev. Mod. Phys.} \textbf{\bibinfo{volume}{83}},
  \bibinfo{pages}{1057} (\bibinfo{year}{2011}).

\bibitem[{\citenamefont{Kane and Mele}(2005{\natexlab{a}})}]{Kane2005:PRL}
\bibinfo{author}{\bibfnamefont{C.~L.} \bibnamefont{Kane}} \bibnamefont{and}
  \bibinfo{author}{\bibfnamefont{E.~J.} \bibnamefont{Mele}},
  \bibinfo{journal}{Phys. Rev. Lett.} \textbf{\bibinfo{volume}{95}},
  \bibinfo{pages}{146802} (\bibinfo{year}{2005}{\natexlab{a}}).

\bibitem[{\citenamefont{Kane and Mele}(2005{\natexlab{b}})}]{Kane2005:PRL2}
\bibinfo{author}{\bibfnamefont{C.~L.} \bibnamefont{Kane}} \bibnamefont{and}
  \bibinfo{author}{\bibfnamefont{E.~J.} \bibnamefont{Mele}},
  \bibinfo{journal}{Phys. Rev. Lett.} \textbf{\bibinfo{volume}{95}},
  \bibinfo{pages}{226801} (\bibinfo{year}{2005}{\natexlab{b}}).

\bibitem[{\citenamefont{Bernevig et~al.}(2006)\citenamefont{Bernevig, Hughes,
  and Zhang}}]{Bernevig2006:Science}
\bibinfo{author}{\bibfnamefont{B.~A.} \bibnamefont{Bernevig}},
  \bibinfo{author}{\bibfnamefont{T.~L.} \bibnamefont{Hughes}},
  \bibnamefont{and} \bibinfo{author}{\bibfnamefont{S.-C.} \bibnamefont{Zhang}},
  \bibinfo{journal}{Science} \textbf{\bibinfo{volume}{314}},
  \bibinfo{pages}{1757} (\bibinfo{year}{2006}).

\bibitem[{\citenamefont{Bernevig and Zhang}(2006)}]{Bernevig2006:PRL}
\bibinfo{author}{\bibfnamefont{B.~A.} \bibnamefont{Bernevig}} \bibnamefont{and}
  \bibinfo{author}{\bibfnamefont{S.-C.} \bibnamefont{Zhang}},
  \bibinfo{journal}{Phys. Rev. Lett.} \textbf{\bibinfo{volume}{96}},
  \bibinfo{pages}{106802} (\bibinfo{year}{2006}).

\bibitem[{\citenamefont{Murakami}(2006)}]{Murakami2006:PRL}
\bibinfo{author}{\bibfnamefont{S.}~\bibnamefont{Murakami}},
  \bibinfo{journal}{Phys. Rev. Lett.} \textbf{\bibinfo{volume}{97}},
  \bibinfo{pages}{236805} (\bibinfo{year}{2006}).

\bibitem[{\citenamefont{Liu et~al.}(2008)\citenamefont{Liu, Hughes, Qi, Wang,
  and Zhang}}]{Liu2008:PRL}
\bibinfo{author}{\bibfnamefont{C.}~\bibnamefont{Liu}},
  \bibinfo{author}{\bibfnamefont{T.~L.} \bibnamefont{Hughes}},
  \bibinfo{author}{\bibfnamefont{X.-L.} \bibnamefont{Qi}},
  \bibinfo{author}{\bibfnamefont{K.}~\bibnamefont{Wang}}, \bibnamefont{and}
  \bibinfo{author}{\bibfnamefont{S.-C.} \bibnamefont{Zhang}},
  \bibinfo{journal}{Phys. Rev. Lett.} \textbf{\bibinfo{volume}{100}},
  \bibinfo{pages}{236601} (\bibinfo{year}{2008}).

\bibitem[{\citenamefont{K\"{o}nig et~al.}(2007)\citenamefont{K\"{o}nig,
  Wiedmann, Br\"{u}ne, Roth, Buhmann, Molenkamp, Qi, and
  Zhang}}]{Koenig2007:Science}
\bibinfo{author}{\bibfnamefont{M.}~\bibnamefont{K\"{o}nig}},
  \bibinfo{author}{\bibfnamefont{S.}~\bibnamefont{Wiedmann}},
  \bibinfo{author}{\bibfnamefont{C.}~\bibnamefont{Br\"{u}ne}},
  \bibinfo{author}{\bibfnamefont{A.}~\bibnamefont{Roth}},
  \bibinfo{author}{\bibfnamefont{H.}~\bibnamefont{Buhmann}},
  \bibinfo{author}{\bibfnamefont{L.~W.} \bibnamefont{Molenkamp}},
  \bibinfo{author}{\bibfnamefont{X.-L.} \bibnamefont{Qi}}, \bibnamefont{and}
  \bibinfo{author}{\bibfnamefont{S.-C.} \bibnamefont{Zhang}},
  \bibinfo{journal}{Science} \textbf{\bibinfo{volume}{318}},
  \bibinfo{pages}{766} (\bibinfo{year}{2007}).

\bibitem[{\citenamefont{K\"{o}nig et~al.}(2008)\citenamefont{K\"{o}nig,
  Buhmann, Molenkamp, Hughes, Liu, Qi, and Zhang}}]{Koenig2008:JPSJ}
\bibinfo{author}{\bibfnamefont{M.}~\bibnamefont{K\"{o}nig}},
  \bibinfo{author}{\bibfnamefont{H.}~\bibnamefont{Buhmann}},
  \bibinfo{author}{\bibfnamefont{L.~W.} \bibnamefont{Molenkamp}},
  \bibinfo{author}{\bibfnamefont{T.}~\bibnamefont{Hughes}},
  \bibinfo{author}{\bibfnamefont{C.-X.} \bibnamefont{Liu}},
  \bibinfo{author}{\bibfnamefont{X.-L.} \bibnamefont{Qi}}, \bibnamefont{and}
  \bibinfo{author}{\bibfnamefont{S.-C.} \bibnamefont{Zhang}},
  \bibinfo{journal}{Journal of the Physical Society of Japan}
  \textbf{\bibinfo{volume}{77}}, \bibinfo{pages}{031007}
  (\bibinfo{year}{2008}).

\bibitem[{\citenamefont{B\"{u}ttner et~al.}(2011)\citenamefont{B\"{u}ttner,
  Liu, Tkachov, Novik, Br\"{u}ne, Buhmann, Hankiewicz, Recher, Trauzettel,
  Zhang et~al.}}]{Buettner2011:NatPhys}
\bibinfo{author}{\bibfnamefont{B.}~\bibnamefont{B\"{u}ttner}},
  \bibinfo{author}{\bibfnamefont{C.~X.} \bibnamefont{Liu}},
  \bibinfo{author}{\bibfnamefont{G.}~\bibnamefont{Tkachov}},
  \bibinfo{author}{\bibfnamefont{E.~G.} \bibnamefont{Novik}},
  \bibinfo{author}{\bibfnamefont{C.}~\bibnamefont{Br\"{u}ne}},
  \bibinfo{author}{\bibfnamefont{H.}~\bibnamefont{Buhmann}},
  \bibinfo{author}{\bibfnamefont{E.~M.} \bibnamefont{Hankiewicz}},
  \bibinfo{author}{\bibfnamefont{P.}~\bibnamefont{Recher}},
  \bibinfo{author}{\bibfnamefont{B.}~\bibnamefont{Trauzettel}},
  \bibinfo{author}{\bibfnamefont{S.~C.} \bibnamefont{Zhang}},
  \bibnamefont{et~al.}, \bibinfo{journal}{Nature Physics}
  \textbf{\bibinfo{volume}{7}}, \bibinfo{pages}{418} (\bibinfo{year}{2011}).

\bibitem[{\citenamefont{Br\"{u}ne et~al.}(2012)\citenamefont{Br\"{u}ne, Roth,
  Buhmann, Hankiewicz, Molenkamp, Maciejko, Qi, and
  Zhang}}]{Bruene2012:NatPhys}
\bibinfo{author}{\bibfnamefont{C.}~\bibnamefont{Br\"{u}ne}},
  \bibinfo{author}{\bibfnamefont{A.}~\bibnamefont{Roth}},
  \bibinfo{author}{\bibfnamefont{H.}~\bibnamefont{Buhmann}},
  \bibinfo{author}{\bibfnamefont{E.~M.} \bibnamefont{Hankiewicz}},
  \bibinfo{author}{\bibfnamefont{L.~W.} \bibnamefont{Molenkamp}},
  \bibinfo{author}{\bibfnamefont{J.}~\bibnamefont{Maciejko}},
  \bibinfo{author}{\bibfnamefont{X.-L.} \bibnamefont{Qi}}, \bibnamefont{and}
  \bibinfo{author}{\bibfnamefont{S.-C.} \bibnamefont{Zhang}},
  \bibinfo{journal}{Nature Physics} \textbf{\bibinfo{volume}{8}},
  \bibinfo{pages}{486} (\bibinfo{year}{2012}).

\bibitem[{\citenamefont{Thouless et~al.}(1982)\citenamefont{Thouless, Kohmoto,
  Nightingale, and den Nijs}}]{Thouless1982:PRL}
\bibinfo{author}{\bibfnamefont{D.~J.} \bibnamefont{Thouless}},
  \bibinfo{author}{\bibfnamefont{M.}~\bibnamefont{Kohmoto}},
  \bibinfo{author}{\bibfnamefont{M.~P.} \bibnamefont{Nightingale}},
  \bibnamefont{and} \bibinfo{author}{\bibfnamefont{M.}~\bibnamefont{den Nijs}},
  \bibinfo{journal}{Phys. Rev. Lett.} \textbf{\bibinfo{volume}{49}},
  \bibinfo{pages}{405} (\bibinfo{year}{1982}).

\bibitem[{\citenamefont{Kohmoto}(1985)}]{Kohmoto1985:AoP}
\bibinfo{author}{\bibfnamefont{M.}~\bibnamefont{Kohmoto}},
  \bibinfo{journal}{Annals of Physics} \textbf{\bibinfo{volume}{160}},
  \bibinfo{pages}{343} (\bibinfo{year}{1985}).

\bibitem[{\citenamefont{Fu and Kane}(2007)}]{Fu2007:PRB}
\bibinfo{author}{\bibfnamefont{L.}~\bibnamefont{Fu}} \bibnamefont{and}
  \bibinfo{author}{\bibfnamefont{C.~L.} \bibnamefont{Kane}},
  \bibinfo{journal}{Phys. Rev. B} \textbf{\bibinfo{volume}{76}},
  \bibinfo{pages}{045302} (\bibinfo{year}{2007}).

\bibitem[{\citenamefont{Wu et~al.}(2006)\citenamefont{Wu, Bernevig, and
  Zhang}}]{Wu2006:PRL}
\bibinfo{author}{\bibfnamefont{C.}~\bibnamefont{Wu}},
  \bibinfo{author}{\bibfnamefont{B.~A.} \bibnamefont{Bernevig}},
  \bibnamefont{and} \bibinfo{author}{\bibfnamefont{S.-C.} \bibnamefont{Zhang}},
  \bibinfo{journal}{Phys. Rev. Lett.} \textbf{\bibinfo{volume}{96}},
  \bibinfo{pages}{106401} (\bibinfo{year}{2006}).

\bibitem[{\citenamefont{Xu and Moore}(2006)}]{Xu2006:PRB}
\bibinfo{author}{\bibfnamefont{C.}~\bibnamefont{Xu}} \bibnamefont{and}
  \bibinfo{author}{\bibfnamefont{J.~E.} \bibnamefont{Moore}},
  \bibinfo{journal}{Phys. Rev. B} \textbf{\bibinfo{volume}{73}},
  \bibinfo{pages}{045322} (\bibinfo{year}{2006}).

\bibitem[{\citenamefont{\ifmmode \check{Z}\else
  \v{Z}\fi{}uti\ifmmode~\acute{c}\else \'{c}\fi{}
  et~al.}(2004)\citenamefont{\ifmmode \check{Z}\else
  \v{Z}\fi{}uti\ifmmode~\acute{c}\else \'{c}\fi{}, Fabian, and
  Das~Sarma}}]{Zutic2004:RMP}
\bibinfo{author}{\bibfnamefont{I.}~\bibnamefont{\ifmmode \check{Z}\else
  \v{Z}\fi{}uti\ifmmode~\acute{c}\else \'{c}\fi{}}},
  \bibinfo{author}{\bibfnamefont{J.}~\bibnamefont{Fabian}}, \bibnamefont{and}
  \bibinfo{author}{\bibfnamefont{S.}~\bibnamefont{Das~Sarma}},
  \bibinfo{journal}{Rev. Mod. Phys.} \textbf{\bibinfo{volume}{76}},
  \bibinfo{pages}{323} (\bibinfo{year}{2004}).

\bibitem[{\citenamefont{Fabian et~al.}(2007)\citenamefont{Fabian,
  Matos-Abiague, Ertler, Stano, and \ifmmode \check{Z}\else
  \v{Z}\fi{}uti\ifmmode~\acute{c}\else \'{c}\fi{}}}]{Fabian2007:APS}
\bibinfo{author}{\bibfnamefont{J.}~\bibnamefont{Fabian}},
  \bibinfo{author}{\bibfnamefont{A.}~\bibnamefont{Matos-Abiague}},
  \bibinfo{author}{\bibfnamefont{C.}~\bibnamefont{Ertler}},
  \bibinfo{author}{\bibfnamefont{P.}~\bibnamefont{Stano}}, \bibnamefont{and}
  \bibinfo{author}{\bibfnamefont{I.}~\bibnamefont{\ifmmode \check{Z}\else
  \v{Z}\fi{}uti\ifmmode~\acute{c}\else \'{c}\fi{}}}, \bibinfo{journal}{Acta
  Phys. Slov.} \textbf{\bibinfo{volume}{57}}, \bibinfo{pages}{565}
  (\bibinfo{year}{2007}).

\bibitem[{\citenamefont{Chadi et~al.}(1972)\citenamefont{Chadi, Walter, Cohen,
  Petroff, and Balkanski}}]{Chadi1972:PRB}
\bibinfo{author}{\bibfnamefont{D.~J.} \bibnamefont{Chadi}},
  \bibinfo{author}{\bibfnamefont{J.~P.} \bibnamefont{Walter}},
  \bibinfo{author}{\bibfnamefont{M.~L.} \bibnamefont{Cohen}},
  \bibinfo{author}{\bibfnamefont{Y.}~\bibnamefont{Petroff}}, \bibnamefont{and}
  \bibinfo{author}{\bibfnamefont{M.}~\bibnamefont{Balkanski}},
  \bibinfo{journal}{Phys. Rev. B} \textbf{\bibinfo{volume}{5}},
  \bibinfo{pages}{3058} (\bibinfo{year}{1972}).

\bibitem[{\citenamefont{Zhu et~al.}(2012)\citenamefont{Zhu, Cheng, and
  Schwingenschl\"ogl}}]{Zhu2012:PRB}
\bibinfo{author}{\bibfnamefont{Z.}~\bibnamefont{Zhu}},
  \bibinfo{author}{\bibfnamefont{Y.}~\bibnamefont{Cheng}}, \bibnamefont{and}
  \bibinfo{author}{\bibfnamefont{U.}~\bibnamefont{Schwingenschl\"ogl}},
  \bibinfo{journal}{Phys. Rev. B} \textbf{\bibinfo{volume}{85}},
  \bibinfo{pages}{235401} (\bibinfo{year}{2012}).

\bibitem[{\citenamefont{D'yakonov and Khaetskii}(1981)}]{Dyakonov1981:JETPL}
\bibinfo{author}{\bibfnamefont{M.~I.} \bibnamefont{D'yakonov}}
  \bibnamefont{and} \bibinfo{author}{\bibfnamefont{A.~V.}
  \bibnamefont{Khaetskii}}, \bibinfo{journal}{JETP Letters}
  \textbf{\bibinfo{volume}{33}}, \bibinfo{pages}{110} (\bibinfo{year}{1981}).

\bibitem[{\citenamefont{Volkov and Pankratov}(1985)}]{Volkov1985:JETPL}
\bibinfo{author}{\bibfnamefont{B.~A.} \bibnamefont{Volkov}} \bibnamefont{and}
  \bibinfo{author}{\bibfnamefont{O.~A.} \bibnamefont{Pankratov}},
  \bibinfo{journal}{JETP Letters} \textbf{\bibinfo{volume}{42}},
  \bibinfo{pages}{178} (\bibinfo{year}{1985}).

\bibitem[{\citenamefont{Pankratov et~al.}(1987)\citenamefont{Pankratov,
  Pakhomov, and Volkov}}]{Pankratov1987:SolidStateCommunications}
\bibinfo{author}{\bibfnamefont{O.}~\bibnamefont{Pankratov}},
  \bibinfo{author}{\bibfnamefont{S.}~\bibnamefont{Pakhomov}}, \bibnamefont{and}
  \bibinfo{author}{\bibfnamefont{B.}~\bibnamefont{Volkov}},
  \bibinfo{journal}{Solid State Communications} \textbf{\bibinfo{volume}{61}},
  \bibinfo{pages}{93 } (\bibinfo{year}{1987}).

\bibitem[{\citenamefont{Rothe et~al.}(2010)\citenamefont{Rothe, Reinthaler,
  Liu, Molenkamp, Zhang, and Hankiewicz}}]{Rothe2010:NJoP}
\bibinfo{author}{\bibfnamefont{D.~G.} \bibnamefont{Rothe}},
  \bibinfo{author}{\bibfnamefont{R.~W.} \bibnamefont{Reinthaler}},
  \bibinfo{author}{\bibfnamefont{C.-X.} \bibnamefont{Liu}},
  \bibinfo{author}{\bibfnamefont{L.~W.} \bibnamefont{Molenkamp}},
  \bibinfo{author}{\bibfnamefont{S.-C.} \bibnamefont{Zhang}}, \bibnamefont{and}
  \bibinfo{author}{\bibfnamefont{E.~M.} \bibnamefont{Hankiewicz}},
  \bibinfo{journal}{New Journal of Physics} \textbf{\bibinfo{volume}{12}},
  \bibinfo{pages}{065012} (\bibinfo{year}{2010}).

\bibitem[{\citenamefont{Reinthaler and Hankiewicz}(2012)}]{Reinthaler2012:PRB}
\bibinfo{author}{\bibfnamefont{R.~W.} \bibnamefont{Reinthaler}}
  \bibnamefont{and} \bibinfo{author}{\bibfnamefont{E.~M.}
  \bibnamefont{Hankiewicz}}, \bibinfo{journal}{Phys. Rev. B}
  \textbf{\bibinfo{volume}{85}}, \bibinfo{pages}{165450}
  (\bibinfo{year}{2012}).

\bibitem[{\citenamefont{Tkachov and Hankiewicz}(2010)}]{Tkachov2010:PRL}
\bibinfo{author}{\bibfnamefont{G.}~\bibnamefont{Tkachov}} \bibnamefont{and}
  \bibinfo{author}{\bibfnamefont{E.~M.} \bibnamefont{Hankiewicz}},
  \bibinfo{journal}{Phys. Rev. Lett.} \textbf{\bibinfo{volume}{104}},
  \bibinfo{pages}{166803} (\bibinfo{year}{2010}).

\bibitem[{\citenamefont{Tkachov and Hankiewicz}(2012)}]{Tkachov2012:PhysicaE}
\bibinfo{author}{\bibfnamefont{G.}~\bibnamefont{Tkachov}} \bibnamefont{and}
  \bibinfo{author}{\bibfnamefont{E.}~\bibnamefont{Hankiewicz}},
  \bibinfo{journal}{Physica E: Low-dimensional Systems and Nanostructures}
  \textbf{\bibinfo{volume}{44}}, \bibinfo{pages}{900 } (\bibinfo{year}{2012}).

\bibitem[{\citenamefont{Chen et~al.}(2012)\citenamefont{Chen, Wang, and
  Sun}}]{Chen2012:PRB}
\bibinfo{author}{\bibfnamefont{J.-C.} \bibnamefont{Chen}},
  \bibinfo{author}{\bibfnamefont{J.}~\bibnamefont{Wang}}, \bibnamefont{and}
  \bibinfo{author}{\bibfnamefont{Q.-F.} \bibnamefont{Sun}},
  \bibinfo{journal}{Phys. Rev. B} \textbf{\bibinfo{volume}{85}},
  \bibinfo{pages}{125401} (\bibinfo{year}{2012}).

\bibitem[{\citenamefont{Zhou et~al.}(2008)\citenamefont{Zhou, Lu, Chu, Shen,
  and Niu}}]{Zhou2008:PRL}
\bibinfo{author}{\bibfnamefont{B.}~\bibnamefont{Zhou}},
  \bibinfo{author}{\bibfnamefont{H.-Z.} \bibnamefont{Lu}},
  \bibinfo{author}{\bibfnamefont{R.-L.} \bibnamefont{Chu}},
  \bibinfo{author}{\bibfnamefont{S.-Q.} \bibnamefont{Shen}}, \bibnamefont{and}
  \bibinfo{author}{\bibfnamefont{Q.}~\bibnamefont{Niu}},
  \bibinfo{journal}{Phys. Rev. Lett.} \textbf{\bibinfo{volume}{101}},
  \bibinfo{pages}{246807} (\bibinfo{year}{2008}).

\bibitem[{\citenamefont{Krueckl and Richter}(2011)}]{Krueckl2011:PRL}
\bibinfo{author}{\bibfnamefont{V.}~\bibnamefont{Krueckl}} \bibnamefont{and}
  \bibinfo{author}{\bibfnamefont{K.}~\bibnamefont{Richter}},
  \bibinfo{journal}{Phys. Rev. Lett.} \textbf{\bibinfo{volume}{107}},
  \bibinfo{pages}{086803} (\bibinfo{year}{2011}).

\bibitem[{\citenamefont{Linder et~al.}(2009)\citenamefont{Linder, Yokoyama, and
  Sudb\o{}}}]{Linder2009:PRB}
\bibinfo{author}{\bibfnamefont{J.}~\bibnamefont{Linder}},
  \bibinfo{author}{\bibfnamefont{T.}~\bibnamefont{Yokoyama}}, \bibnamefont{and}
  \bibinfo{author}{\bibfnamefont{A.}~\bibnamefont{Sudb\o{}}},
  \bibinfo{journal}{Phys. Rev. B} \textbf{\bibinfo{volume}{80}},
  \bibinfo{pages}{205401} (\bibinfo{year}{2009}).

\bibitem[{\citenamefont{Liu et~al.}(2010)\citenamefont{Liu, Zhang, Yan, Qi,
  Frauenheim, Dai, Fang, and Zhang}}]{Liu2010:PRB}
\bibinfo{author}{\bibfnamefont{C.-X.} \bibnamefont{Liu}},
  \bibinfo{author}{\bibfnamefont{H.}~\bibnamefont{Zhang}},
  \bibinfo{author}{\bibfnamefont{B.}~\bibnamefont{Yan}},
  \bibinfo{author}{\bibfnamefont{X.-L.} \bibnamefont{Qi}},
  \bibinfo{author}{\bibfnamefont{T.}~\bibnamefont{Frauenheim}},
  \bibinfo{author}{\bibfnamefont{X.}~\bibnamefont{Dai}},
  \bibinfo{author}{\bibfnamefont{Z.}~\bibnamefont{Fang}}, \bibnamefont{and}
  \bibinfo{author}{\bibfnamefont{S.-C.} \bibnamefont{Zhang}},
  \bibinfo{journal}{Phys. Rev. B} \textbf{\bibinfo{volume}{81}},
  \bibinfo{pages}{041307} (\bibinfo{year}{2010}).

\bibitem[{\citenamefont{Lu et~al.}(2010)\citenamefont{Lu, Shan, Yao, Niu, and
  Shen}}]{Lu2010:PRB}
\bibinfo{author}{\bibfnamefont{H.-Z.} \bibnamefont{Lu}},
  \bibinfo{author}{\bibfnamefont{W.-Y.} \bibnamefont{Shan}},
  \bibinfo{author}{\bibfnamefont{W.}~\bibnamefont{Yao}},
  \bibinfo{author}{\bibfnamefont{Q.}~\bibnamefont{Niu}}, \bibnamefont{and}
  \bibinfo{author}{\bibfnamefont{S.-Q.} \bibnamefont{Shen}},
  \bibinfo{journal}{Phys. Rev. B} \textbf{\bibinfo{volume}{81}},
  \bibinfo{pages}{115407} (\bibinfo{year}{2010}).

\bibitem[{\citenamefont{Grigoryan et~al.}(2009)\citenamefont{Grigoryan,
  Matos-Abiague, and Badalyan}}]{Grigoryan2009:PRB}
\bibinfo{author}{\bibfnamefont{V.~L.} \bibnamefont{Grigoryan}},
  \bibinfo{author}{\bibfnamefont{A.}~\bibnamefont{Matos-Abiague}},
  \bibnamefont{and} \bibinfo{author}{\bibfnamefont{S.~M.}
  \bibnamefont{Badalyan}}, \bibinfo{journal}{Phys. Rev. B}
  \textbf{\bibinfo{volume}{80}}, \bibinfo{pages}{165320}
  (\bibinfo{year}{2009}).

\bibitem[{\citenamefont{Badalyan and Fabian}(2010)}]{Badalyan2009:PRL}
\bibinfo{author}{\bibfnamefont{S.~M.} \bibnamefont{Badalyan}} \bibnamefont{and}
  \bibinfo{author}{\bibfnamefont{J.}~\bibnamefont{Fabian}},
  \bibinfo{journal}{Phys. Rev. Lett.} \textbf{\bibinfo{volume}{105}},
  \bibinfo{pages}{186601} (\bibinfo{year}{2010}).

\bibitem[{\citenamefont{Schmidt et~al.}(2009)\citenamefont{Schmidt, Novik,
  Kindermann, and Trauzettel}}]{Schmidt2009:PRB}
\bibinfo{author}{\bibfnamefont{M.~J.} \bibnamefont{Schmidt}},
  \bibinfo{author}{\bibfnamefont{E.~G.} \bibnamefont{Novik}},
  \bibinfo{author}{\bibfnamefont{M.}~\bibnamefont{Kindermann}},
  \bibnamefont{and}
  \bibinfo{author}{\bibfnamefont{B.}~\bibnamefont{Trauzettel}},
  \bibinfo{journal}{Phys. Rev. B} \textbf{\bibinfo{volume}{79}},
  \bibinfo{pages}{241306} (\bibinfo{year}{2009}).

\bibitem[{\citenamefont{Olver et~al.}(2010)\citenamefont{Olver, Lozier,
  Boisvert, and Clark}}]{Olver2010}
\bibinfo{author}{\bibfnamefont{F.~W.~J.} \bibnamefont{Olver}},
  \bibinfo{author}{\bibfnamefont{D.~W.} \bibnamefont{Lozier}},
  \bibinfo{author}{\bibfnamefont{R.~F.} \bibnamefont{Boisvert}},
  \bibnamefont{and} \bibinfo{author}{\bibfnamefont{C.~W.} \bibnamefont{Clark}},
  \emph{\bibinfo{title}{NIST Handbook of Mathematical Functions}}
  (\bibinfo{publisher}{Cambridge Univ. Press, New York}, \bibinfo{year}{2010}).

\bibitem[{\citenamefont{Datta}(2007)}]{Datta2007}
\bibinfo{author}{\bibfnamefont{S.}~\bibnamefont{Datta}},
  \emph{\bibinfo{title}{Electronic transport in mesoscopic systems}}
  (\bibinfo{publisher}{Cambridge Univ. Press, Cambridge},
  \bibinfo{year}{2007}).

\bibitem[{\citenamefont{Peierls}(1933)}]{Peierls1933:ZfP}
\bibinfo{author}{\bibfnamefont{R.}~\bibnamefont{Peierls}},
  \bibinfo{journal}{Zeitschrift f{\"u}r Physik A Hadrons and Nuclei}
  \textbf{\bibinfo{volume}{80}}, \bibinfo{pages}{763} (\bibinfo{year}{1933}).

\bibitem[{\citenamefont{Meyer et~al.}(1990)\citenamefont{Meyer, Wagner,
  Bartoli, Hoffman, Dobrowolska, Wojtowicz, Furdyna, and
  Ram-Mohan}}]{Meyer1990:PRB}
\bibinfo{author}{\bibfnamefont{J.~R.} \bibnamefont{Meyer}},
  \bibinfo{author}{\bibfnamefont{R.~J.} \bibnamefont{Wagner}},
  \bibinfo{author}{\bibfnamefont{F.~J.} \bibnamefont{Bartoli}},
  \bibinfo{author}{\bibfnamefont{C.~A.} \bibnamefont{Hoffman}},
  \bibinfo{author}{\bibfnamefont{M.}~\bibnamefont{Dobrowolska}},
  \bibinfo{author}{\bibfnamefont{T.}~\bibnamefont{Wojtowicz}},
  \bibinfo{author}{\bibfnamefont{J.~K.} \bibnamefont{Furdyna}},
  \bibnamefont{and} \bibinfo{author}{\bibfnamefont{L.~R.}
  \bibnamefont{Ram-Mohan}}, \bibinfo{journal}{Phys. Rev. B}
  \textbf{\bibinfo{volume}{42}}, \bibinfo{pages}{9050} (\bibinfo{year}{1990}).

\bibitem[{\citenamefont{von Truchsess et~al.}(1997)\citenamefont{von Truchsess,
  Pfeuffer-Jeschke, Latussek, Becker, and Batke}}]{Truchsess1997:HMFPS}
\bibinfo{author}{\bibfnamefont{M.}~\bibnamefont{von Truchsess}},
  \bibinfo{author}{\bibfnamefont{A.}~\bibnamefont{Pfeuffer-Jeschke}},
  \bibinfo{author}{\bibfnamefont{V.}~\bibnamefont{Latussek}},
  \bibinfo{author}{\bibfnamefont{C.~R.} \bibnamefont{Becker}},
  \bibnamefont{and} \bibinfo{author}{\bibfnamefont{E.}~\bibnamefont{Batke}}, in
  \emph{\bibinfo{booktitle}{High Magnetic Fields in the Physics of
  Semiconductors II}}, edited by
  \bibinfo{editor}{\bibfnamefont{G.}~\bibnamefont{Landwehr}} \bibnamefont{and}
  \bibinfo{editor}{\bibfnamefont{W.}~\bibnamefont{Ossau}}
  (\bibinfo{publisher}{World Scientific, Singapore}, \bibinfo{year}{1997}),
  vol.~\bibinfo{volume}{2}, p. \bibinfo{pages}{813}.

\bibitem[{\citenamefont{Schultz et~al.}(1998)\citenamefont{Schultz, Merkt,
  Sonntag, R\"ossler, Winkler, Colin, Helgesen, Skauli, and
  L\o{}vold}}]{Schultz1998:PRB}
\bibinfo{author}{\bibfnamefont{M.}~\bibnamefont{Schultz}},
  \bibinfo{author}{\bibfnamefont{U.}~\bibnamefont{Merkt}},
  \bibinfo{author}{\bibfnamefont{A.}~\bibnamefont{Sonntag}},
  \bibinfo{author}{\bibfnamefont{U.}~\bibnamefont{R\"ossler}},
  \bibinfo{author}{\bibfnamefont{R.}~\bibnamefont{Winkler}},
  \bibinfo{author}{\bibfnamefont{T.}~\bibnamefont{Colin}},
  \bibinfo{author}{\bibfnamefont{P.}~\bibnamefont{Helgesen}},
  \bibinfo{author}{\bibfnamefont{T.}~\bibnamefont{Skauli}}, \bibnamefont{and}
  \bibinfo{author}{\bibfnamefont{S.}~\bibnamefont{L\o{}vold}},
  \bibinfo{journal}{Phys. Rev. B} \textbf{\bibinfo{volume}{57}},
  \bibinfo{pages}{14772} (\bibinfo{year}{1998}).

\bibitem[{\citenamefont{Sharapov et~al.}(2004)\citenamefont{Sharapov, Gusynin,
  and Beck}}]{Sharapov2004:PRB}
\bibinfo{author}{\bibfnamefont{S.~G.} \bibnamefont{Sharapov}},
  \bibinfo{author}{\bibfnamefont{V.~P.} \bibnamefont{Gusynin}},
  \bibnamefont{and} \bibinfo{author}{\bibfnamefont{H.}~\bibnamefont{Beck}},
  \bibinfo{journal}{Phys. Rev. B} \textbf{\bibinfo{volume}{69}},
  \bibinfo{pages}{075104} (\bibinfo{year}{2004}).

\bibitem[{\citenamefont{Koshino and Ando}(2007)}]{Koshino2007:PRB}
\bibinfo{author}{\bibfnamefont{M.}~\bibnamefont{Koshino}} \bibnamefont{and}
  \bibinfo{author}{\bibfnamefont{T.}~\bibnamefont{Ando}},
  \bibinfo{journal}{Phys. Rev. B} \textbf{\bibinfo{volume}{75}},
  \bibinfo{pages}{235333} (\bibinfo{year}{2007}).

\bibitem[{\citenamefont{Koshino and Ando}(2010)}]{Koshino2010:PRB}
\bibinfo{author}{\bibfnamefont{M.}~\bibnamefont{Koshino}} \bibnamefont{and}
  \bibinfo{author}{\bibfnamefont{T.}~\bibnamefont{Ando}},
  \bibinfo{journal}{Phys. Rev. B} \textbf{\bibinfo{volume}{81}},
  \bibinfo{pages}{195431} (\bibinfo{year}{2010}).

\bibitem[{\citenamefont{Ominato and Koshino}(2012)}]{Ominato2012:PRB}
\bibinfo{author}{\bibfnamefont{Y.}~\bibnamefont{Ominato}} \bibnamefont{and}
  \bibinfo{author}{\bibfnamefont{M.}~\bibnamefont{Koshino}},
  \bibinfo{journal}{Phys. Rev. B} \textbf{\bibinfo{volume}{85}},
  \bibinfo{pages}{165454} (\bibinfo{year}{2012}).

\bibitem[{\citenamefont{Landau and Lifshitz}(1999)}]{LandauLifshitz1999}
\bibinfo{author}{\bibfnamefont{L.~D.} \bibnamefont{Landau}} \bibnamefont{and}
  \bibinfo{author}{\bibfnamefont{E.~M.} \bibnamefont{Lifshitz}},
  \emph{\bibinfo{title}{Statistical Physics (3rd Edition Part 1)}}
  (\bibinfo{publisher}{Butterworth-Heinemann, Oxford}, \bibinfo{year}{1999}).

\bibitem[{\citenamefont{{Cheremisin}}(2011)}]{Cheremisin2011:arXiv}
\bibinfo{author}{\bibfnamefont{M.~V.} \bibnamefont{{Cheremisin}}},
  \bibinfo{journal}{ArXiv e-prints}  (\bibinfo{year}{2011}),
  \eprint{1110.5778}.

\end{thebibliography}

\end{document}